\documentclass[final,12pt]{clear_arxiv} 

\usepackage[utf8]{inputenc} 
\usepackage[T1]{fontenc}    
\usepackage{hyperref}       
\usepackage{booktabs}       
\usepackage{amsfonts}       
\usepackage{nicefrac}       
\usepackage{microtype}      
\usepackage{xcolor}         
\usepackage{hyperref}
\usepackage{mwe}
\usepackage{float}
\usepackage{adjustbox}
\usepackage{mwe}
\usepackage{mathtools}
\usepackage{bbm}
\usepackage{times}

\title[Hyperparameter Tuning with Double Machine Learning]{Hyperparameter Tuning for Causal Inference with Double Machine Learning: A Simulation Study}

\clearauthor{\Name{Philipp Bach} \Email{philipp.bach@uni-hamburg.de}\\
 \addr University of Hamburg, Germany \\
\Name{Oliver Schacht} \Email{oliver.schacht@uni-hamburg.de}\\
 \addr University of Hamburg, Germany \\
\Name{Victor Chernozhukov} \Email{vchern@mit.edu}\\
\addr Massachusetts Institute of Technology \\
\Name{Sven Klaassen} \Email{sven.klaassen@uni-hamburg.de}\\
 \addr University of Hamburg, Germany \\
 \addr Economic AI \\
 \Name{Martin Spindler} \Email{martin.spindler@uni-hamburg.de}\\
 \addr University of Hamburg, Germany\\
 \addr Economic AI }
 
\begin{document}
\maketitle

\begin{abstract}
    Proper hyperparameter tuning is essential for achieving optimal performance of modern machine learning (ML) methods in predictive tasks. While there is an extensive literature on tuning ML learners for prediction, there is only little guidance available on tuning ML learners for causal machine learning and how to select among different ML learners. In this paper, we empirically assess the relationship between the predictive performance of ML methods and the resulting causal estimation based on the Double Machine Learning (DML) approach by Chernozhukov et al.~(2018). DML relies on estimating so-called nuisance parameters by treating them as supervised learning problems and using them as plug-in estimates to solve for the (causal) parameter. We conduct an extensive simulation study using data from the 2019 Atlantic Causal Inference Conference Data Challenge. We provide empirical insights on the role of hyperparameter tuning and other practical decisions for causal estimation with DML. First, we assess the importance of data splitting schemes for tuning ML learners within Double Machine Learning. Second, we investigate how the choice of ML methods and hyperparameters, including recent AutoML frameworks, impacts the estimation performance for a causal parameter of interest. Third, we assess to what extent the choice of a particular causal model, as characterized by incorporated parametric assumptions, can be based on predictive performance metrics.
\end{abstract}


\section{Introduction}
Double/Debiased machine learning (DML) is an estimation framework for causal parameters based on ML-estimated nuisance functions that has been established by \citet{chernozhukov2018double}. DML combines the strengths of machine learning for prediction with estimation and inference of causal parameters. The major idea of the double machine learning framework is to make the estimation framework robust to inherent biases of ML estimation: To address the bias-variance tradeoff, ML methods introduce some regularization. Without any further adaption of the estimation procedure this will effectively translate into a bias of the causal parameter of interest. To overcome these shortcomings, the double machine learning approach combines three key ingredients \citep{bach2022, bach2021R}: (i) Identification of causal parameters through Neyman-orthogonal moment conditions, (ii) high-quality machine learning estimators and (iii) sample splitting. Incorporating these ingredients makes it possible to establish $\sqrt{N}$ convergence and asymptotic normality of the causal estimator. In recent years, the DML framework has become popular in various disciplines, including econometrics \citep{Knaus_2022}, reinforcement learning \citep{narita2020off} and management science \citep{schacht2023}. Whereas there is an extensive literature and ample benchmark studies on hyperparameter tuning approaches in predictive ML tasks, for example see \citet{bischl2021}, there is a considerable gap on the interaction between the predictive performance and the quality in terms of causal estimation. Several studies investigate model selection strategies for different estimation approaches that are based on ML\textcolor{black}{, mostly with a focus on heterogeneous treatment effects} \citep{schuler2018comparison, mahajan2022empirical}. 
\begin{figure}[t]
    \centering
    \subfigure{\includegraphics[width=.49\textwidth]{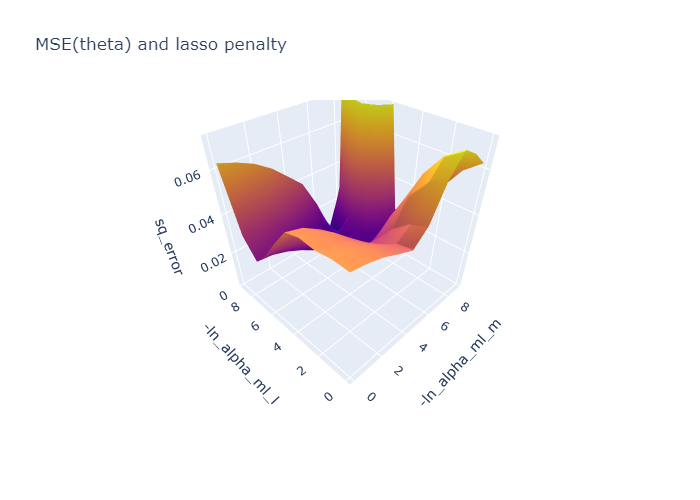}}
    \subfigure{\includegraphics[width=.49\textwidth]{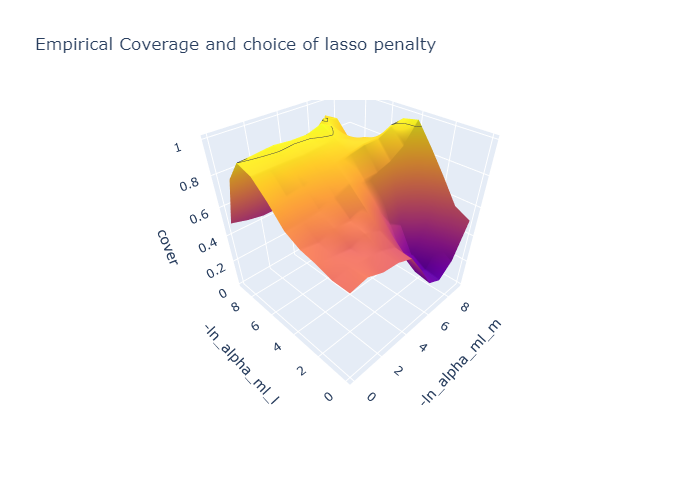}}
    \subfigure{\includegraphics[width=.45\textwidth]{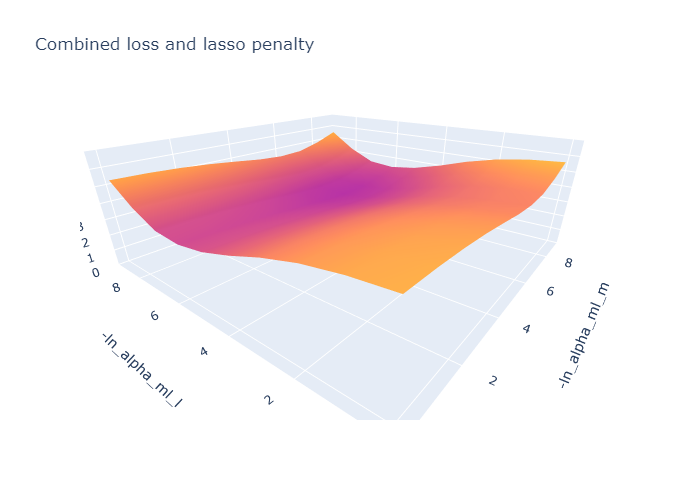}}
    \caption{Average results from 100 repetitions for a simulated data example, DGP from \citet{belloni2014} with $n=100$ observations and $p=200$ explanatory variables. The figures show various metrics as obtained for different combinations of the lasso penalty for the nuisance components $\eta_0=(\ell_0, m_0)$. Panel (a): Mean squared error, Panel (b): Empirical coverage of true parameter, Panel (c): Combined loss for PLR. An ineractive version of the figures is available at \url{https://docs.doubleml.org/stable/examples/py_double_ml_firststage.html}}\label{fig:3d}
\end{figure}
\textcolor{black}{Previous studies with a specific focus on DML compare different sample splitting schemes \citep{okasa2022metalearners} or different ML learners \citep{knausMC} for estimation of heterogeneous treatment effects based on synthetic or semi-synthetic data}. However, the question of how to select learners within the DML framework remains unclear in the existing literature. Few studies address the aspect of hyperparameter tuning in the context of double machine learning.
The approach by \cite{cdml} is based on a target metric that combines the predictive losses from two predictive tasks incorporated in partially linear regression models.  \cite{chernozhukov2022riesznet} provide a multitask framework for neural networks that is based on a general Riesz representation.
In this work, we will perform a thorough empirical study to show that hyperparameter tuning is required when using DML in order to achieve optimal inference on the target parameter. We assess different data splitting schemes for the tuning process, draw connections between the predictive performance of the ML learners and the error of the causal estimate and give recommendations on the selection of the causal model.
\section{Problem Setting: Learners, Hyperparameters and Sample Splitting}  \label{problem}
\subsection{The role of learners in double machine learning} \label{dmlintro}
In DML, the goal is estimation and inference on a target parameter $\theta_0$ in the presence of a high-dimensional nuisance parameter. Importantly, identification in DML is based on an orthogonal moment condition with a score function $\psi(W;\theta,\eta)$, namely
\begin{equation*}
    E[\psi(W;\theta_0,\eta_0)] = 0,
\end{equation*}
with data $W$, true value of the causal parameter $\theta_0$ is and nuisance function $\eta_0$. The general idea is to use ML methods to estimate each of the nuisance parameters, plug them into the score function and solve for the target parameter. A key condition for valid inference is so-called Neyman-orthogonality
\begin{equation*}
    \partial_\eta \mathbb{E}[\psi(W;\theta_0,\eta)]|_{\eta = \eta_0} = 0.
\end{equation*}
$\partial_\eta$ denotes the pathwise Gateaux derivative operator. Neyman-orthogonality ensures that the moment condition identifying $\theta_0$ is insensitive to small perturbations of the nuisance function $\eta$ around $\eta_0$ under relatively general assumptions. This eliminates the first-order bias which might stem from regularization, when replacing $\eta_0$ by the ML estimate $\hat{\eta}_0$. Formally, the learner requirement applies to the rates of convergence of the employed ML prediction methods. If the used ML estimators do  converge sufficiently fast, the bias in $\hat{\eta}_0$ will eventually vanish in the limit, giving rise to a $\sqrt{N}$-consistent and normally distributed estimate $\hat{\theta}_0$. The theoretical framework of \citet{chernozhukov2018double} specifies structural assumptions that guides the choice of the ML learners. For example, under the assumption of sparsity, $\ell_1$ penalized estimators such as the lasso \citep{tibshirani1996} are known to satisfy the rate criterion for estimating $\eta_0$, such that it can be used for causal estimation of $\theta_0$. Generally, the theoretical criterion on the learners refers to the error in the composed nuisance term $\eta_0$, which collects multiple prediction problems in a so-called \textit{combined loss}, which is specific to a particular causal model and orthogonal score formulation \citep{chernozhukov2018double}.
\begin{figure}[t]
    \centering
     \subfigure{\includegraphics[width=0.45\linewidth]{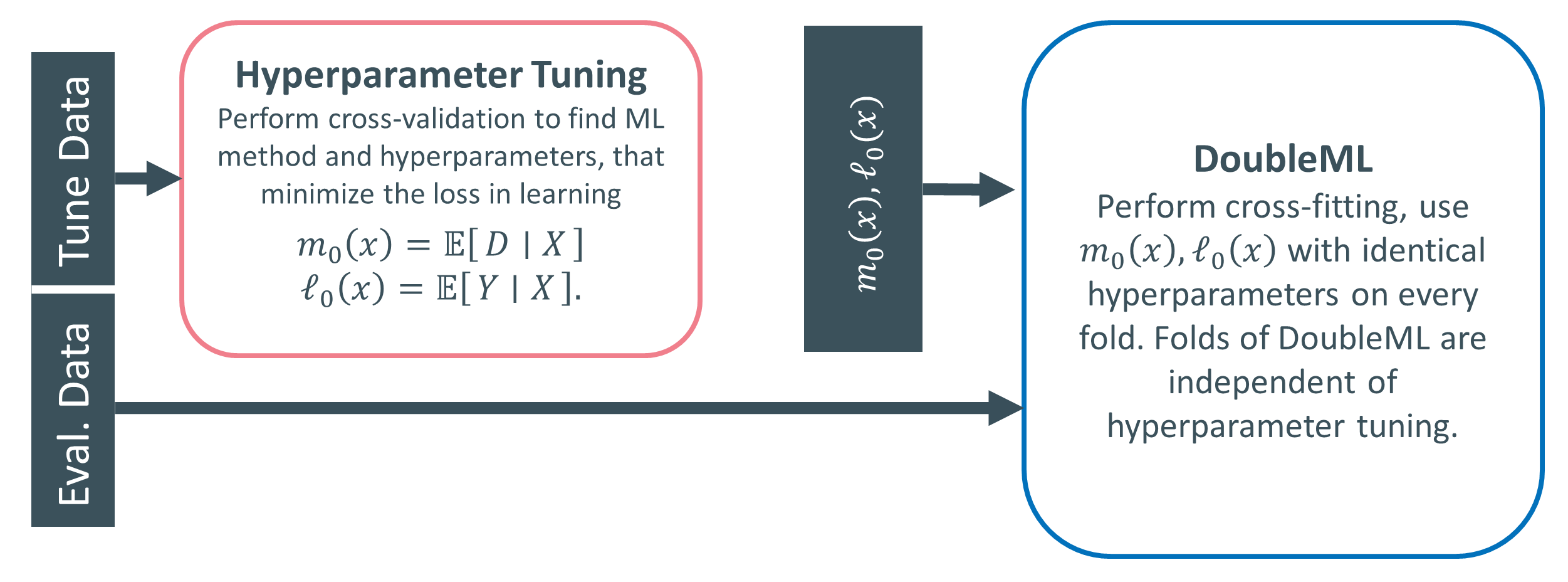}}\hfill
     \subfigure{\includegraphics[width=0.45\linewidth]{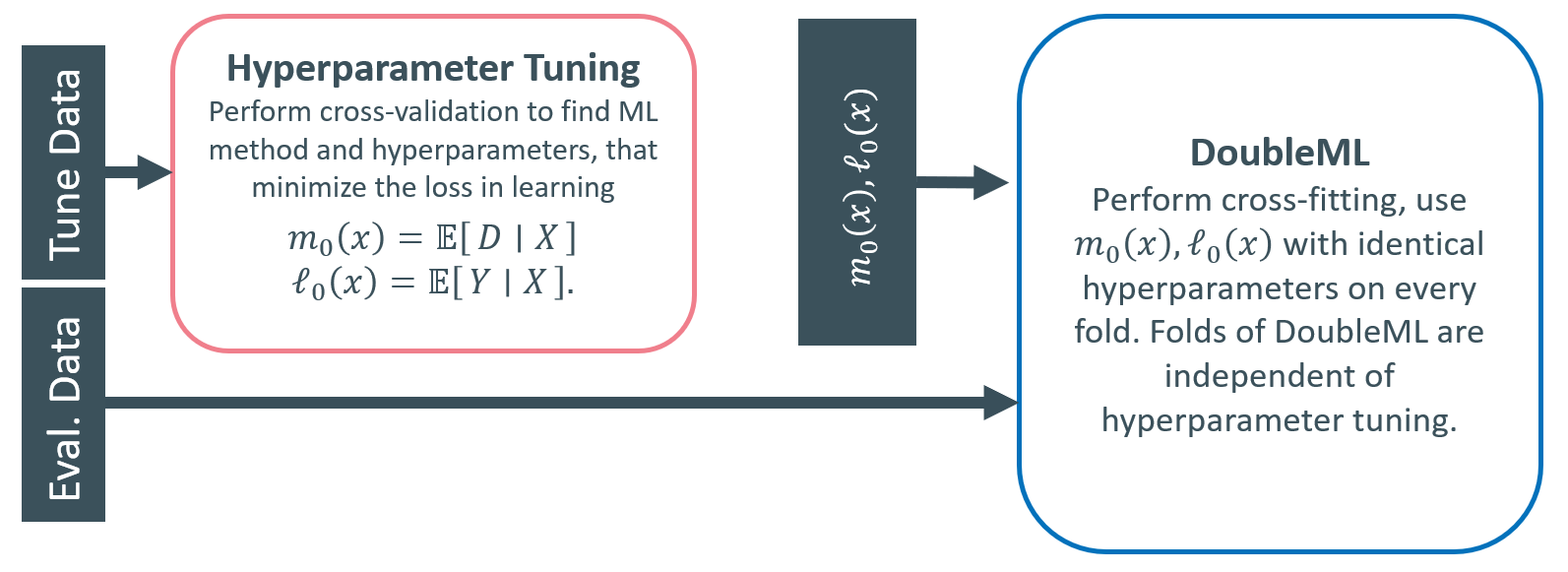}}
     \subfigure{\includegraphics[width=0.4\linewidth]{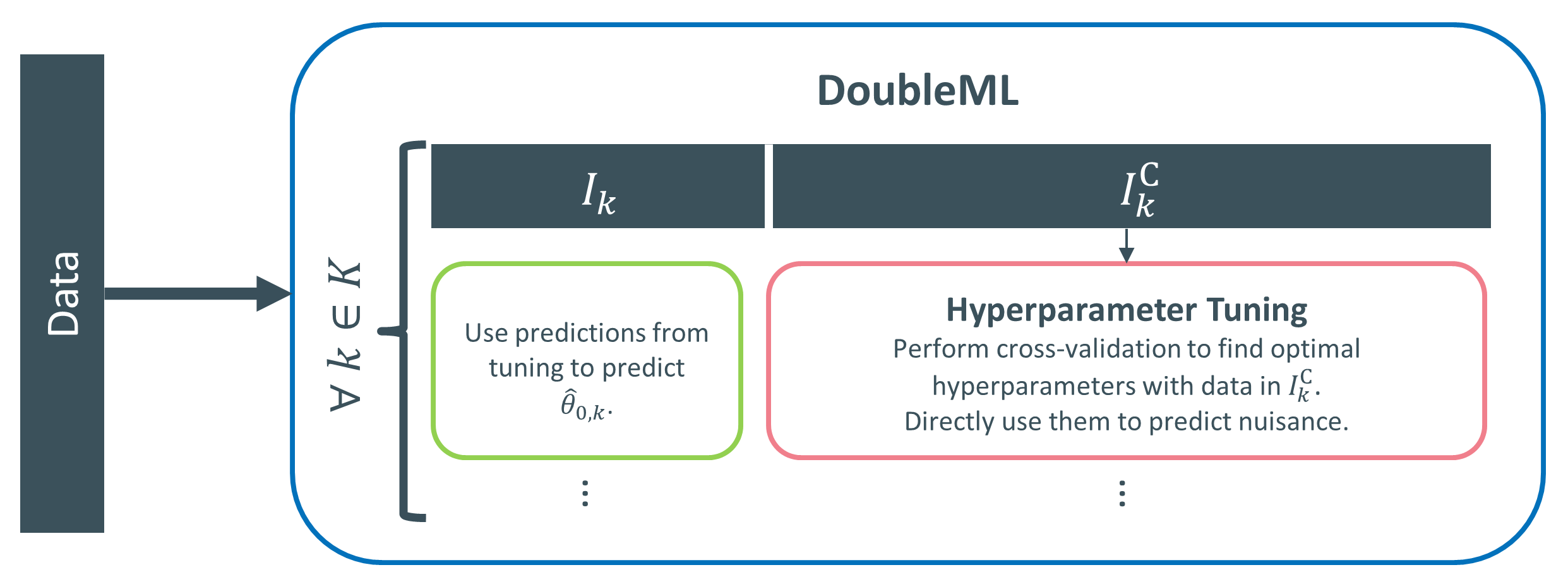}}    
    \caption{Schematic representation of sample splitting approaches for hyperparameter tuning: Panel (a) \textit{full sample}, (b) \textit{split sample}, (c) \textit{on folds}.}
    \label{fig:schemes}
\end{figure}
\subsection{Choice of hyperparameters in double machine learning}
\textcolor{black}{How to select among different ML learners for the nuisance part, is key for practical applications and understanding this better has been a main motivation for this study.}\\
To make the role of the learners in DML more tractable, we will briefly introduce two causal models that will also be used in our simulation study: The Partially Linear Regression (PLR) model and the Interactive Regression Model (IRM). Under common assumptions for identification such as unconfoundedness, the PLR can be used to estimate the Average Treatment Effect (ATE) of a binary or continuous treatment variable $D$ on an outcome $Y$. The PLR incorporates a partially linear additive structure, i.e., the causal effect is constant for all individuals
\begin{align*}
    Y &= D\theta_0 + \ell_0(X) + \xi, \quad \mathbb{E}[\xi \mid X,D] = 0\\
    D &= m_0(X) + V , \quad\mathbb{E}[V\mid X] = 0
\end{align*}
Using the so-called partialling out score, the corresponding nuisance component comprises two functions, $\eta_0 = (\ell_0, m_0) = \left(\mathbb{E}[Y|X], \mathbb{E}[D|X]\right)$.
The IRM imposes weaker functional form assumptions and makes it possible to estimate heterogeneous treatment effects. Thus, the nuisance component $g_0$ can also model interactions between $D$ and $Y$.
\begin{align*}
    Y &= g_0(D, X) + \xi , \quad \mathbb{E}[\xi \mid X,D] = 0\\
    D &= m_0(X) + V , \quad \mathbb{E}[V\mid X] = 0
\end{align*}
Using the doubly robust score which identifies the ATE of a binary treatment variable $D$ on $Y$, the nuisance parameter comprises two functions $\hat{\eta}_0 = (g_0, m_0) = \left(\mathbb{E}[Y|D,X], \mathbb{E}[D|X]\right)$. 
\textcolor{black}{Double machine learning is closely related to “double robustness” and DML-based estimators build on doubly robust scores (\cite{Robins1994}) and their augmented inverse probability weighting estimators. Doubly robust estimators, which rely on two nuisance parameters (outcome regression and propensity score), are robust against misspecification of one of the nuisance parameters. Hence, whenever it is possible to estimate one nuisance component at the parametric rate, DR estimators tolerate slower rates for the other nuisance component. DML generalizes this idea to a broader class of causal models and formulates the rate requirement in terms of a model-specific \textit{combined loss}. For the PLR and IRM, their definition is provided in Table \ref{tab:metrics}}. For a detailed discussion we refer to \cite{Knaus_2022} and \cite{Smucler2019}. Finally, we would like to note that DML is also closely related to targeted maximum likelihood learning \citep{vanderLaanRubin+2006}.\\
Figure \ref{fig:3d} serves as a simplified illustration of the role of an appropriate hyperparameter choice for the causal estimation quality. It shows the mean squared error with respect to the true causal parameter $\theta_0$ in Panel (a), \textcolor{black}{the empirical coverage in Panel (b) and the combined loss in the PLR in Panel (c) as obtained for a grid of lasso-penalty values}. The example is based on a high-dimensional sparse data generating process with $n=100$ observations and $p=200$ explanatory variables as presented in \citet{belloni2014}, which we refer to as BCH. In a PLR model, the lasso can be used as an appropriate learner for the sparse nuisance relationships $\eta_0 = (\ell_0, m_0)$. The surface plots illustrate that an appropriate choice of the hyperparameter, i.e., the lasso penalty $\lambda$, is essential for a precise estimator of $\theta_0$.
The figures show that it is important to have suitable estimation performance in both nuisance components $\ell_0$ and $m_0$ in order to obtain a low mean squared error for the causal parameter as well as to achieve the nominal coverage of the confidence intervals. \textcolor{black}{In line with the theoretical foundation of the DML approach, Panel (c) of Figure \ref{fig:3d} reveals that a lower combined loss for the nuisance component corresponds to a better estimation performance and higher coverage in Panel (a) and (b). An additional motivating example that highlights the importance of an appropriate hyperparameter choice in terms of the empirical distribution of the DML estimator is provided in Appendix \ref{app:addmotiv}.}
\begin{table}[t]
    \centering
    \resizebox{.75\linewidth}{!}{\begin{tabular}{p{0.2\linewidth}|p{0.4\linewidth}|p{0.4\linewidth}|p{0.4\linewidth}}
        Learner & Implementation & \textcolor{black}{Tuning Procedure and Parameters}& \textcolor{black}{Fixed Parameters} \\
        \hline
        Lasso & \texttt{Scikit-learn} \citep{scikit-learn} & \textcolor{black}{Internal} cross-validated fitting of $\ell_1$ regularization parameter & \textcolor{black}{5-fold CV}\\
        Random Forest & \texttt{Scikit-learn} \citep{scikit-learn} & \textcolor{black}{Cross-validated grid search for tree depth (4,5,6).} & $100$ trees, \textcolor{black}{5-fold CV}\\
        Extreme Gradient Boosting & \texttt{xgboost} \citep{xgboost} & \textcolor{black}{Cross-validated grid search for number of trees (up to 100 with early stopping)} & Tree-depth $2$, learning rate $0.1$, \textcolor{black}{5-fold CV} \\
        AutoML & \texttt{FLAML} \citep{wang2019flaml} & Computation time of 60s\footnote{For the ''on folds'' tuning scheme, we equally devide the tuning time on all folds.} & 
    \end{tabular}}
    \caption{Overview of the learners and tuning grids used in the simulation study.}
    \label{tab:learneroverview}
\end{table}
\subsection{Cross-fitting and forms of data splitting} \label{splits}
In addition to using an orthogonal score function for identification and appropriate ML estimation of the nuisance part, DML incorporates sample-splitting as a third ingredient. Splitting the sample in different partitions makes it possible to abstract from (small) overfitting biases. Sample splitting guarantees the independence of the samples that are (i) used to learn the nuisance functions $\eta_0$ (training data) and (ii) used to solve the orthogonal score with regard to the causal parameter $\theta_0$ (holdout data). Cross-fitting is an efficient form of data splitting, that uses all the available data by swapping the roles of the training and holdout data in a cross-validated manner. Cross-fitting proceeds as \citep{chernozhukov2018double}: (1) Take a $K$-fold random partition $(I_k)_{k=1}^K$ of observation indices $\{1,\dots,N\}$ such that the size of each fold $I_k$ is $n=N/K$. For each $\{1,\dots,K\}$, define $I^{\textrm{C}}_k\coloneqq\{1,\dots,N\}\setminus I_k$. (2) For each $k \in K$, construct an ML estimator $\hat{\eta}_{0,k}$ based on the observations $i \in I^{\textrm{C}}_k$. (3) For each $k \in K$, construct the estimator $\hat{\theta}_0$ using the empirical expectation of the estimator using the observations $i \in I_k$ and average over the $K$ folds. This procedure refers to fitting the nuisance learners and does not explicitly address hyperparameter tuning and similar considerations.
Sample splitting and resampling is crucial to avoid overfitting in predictive machine learning \citep{bischl2023hyperparameter}. There are different ways to combine the DML cross-fitting procedure with ML tuning cross-validation schemes in practice.  In our simulation study, we investigate three sample splitting approaches when tuning the nuisance estimators, which are illustrated in Figure \ref{fig:schemes}. \textbf{Full sample}: We apply cross-validated tuning on the whole sample and find the best learner and the best hyperparameter configuration by minimizing a loss target for each of the nuisance prediction tasks. Causal estimation by DML is also performed on all observations according to the cross-fitting algorithm, using the same learner and set of hyperparameters on all folds. The splits in the cross-validation scheme for tuning and the splits in the DML cross-fitting algorithms are independent of each other. 
\textbf{Split sample}: Splitting the data into a tuning and an inference fold is frequently employed in predictive machine learning. We adapt it by splitting the sample in half and tune the hyperparameters using cross-validation on $50\%$ of the data only. We choose the optimal learner and hyperparameter set and use them for fitting the nuisance functions in the remaining $50\%$ observations using DML cross-fitting. In this schedule, the samples used for tuning and causal estimation do not overlap. Hence, fitting the ML learners only uses data that has not been used for tuning. \textbf{On folds}: We perform sample splitting as in the cross-fitting algorithm and perform cross-validated hyperparameter tuning within each of the cross-fitting folds: We take a $K$-fold random partition $(I_k)_{k=1}^K$ of observation indices $[N] =\{1,\dots,N\}$ such that the size of each fold $I_k$ is $n=N/K$. For each $k \in [K]$, we perform cross-validation based hyperparameter optimization with the observations $i \notin I_k$ directly on the same partition (or fold) which is used for predicting the nuisance components. This approach takes $K$-times the calculation power of the other two. \textcolor{black}{With our empirical results, we would like to investigate how the cross-fitting scheme of DML and the cross-validation splitting schemes for hyperparameter tuning interact, for example, comparing the split sample approach with non-overlapping subsamples to the overlapping, but independent sample splits of the full sample and on folds approach.}
\begin{table}[t]
\resizebox{\linewidth}{!}{
\begin{tabular}{llllllllll}
DGP & $D$ on $Y$                                & $X$ on $D$     & $X$ on $Y$     & $\varepsilon$    & $\theta$ & oracle bias (\%) & oracle std & SNR     & n    \\
\hline
1   & linear, additive                          & linear         & linear         & $N(0,2)$         & 0.2      & 7.4472           & 0.1318     & 0.404   & 1000 \\
2   & linear, additive                          & non-linear     & non-linear     & $N(0,1)     $    & 0.8      & 0.0289           & 0.0703     & 0.462   & 1000 \\
3   & linear, additive                          & linear, sparse & linear, sparse & $N(0,1)     $    & -0.8     & 0.0039           & 0.0738     & 0.985   & 1000 \\
4   & linear, additive                          & non-linear     & non-linear     & $t(5)      $     & 2.1      & 0.4789           & 0.0541     & 0.842   & 2000 \\
5   & interacts with 11 $X_i$ and $\varepsilon$ & non-linear     & non-linear     & $t(12)     $     & -0.3429  & 2.2763           & 0.0861     &         & 2000 \\
6   & interacts with 1 $X_i$ and $\varepsilon$  & linear         & linear         & $t(19)     $     & -1.1039  & 0.0021           & 0.0718     &         & 1000 \\
7   & No effect                                 & non-linear     & non-linear     & $N(0,1)    $     & 0        &                  & 0.0511     & 0.905   & 2000 \\
8   & interacts with 5 $X_i$                    & non-linear     & non-linear     & $N(0,0.02)  $    & -1.432   & 0.0065           & 0.0143     & 12.463  & 1000 \\
9   & interacts with 2 $X_i$                    & non-linear     & non-linear     & $N(0,3)  $       & 12.62    & -0.0181          & 0.0952     & 12.284  & 2000 \\
10  & interacts with 3 $X_i$                    & non-linear     & linear         & $t(4) $          & 9.134    & 0.0057           & 0.1206     & 11.676  & 2000 \\
11  & interacts with 23 $X_i$                   & linear         & non-linear     & $N(0,2)$         & 10.77    & 0.0338           & 0.0547     & 61.092  & 2000 \\
12  & interacts with 7 $X_i$                    & non-linear     & non-linear     & $N(0,2)$         & -3.159   & -0.0011          & 0.1009     & 12.820  & 2000 \\
13  & interacts with 1 $X_i$                    & non-linear     & non-linear     & dependent on $D$ & -0.8486  & 0.0871           & 0.0402     &         & 2000 \\
14  & interacts with 4 $X_i$                    & non-linear     & non-linear     & $N(0,4)$         & 61.11    & -0.0074          & 1.2539     & 219.221 & 1000 \\
15  & interacts with 2 $X_i$                    & non-linear     & non-linear     & dependent on $D$ & -0.1606  & -0.3792          & 0.0254     &         & 2000 \\
16  & linear, additive                          & linear         & linear         & $t(8)$           & 1        & -0.0316          & 0.0553     & 10.934 & 2000
\end{tabular}}
\caption{Details on the data generating processes, with description of relationships and error term, true parameter, oracle estimate and standard error, sample size and signal to noise ratio (SNR). The SNR is calculated empirically from the DGPs, by calculating $Var(Y)/Var(\varepsilon)$ for each repetition and averaging. Full description of the DGPs can be found in Appendix \ref{app:dgp}.}\label{tab:dgp}
\end{table}

\section{Set-up of the Simulation Study}
The empirical assessment of causal estimators is generally limited by the fact that the ground truth (i.e., the true value of the causal parameter $\theta_0$) is not observable in real data sets. Hence, empirical studies have to rely on simulated data sets. To investigate the role of different sample splitting schemes, the choice of learners and hyperparameters as well as the scope to which predictive metrics can be used to motivate the choice of a causal model (PLR vs. IRM), we use two sources of data. Our major analysis will be based on data originally introduced for the 2019 Atlantic Causal Inference Conference (ACIC) data challenge.\footnote{For more info, see \url{https://sites.google.com/view/acic2019datachallenge/data-challenge}.} We choose this data \textcolor{black}{source} because it \textcolor{black}{generates designs with common and relatively complex empirical patterns} under unconfoundedness.\footnote{There are some scenarios, for which the unconfoundness assumption will not hold exactly, see Table \ref{tab:dgp}.} Moreover, we focused on an external resource for the DGP as this helps to abstract from biases that might stem from using self-defined simulation scenarios. To gain more detailed insights, we used the BCH DGP by \citet{belloni2014} as a second source. The advantage of this design is its clear structure, which makes it possible to match the empirical results with general theoretical reasoning.
\begin{figure}[t]
    \centering
    \subfigure[]{\includegraphics[width=0.3\textwidth]{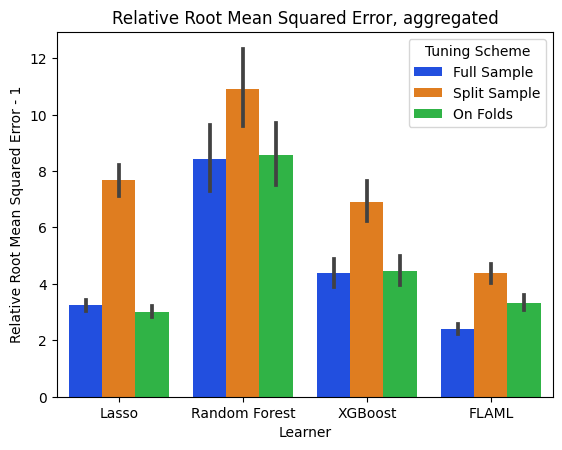}\label{fig:aggrRMSE}}
    \hfill 
    \subfigure[]{\includegraphics[width=0.3\textwidth]{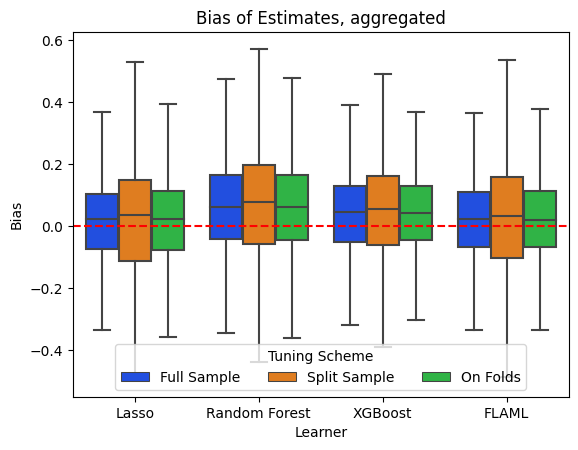}\label{fig:aggbias}}
    \hfill 
    \subfigure[]{\includegraphics[width=0.3\textwidth]{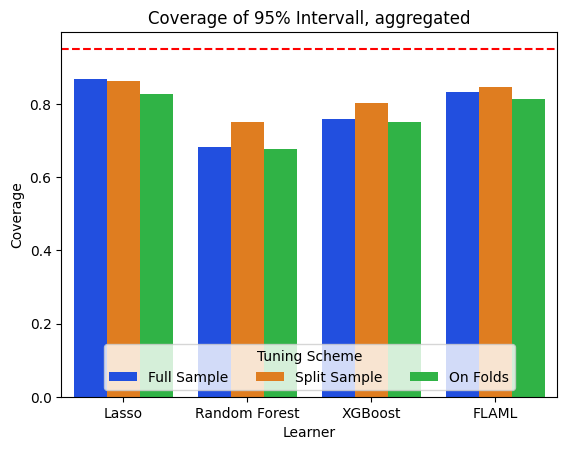}\label{fig:aggcov}}
    \caption{Average results on the ATE estimation over all ACIC DGPs, grouped by learner and sampling scheme. The groups show the learner, the colors the tuning scheme. The results refer to the causal model that is suitable for the underlying assumptions of the DGP, see Table \ref{tab:dgp}. (a) Root mean squared error on the ATE parameter, relative to oracle and subtracted by $-1$. (b) Boxplot of bias on ATE parameter. (c) Coverage of the estimated $95\%$ confidence interval with the true ATE parameter.}
    \label{fig:aggregated}
\end{figure}
The ACIC data consist of $1600$ data sets from $16$ data generating processes (DGP) with $n=1000$ or $n=2000$ observations \textcolor{black}{and} $p=200$ covariates, a binary treatment variable and a continuous outcome. Table \ref{tab:dgp} contains a brief overview of the processes. A more detailed description can be found in Appendix \ref{app:dgp}. The data generating processes are mostly in line with the unconfoundedness or conditional exogeneity assumption, $Y(d)\perp D \mid X$, with $Y(d)$ indicating the potential outcome under treatment $D=d$. Hence, these settings allow for identification of the average treatment effect, $\theta_0 = \mathbb{E}[Y(d=1)-Y(d=0)]$. Overall, the summary in Table \ref{tab:dgp} makes clear that the data generating processes of the ACIC challenge are very disparate, including settings with homogeneous and heterogeneous treatment effects, sparse and non-sparse, linear and non-linear settings, different values for the signal-to-noise ratio and varying values for the ATE across settings. In DGP $5$ and $6$, the unconfoundedness assumption is violated, which complicates estimation of the ATE. The information on the effect of $D$ on $Y$ is informative for the assumptions in the causal models introduced in Section \ref{problem}. In cases of a linear and additive treatment effect of $D$ on $Y$, the assumptions of the PLR are satisfied, such that it should allow for consistent estimation of $\theta_0$. In case of interactions of $D$ with other covariates, $X$, the functional form assumption of the PLR is violated, such that the resulting estimator is likely to be affected by a misspecification bias. The IRM, however, is more general, such that it should basically be suitable for causal estimation of $\theta_0$ in both cases. As the PLR incorporates more structure, it is expected to lead to more efficient estimation in cases the underlying assumptions hold in the data generating process. \\
We study hyperparameter optimization with four different learners, including lasso, random forest \citep{breiman2001}, extreme gradient boosting \citep{xgboost} and the AutoML learning framework FLAML \citep{wang2019flaml}. \textcolor{black}{We focus on these learners as they are widely used in empirical research and industry practice. Moreover, the learners differ substantially in terms of the modeling assumptions and the way they introduce regularization to avoid overfitting. For example, the linear lasso learner uses an explicit penalty, whereas the nonlinear random forests and gradient boosting learners are based on ensembles of tree learners. As a result, these learners have different hyperparameters, which in turn affect the estimation performance in a different way \citep{probst2019tunability}. For random forests\footnote{For random forests, 100 trees were chosen for computational reasons. However, the performance did not significantly alter in trials with 300 or 500 trees.} and gradient boosting, the hyperparameter choice is based on cross-validated search over a grid of values presented in Table \ref{tab:learneroverview}. We use cross-validated lasso with the built-in penalty selection and the automated search for FLAML with a time budget of 60 seconds. As a tuning-free, but linear and low-dimensional benchmark, we compare our results to causal models that are based on linear and logistic regression.}
\section{Results} \label{results}
In this section we present results on different aggregation levels. Table \ref{tab:metrics} in Appendix \ref{app:acic} presents an overview on the different metrics\footnote{Full results and replication code are available at \url{https://github.com/DoubleML/DML-Hyperparameter-Tuning-Replication}.}.
\subsection{Sample Splitting: Tuning and Causal Estimation}
\begin{figure}[]
    \centering
    \subfigure{ \includegraphics[width=0.35\textwidth]{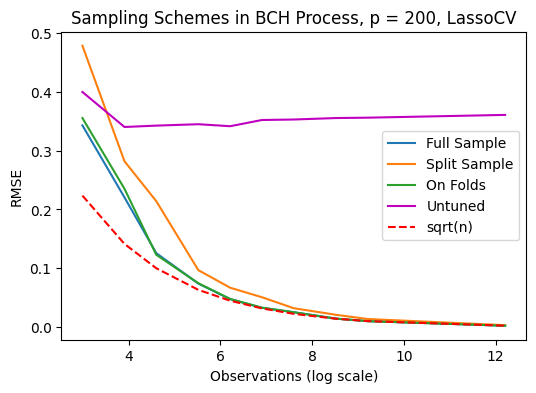}\label{fig:scalenlas}} 
    \subfigure{\includegraphics[width=0.35\textwidth]{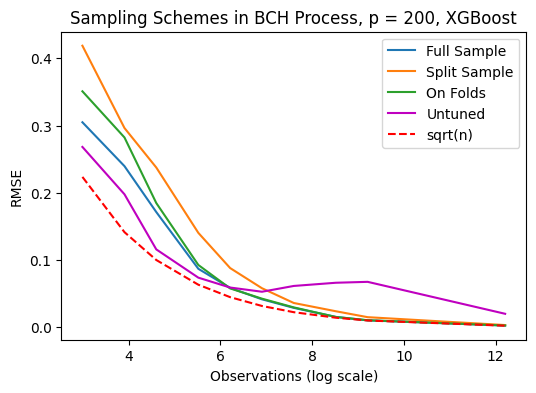}\label{fig:scalenxg}}
    \caption{Comparison of the RMSE on the ATE parameter of different sample splitting strategies in the BCH data generating process for (a) Lasso and (b) Boosting. Number of covariates in DGP is fixed, but $n$ is increasing.}\label{fig:scalen}
\end{figure}
\textcolor{black}{\textbf{Tuning on the full sample or on folds exhibit similar performance, which is superior to the split sample approach in small samples.}}
The first set of our results refer to the role of the sample splitting scheme that is used for hyperparameter optimization as described in Section \ref{splits}. Figure \ref{fig:aggregated} provides a summary of our results across all ACIC DGPs as obtained from 100 repetitions of every DGP. It presents an evaluation of the performance for different learners and sampling schemes based on a selection of the causal model (PLR, IRM) according to the underlying assumptions of the DGPs presented in Table \ref{tab:dgp}.\footnote{If $\theta_0$ is additively separable, i.e., constant for all individuals, we use the PLR and the IRM otherwise. We select the results according to the causal model in order to abstract from model misspecification considerations.} \footnote{\textcolor{black}{Due to the considerable differences in the simulation designs, we would like to note that Figure \ref{fig:aggregated} is not entirely able to present a comprehensive picture of our extensive simulation results. We would like to refer readers to the more detailed results for every learner, sample splitting scheme and each causal model for each DGP in Appendix \ref{app:rrmsefig} and \ref{app:biasfig}.}} Panel (a) in Figure \ref{fig:aggregated} shows the average root mean squared error relative to an oracle estimator for each of the employed learners and each sampling scheme as averaged over all DGPs. 
Panel (b) shows a boxplot of the bias of the corresponding causal estimates and the empirical coverage of $95\%$ confidence intervals is depicted in Panel (c). The results reflect the main empirical finding that the \textit{full sample} and \textit{on folds} scheme perform similarly well. The only exception from this is the result on the AutoML learner, for which the adjusted tuning time for the on-folds scheme might not have been sufficient. The causal estimates obtained from the \textit{split sample} approach are on average less accurate (in terms of the RMSE relative to an oracle) and more variable (cf. the boxplots in Panel (b)). This observation holds for virtually all learners and simulation settings considered as can be recognized from the Figures in \ref{app:rrmsefig} and \ref{app:biasfig} as well.  We interpret these results as evidence of a considerable efficiency loss in finite samples by using only $50\%$ of the observations for hyperparameter tuning. The empirical coverage of the confidence intervals are below the nominal level, which mainly reflects the differences in the underlying simulation settings. Due to the different structural assumptions across DGPs, it is unlikely that one learner will perform well in all settings.\\
\textcolor{black}{\textbf{The efficiency loss of the split sample approach vanishes with increasing sample size.}}
We investigate to what extent the efficiency loss of the \textit{split sample} approach persists with increasing samples size in Figure \ref{fig:scalen}, which is based on data from the BCH DGP with $p=200$ explanatory variables. In line with our previously reported results, the split sample scheme provides a higher RMSE on the causal parameter than the other two options. However, with growing number of observations, we observe that the RMSE of all tuning approaches converge with $\sqrt{n}$-rate. Thus, the difference between the schemes diminishes for larger samples. 
\begin{figure}[t]
    \centering
    \includegraphics[width=\linewidth]{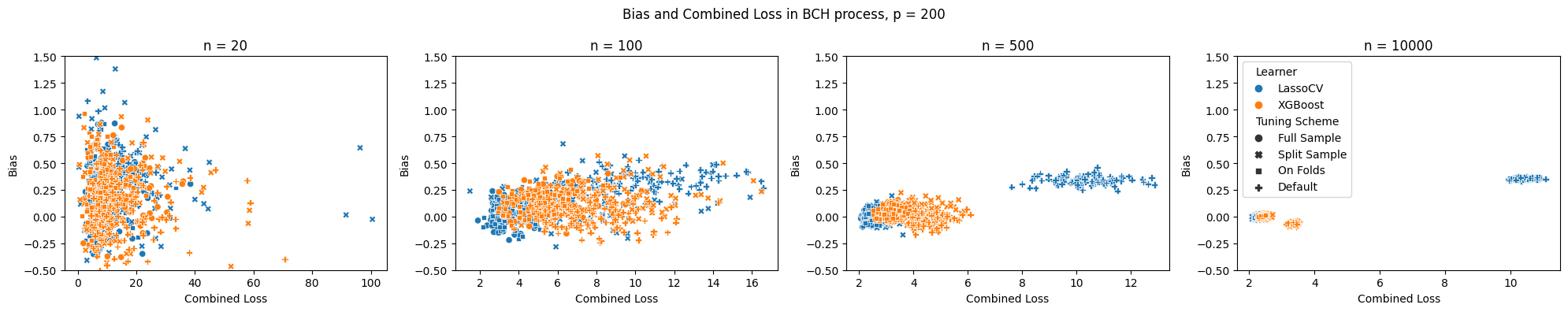}
    \caption{Combined nuisance loss and bias in the BCH DGP with increasing sample size $n$ for $p=200$ explanatory variables. For large $n/p$, the tuned learner becomes more distinct in terms of both bias and combined nuisance loss.}
    \label{fig:combbch}
\end{figure}
\subsection{Selection of Learners}\label{sec:learners}
\begin{figure}[t]
    \centering   \includegraphics[width=.8\textwidth]{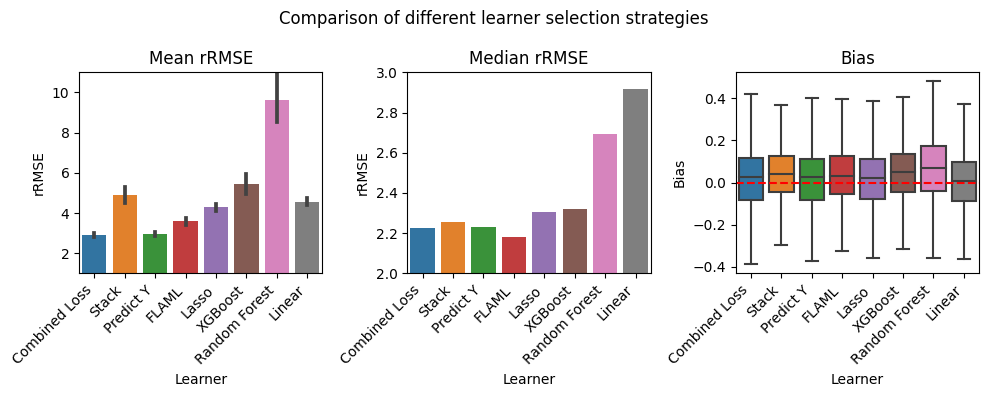}
    \caption{rRMSE of ATE parameter aggregated over learners and learner selection strategies for the \textit{full sample} tuning scheme.}
    \label{fig:learners}
\end{figure}

\textcolor{black}{\textbf{The performance of the causal estimation depends on structural assumptions and appropriate learner choice.}}
Comparing the four learners in the study, overall we find that the tuned AutoML and lasso learners perform very well across multiple settings (see Appendix \ref{app:rrmsefig} and \ref{app:biasfig}). In line with the theoretical formulation of the learner criterion in the DML approach, we observe that the causal estimation performance depends on the validity of structural form assumptions such as sparsity and linearity. For example, in linear settings (i.e., DGP 1, 3 or 16) lasso performs well in terms of the rRMSE for the causal estimate as well as the empirical coverage. We believe that the reason why the AutoML learner performs well is its ability to adapt to various settings.\\
\textcolor{black}{\textbf{A lower combined nuisance loss is associated with better causal estimation. Low signal-to-noise ratios create challenges in small samples.}}
In our simulation study, we investigate the relationship of the predictive performance for the nuisance functions, $\eta_0$, and the quality of the resulting causal estimate. A practical strategy for selecting the learners which would be aligned with the theoretical reasoning would be to pick the learner with the smallest out-of-sample loss for the nuisance predictions. In our simulations, however, we noticed that greedily minimizing the combined loss might not always deliver the lowest error (bias) on the causal parameter. \textcolor{black}{We provide more insights on this in Appendix \ref{app:learners}.}\\
As we compare tuned algorithms, for most DGPs we only observe an area of combined loss, where the fit is sufficient for a good causal inference performance. Still, considering the plots in Appendix \ref{app:absbiasvloss}, it is valid to say that the possibility of a high bias decreases with a low combined loss. Considering the different scenarios of the ACIC data, it is important to relate the predictive performance for the nuisance function, $\eta_0$, to the noise in the prediction problems. Whenever it is possible to approximate $\eta_0$ relatively well, it is possible to obtain an accurate causal estimator. However, the ability to find such a predictive model depends on the noisiness of the incorporated predictive tasks, see the information in Table \ref{tab:dgp}. This is further highlighted by Figure \ref{fig:combbch}, which illustrates the relationship of the average bias and the combined predictive loss for increasing samples based on the BCH DGP: The larger the $n/p$-ratio, the higher is the predictive quality of the nuisance estimators in terms of the combined loss. The scatter plots also include the comparison to the default settings for the parameters of lasso and xgboost, which basically implement untuned learners. With increasing sample size, it is easier to detect the default parameter learners by considering their higher combined nuisance loss. It can be observed that the estimation quality with respect to $\theta_0$ is worse for these learners. The results for the tuned estimators show a similar performance in terms of the combined nuisance loss as well as the \textcolor{black}{bias of the causal estimator}.\\
\textcolor{black}{\textbf{Combined nuisance loss serves as a good learner selection metric for causal estimation.}}
We briefly evaluate practical selection strategies for the employed learners based on the ACIC data. To do so, we focus on the \textit{full sample} scheme and pick the causal model with the lowest predictive loss on $Y$ (see Section \ref{sec:causalmodel}). We investigate the following approaches: (1) Pick the single best learner for each data set based on the best predictive performance for $Y$, (2) Pick the single best learner for each iteration based on the combined out-of-sample nuisance loss, (3) Stack all tuned learners. We present average results for the rRMSE as well as a boxplot for the bias as aggregated over all DGP in Figure \ref{fig:learners}. 
It is visible that using the learner with the lowest combined loss leads to the best overall rRMSE (mean). The strategies that select using the predictive error on $Y$ or always using \textit{FLAML} appear to perform also well, especially concerning the median rRMSE. The performance of basic stacking \citep{wolpert1992} is overall competitive in terms of the median rRMSE, but is affected by a weak performance of random forests in several scenarios (high mean rRMSE). \textcolor{black}{The linear/logistic benchmark model exhibits an overall inferior performance as compared most other learners.}
\subsection{Causal Models} \label{sec:causalmodel}
\begin{figure}[t]
    \centering
    \includegraphics[width=0.8\linewidth]{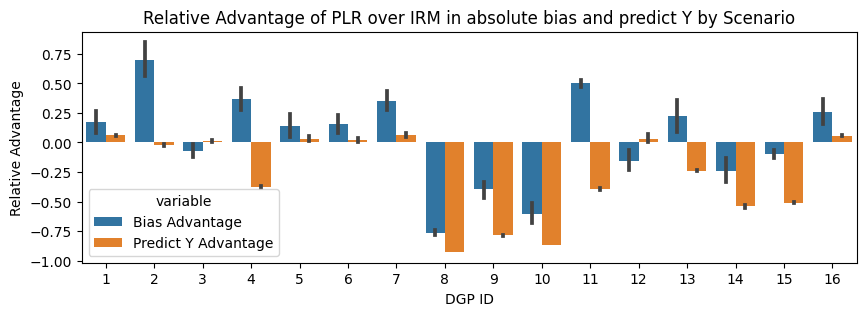}
    \caption{Relative advantage (or disadvantage) of PLR over IRM for all 16 ACIC processes. In many cases, the mean relative advantage in prediction error on the outcome $Y$ coincides with the mean relative advantage in estimation of $\theta$.}
    \label{fig:modeladv}
\end{figure}

In this section we assess the relationship of the structural assumptions encoded in the causal models (IRM and PLR) and the predictive performance of these models.\\
\textbf{The performance of the causal estimators depends on the structural assumptions encoded in causal models.} As described in Section \ref{dmlintro}, the PLR model incorporates stronger structural assumptions than the IRM. Hence, in cases where these assumptions hold in the underlying DGP, we expect the PLR to benefit from that additional structure in terms of more precise estimation performance as compared to IRM estimates. However, in cases with a non-additive heterogeneous causal effect, the PLR estimate on $\theta_0$ will be biased due to model misspecification. For our investigation, we split the DGPs into two categories: DGPs with a linear, additive effect and in DGPs with a heterogenous effect.\footnote{We drop some DGPs, in which we observe both models struggle with the estimation, i.e. because model assumptions (conditional exogeneity) fail.}
\begin{figure}[]
    \centering
    \subfigure{\includegraphics[width=0.45\textwidth]{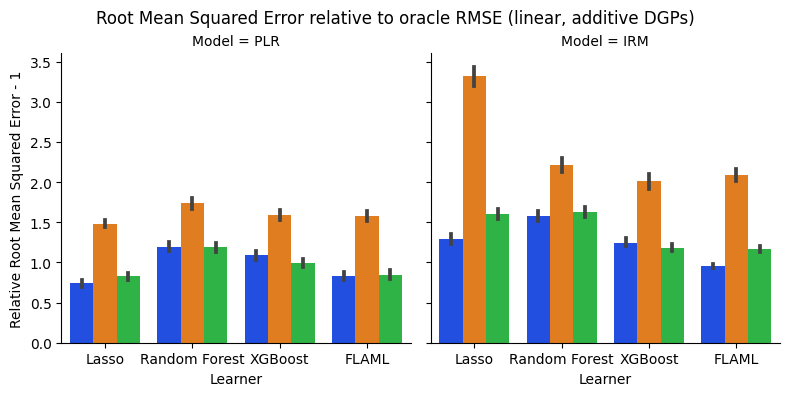}\label{fig:linadd}}
    \hfill
    \subfigure{\includegraphics[width=0.45\textwidth]{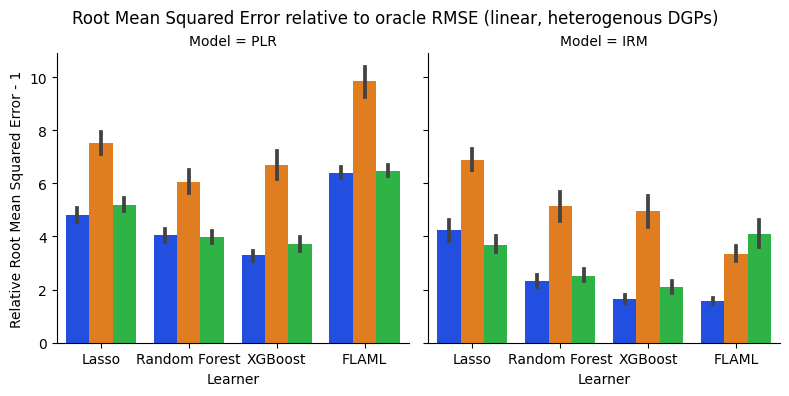}\label{fig:het}}
    \caption{Root mean squared error on the ATE parameter, relative to oracle and subtracted by $-1$. The groups show the learner, the colors the tuning scheme. Results are aggregated over different groups of DGP. The columns show the two different causal models. (a) Aggregated over group of linear, additive processes. (b) Aggregated over group of interactive processes.}
    \label{fig:modelcomparison}
\end{figure}
The results in Figure \ref{fig:linadd} are in line with our expectations, i.e., in linear additive DGPs, the PLR performs better than the IRM as it exhibits a lower rRMSE on average. Figure \ref{fig:het} illustrates that violations of these assumptions lead to a bias and, thus, less precise estimation of the ATE with the PLR. In contrast, the IRM is found to still accurately estimate $\theta_0$. These results hold generally for all considered learners.\\
\textcolor{black}{\textbf{The predictive performance of the causal models provides guidance on the choice of the causal model.}}
In practice, researchers are typically concerned with the question of which causal model to choose when multiple models are basically applicable, i.e., to what extent one is willing to make the assumption of a constant treatment effect. We investigate whether the predictive performance of the causal models with regard to the outcome variable $Y$ might be informative for that choice. To do so, we first estimate the causal model and then, generate predictions for the outcome variable $Y$ as based on the PLR model (i.e., plugging in the observed values for $D$ which are multiplied by the estimate $\hat{\theta_0}$) and the IRM (using predictions as generated by $g(D,X)$), see Table \ref{tab:metrics}. The results are displayed in Figure \ref{fig:modeladv}. We find that in many of the settings, the relative advantage in the error predicting $Y$ is associated with a better estimation performance for the causal parameter $\theta_0$. However, as displayed in Table \ref{tab:linadd} presented in Appendix \ref{app:fullrestab}, this measure does not always prove reliable. In the group of linear, additive DGPs, only in $33.42\%$ of the iterations, the predictive loss using the PLR was lower, although it would yield the lower rRMSE on $\theta$ in overall nearly $60\%$ . In terms of the IRM, this is a bit more straight forward: In $75.65\%$ of the cases, the predictive loss accurately points towards the better causal model to use (see Table \ref{tab:het}). \textcolor{black}{These results can be interpreted to reflect a conservative modeling approach. The prediction-based selection of the causal rather tends to recommend the more general IRM unless there is stronger evidence for validity of the PLR assumptions.}
\section{Application}
We apply the tuning methodology to the infant health and development program (IHDP) semi-synthetic data, which was previously used as a benchmark of causal ML methods in \cite{chernozhukov2022riesznet} and \cite{shi2019adapting}, see Appendix \ref{app:ihdp} for more information. We use the IRM model and the full sample tuning approach, as well as \texttt{FLAML}, as this combination has performed well in our ACIC analysis.
The results are displayed in Table \ref{tab:IHDP}. It is visible that the standard DML approach is very competitive with other state-of-the art appraoches such as RieszNets and DragonNet of \cite{shi2019adapting}.
\begin{table}[H]
    \centering
\begin{tabular}{ll}
Method                                 & MAE $\pm$ std. err. \\ \hline
DML with \texttt{FLAML}                         & 0.110 $\pm$ 0.009    \\
RieszNet (Chernozhukov et al.,  2021)  & 0.110 $\pm$ 0.003     \\
Dragonnet (Shi et al., 2019)           & 0.146 $\pm$ 0.010        
\end{tabular}
    \caption{Comparison of the tuning strategies discussed in this work and other recent approaches.}
    \label{tab:IHDP}
\end{table}
\section{Conclusion}
Our results underscore the importance of proper hyperparameter tuning and selection of the ML estimators \textcolor{black}{and the proper choice of DML related parameters}. Particularly, the use of default parameter estimators can lead to significant bias in the causal estimate. Hence, we would like to encourage analysts working with causal and double machine learning approaches to transparently communicate their tuning strategies as well as information on the nuisance prediction loss, \textcolor{black}{in particular when also employing several ML learners for nuisance estimation}. Our results  show that in general, tuning on the full data or on the folds is preferable over splitting the sample. In larger samples the difference of these approaches is found to vanish. Moreover, the choice between PLR and IRM was crucial in our results to achieve a low causal estimation error. To find the proper causal model, monitoring the predictive performance of the nuisance learners on $Y$ can serve as a helpful guidance for the choice of the causal model. Finally, we generally find evidence supporting a relationship between the out-of-sample prediction error in the nuisance components and the error on the target parameter. This is in line with theoretic results. However, our results also suggest the measure of combined loss can be noisy, especially when dealing with small sample sizes and low signal-to-noise ratios. In our study, the lowest nuisance prediction error did not always translate in the optimal ML configuration for DML. As guidance for empirical applications is crucial, we would like to analyse the use of neural networks \citep{klaassen2024doublemldeep}, more advanced stacking algorithms \citep{ahrens2022pystacked, van2007super}, more elaborate tuning approaches, and extend the simulation framework to conditional average treatment effects (CATE).

\acks{We would like to thank the organizers of the 2019 ACIC data challenge, namely Susan Gruber, for sharing the data generating processes as well as the results of the competing teams. Our gratitude goes to the participants of the 2022 Causal Data Science Meeting as well as the 2023 YES Causal Inference Workshop for their feedback on the presentation of this paper.}

\clearpage
\bibliography{sample.bib}

\clearpage
\appendix
\section{Motivation for Hyperparameter Tuning for Double Machine Learning}  \label{app:addmotiv}

\begin{figure}[t]
    \centering
    \subfigure{\includegraphics[width=.4\textwidth]{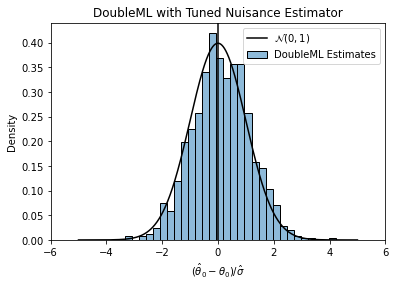}} 
    \subfigure{\includegraphics[width=.4\textwidth]{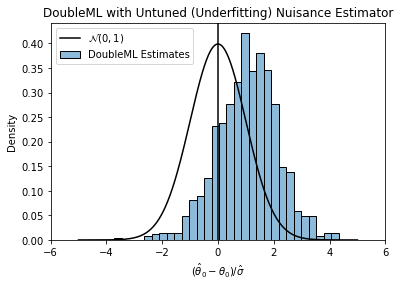}}
    \caption{Histogram of studentized DML estimator corresponding to a properly tuned ML estimator in Panel (a) and to an underfitting learner in Panel (b).}\label{fig:dml}
\end{figure}
Figure \ref{fig:dml} highlights the importance of appropriately choosing the ML predictors using a simulated data from a data generating process (DGP) from \citet{chernozhukov2018double} with $1000$ repetitions. The left plot shows the empirical distribution corresponding to a DML estimate on $\theta_0$ when estimating the nuisance with optimally tuned ML methods (random forest). The distribution of the studentized estimator is close to a normal distribution. However, the right figure shows the analogous results for regression stumps, which implies underfitting of the nuisance components, $\eta_0$. Even though a Neyman-orthogonal scoring function as well as sample-splitting are used in the DML procedure, the causal estimate exhibits a severe bias and thus the inference becomes invalid.

\section{Additional Information on Learner Selection}\label{app:learners}
\begin{figure}[h]
    \centering
    \includegraphics[width=.6\linewidth]{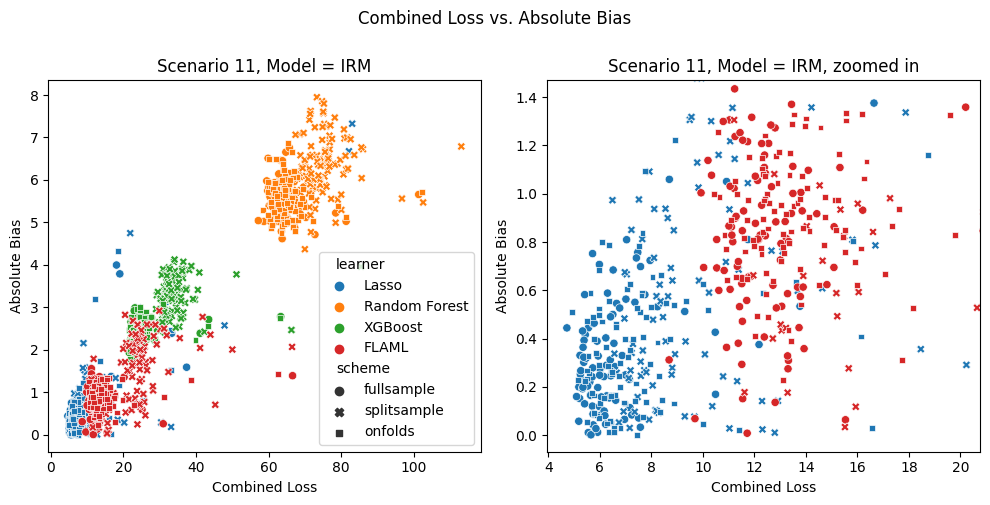}
    \caption{Absolute bias on causal estimate and combined loss of nuisance prediction for DGP 11 in IRM model.}
    \label{fig:absbiasvloss}
\end{figure}
As motivated in Section \ref{sec:learners}, based on our results, we find it not straight forward to select the best learner and tuning scheme purely on the combined nuisance loss. We showcase this in Figure \ref{fig:absbiasvloss} for DGP number 11, in which tuning of the nuisance estimators is particularly important, as the signal-to-noise ratio is high. On the left panel, we see that there is clearly a relationship that a lower combined loss implies a lower absolute bias. In the group of very well fitted learners (right panel), this difference becomes marginal and random statistical deviations are becoming more important.

\section{Information on Metrics and ACIC 2019 Competition}  \label{app:acic}
\begin{table}[]
    \centering
    \resizebox{0.75\textwidth}{!}{
    \begin{tabular}{p{0.25\linewidth}|p{0.75\linewidth}}
       \textbf{Name of Metric} & \textbf{Explanation}  \\
       \hline
       Relative root mean squared error (rRMSE)  & $\sqrt{\frac{MSE(\hat{\theta})}{MSE(\hat{\theta}_{oracle})}}$. The rRMSE was used as a metric in the ACIC 2019 challenge.\\
       rRMSE-1 & We subtract the rRMSE by one for clearer barplots.\\
       Mean bias & $\Bar{\theta} = \frac{1}{R}\sum^R_r(\hat{\theta}_r-\theta)$ with $r= 1, ..., R$ being the repetitions. \\
       Standard deviation & $s= \sqrt{\frac{1}{R-1}\sum^R_r(\hat{\theta}_r-\Bar{\theta})^2}$ \\
       Coverage & The coverage of the estimated $95\%$ confidence band with the true parameter. \\
       Predictive loss & $MSE(\hat{Y})$. In case of an IRM, this is equivalent to the out-of-sample loss of $\hat{g}_0(D,X)$. In case of an PLR, this is $\frac{1}{n}\sum^n_i(Y_i-\hat{\theta}D_i-\hat{\ell}(X_i))^2$. \\
       Combined loss & We combine the out-of-sample error for the prediction of both nuisance components. Identically to \citet{chernozhukov2018double}, we use for the PLR $RMSE(\hat{m}_0) \cdot \left(RMSE(\hat{m}_0) + RMSE(\hat{\ell}_0)\right)$ and for the IRM $RMSE(\hat{m}_0) \cdot RMSE(\hat{g}_0)$. This serves a measure to assess the overall quality of predictive fit for the nuisance function. \\
    \end{tabular}}
    \caption{Overview of metrics used in the analysis of the results.}
    \label{tab:metrics}
\end{table}

Based on the learnings from above, we pick the tuning scheme, learner and causal model with different approaches to compare our results to the results of some selected top challenge participants.
\textit{Strategy 1} refers to the most sophisticated strategy: Per each repitition first selecting the causal model by the lowest median of MSE on $Y$, then selecting learner and tuning scheme by the lowest combined loss. \textit{Always FLAML} uses the AutoML tuning on the fullsample. \textit{Untuned RF} is a comparison to the straight-away approach to always use the more flexible model (IRM) with Random Forest in standard hyperparameter configurations. We would like to note that contrary to the participants of the original challenge, our goal was not to achieve an optimal result in terms of a good rank. Rather we used the DGPs from the ACIC 2019 challenge as realistic benchmark scenarios to investigate the performance of DML with regard to various parameter choices.  We provide the comparison to the challenge participants in order to relate the performance of our strategies presented in the main paper to the original results. 
\begin{table}[H]
    \centering
\begin{tabular}{lllll}
\textbf{Approach}     & \textbf{Relative Bias} & \textbf{RRMSE} & \textbf{Coverage} & \textbf{Rel. Std.} \\
\textit{Strategy 1}   & 1372.1 (10)            & 2.932 (11)     & 0.813 (14)         & 2.407 (16)         \\
\textit{Always \texttt{FLAML}} & 1457.1 (15)            & 2.971 (13)     & 0.815 (8)         & 2.441 (17)         \\
\textit{Untuned RF}   & 1735.0 (17)            & 9.256 (16)     & 0.665 (20)        & 2.559 (12)         \\
\textit{BART*}        & 1628.8 (11)            & 2.257 (3)      & 0.750 (11)        & 1.244 (1)          \\
\textit{Std*}         & 1105.0 (9)             & 3.428 (9)      & 0.821 (4)         & 2.509 (11)         \\
\textit{BART TMLE*}   & 1301.5 (2)             & 2.192 (1)      & 0.811 (2)         & 1.356 (4)          \\
\textit{eb*}          & 988.04 (1)             & 3.567 (15)     & 0.850 (7)         & 2.748 (21)         
\end{tabular}
    \caption{Data challenge results for selected tuning strategies. Approaches marked with an * are original challenge participants. The number in parenthesis is the rank of that method in the competition. The ranks were assigned on a weighted average on the scenarios, which explains deviations between the absolute mean and the rank.}
    \label{tab:challenge}
\end{table}

\subsection{Information on IHDP Benchmark} \label{app:ihdp}
IHDP was a randomized control trial, studying the effect of certain measures (i.e. attendance at specialized clinics) of low birth weight infants on future developmental and health outcomes. We use $1000$ generated datasets of 747 observations as in \cite{chernozhukov2022riesznet}. Each observation consists of an outcome $Y$, a binary treatment $T$ and $25$ confounders $X$ that can are continuous or binary. 
\section{Full Results in Tables}\label{app:fullrestab}
\subsection{Results for ACIC in PLR} 
\begin{table}[H]
  \centering
  \resizebox{\linewidth}{!}{
\begin{tabular}{lrrrrrrrrrrrrr}
\toprule
Learner & \multicolumn{3}{c}{FLAML} & \multicolumn{3}{c}{Lasso} & \multicolumn{3}{c}{Random Forest} & \multicolumn{3}{c}{XGBoost} & Linear \\
Scheme & Full Sample &    On Folds & Split Sample & Full Sample &    On Folds & Split Sample &    Full Sample &    On Folds & Split Sample & Full Sample &    On Folds & Split Sample & Full Sample \\
DGP \# &            &            &             &            &            &             &               &            &             &            &            &             \\
\midrule
1  &   1.0956 &  \textbf{1.0711} &   1.5503 &   1.0950 &           1.0834 &   1.5079 &   1.1037 &           1.1243 &   1.5464 &           1.1211 &           1.1007 &   1.5178 &           1.1316 \\
2  &   1.2170 &  \textbf{1.1127} &   1.9287 &   1.3736 &           1.3645 &   1.9830 &   1.5906 &           1.6105 &   2.1073 &           1.4099 &           1.4231 &   1.7750 &           1.2233 \\
3  &   2.5519 &           2.7523 &   3.0255 &   1.8610 &           1.9886 &   2.6898 &   2.6640 &           2.6650 &   2.9801 &           2.6569 &           2.7180 &   3.0301 &  \textbf{1.2472} \\
4  &   1.5408 &           1.2863 &   2.0526 &   1.2975 &           1.7261 &   2.5651 &   1.9560 &           1.9304 &   2.2321 &           1.7138 &           1.6899 &   2.1581 &  \textbf{1.2317} \\
5  &   5.8245 &           7.4997 &  11.1874 &   5.2938 &           4.9919 &  10.2095 &   4.7933 &           4.0252 &   8.1064 &  \textbf{3.9275} &           4.5782 &   5.6138 &           7.7705 \\
6  &   5.2872 &  \textbf{5.2690} &   7.2466 &   6.3771 &           6.6130 &   9.1519 &   6.5917 &           6.5087 &   7.9793 &           6.5679 &           6.2539 &   8.2382 &           6.0200 \\
7  &   1.9908 &           1.9907 &   3.0353 &   2.7420 &           2.7849 &   3.1009 &   2.7755 &           2.6055 &   3.6463 &           3.1572 &           2.6295 &   3.6171 &  \textbf{1.7489} \\
8  &  11.5467 &          10.6801 &   8.0848 &   7.2969 &           7.2996 &   8.0905 &   6.2807 &           6.1882 &   6.7360 &           6.8165 &           6.8931 &   8.3904 &  \textbf{5.9193} \\
9  &   7.0943 &           6.0958 &  11.1786 &   4.7796 &           5.7621 &   7.5124 &   2.5647 &  \textbf{2.4356} &   3.6759 &           2.7096 &           2.7590 &   4.1712 &           6.3971 \\
10 &   4.7436 &           4.5514 &   7.4038 &   1.9442 &           1.9287 &   2.6623 &   2.9929 &           3.0181 &   2.8375 &  \textbf{1.8525} &           1.8546 &   2.4634 &           5.1557 \\
11 &   8.3446 &  \textbf{5.5440} &  15.8245 &   9.0712 &           9.7453 &  12.9054 &  73.4896 &          73.7701 &  88.6298 &          28.7352 &          29.0876 &  32.8459 &          28.2037 \\
12 &   7.2101 &           9.9652 &  22.9163 &  10.8196 &          10.4065 &  17.0188 &   9.7530 &           9.5929 &  17.2773 &  \textbf{7.0729} &           9.1706 &  18.9174 &          10.5871 \\
13 &   1.5778 &           1.5353 &   2.6675 &   1.1738 &           1.2183 &   2.5256 &   1.3325 &           1.3368 &   2.0142 &  \textbf{1.1345} &           1.1930 &   1.7841 &           1.2574 \\
14 &   6.4037 &           6.0266 &   4.6313 &   4.1826 &           5.6041 &   7.2586 &   3.6332 &           3.6325 &   4.7675 &           2.9706 &  \textbf{2.9178} &   4.4545 &           7.3758 \\
15 &   2.3446 &           2.4324 &   3.7893 &   1.5638 &  \textbf{1.5424} &   2.6860 &   8.3167 &           8.4921 &   8.8339 &           3.4202 &           3.4428 &   4.2264 &           1.6678 \\
16 &   2.6017 &           2.8543 &   3.8674 &   2.0663 &  \textbf{2.0125} &   3.0523 &   3.0596 &           3.1965 &   3.9160 &           2.4614 &           2.3995 &   3.4155 &           2.3445 \\
\bottomrule
\end{tabular}
  }
  \caption{Relative Root Mean Squared Error of estimates}
  \label{tab:rrmseplr}
\end{table}

\begin{table}[H]
  \centering
  \resizebox{\linewidth}{!}{
\begin{tabular}{lrrrrrrrrrrrrr}
\toprule
Learner & \multicolumn{3}{c}{FLAML} & \multicolumn{3}{c}{Lasso} & \multicolumn{3}{c}{Random Forest} & \multicolumn{3}{c}{XGBoost} & Linear \\
Scheme & Full Sample &    On Folds & Split Sample & Full Sample &    On Folds & Split Sample &    Full Sample &    On Folds & Split Sample & Full Sample &    On Folds & Split Sample & Full Sample \\
DGP \# &            &            &             &            &            &             &               &            &             &            &            &             \\
\midrule
1  &  \textbf{0.0008} &           0.0014 &   0.0162 &          -0.0149 &          -0.0144 &  -0.0059 &   0.0045 &   0.0034 &   0.0196 &           0.0021 &   0.0105 &   0.0242 &           0.0131 \\
2  &  \textbf{0.0123} &           0.0152 &   0.0976 &           0.0627 &           0.0624 &   0.0878 &   0.0872 &   0.0866 &   0.1187 &           0.0684 &   0.0662 &   0.0873 &           0.0226 \\
3  &           0.1690 &           0.1863 &   0.1879 &           0.1097 &           0.1220 &   0.1467 &   0.1777 &   0.1788 &   0.1817 &           0.1778 &   0.1809 &   0.1874 &  \textbf{0.0357} \\
4  &          -0.0029 &           0.0099 &  -0.0310 &           0.0351 &           0.0636 &   0.1096 &   0.0886 &   0.0880 &   0.0905 &           0.0738 &   0.0721 &   0.0843 &          -0.0000 \\
5  &          -0.0583 &  \textbf{0.0043} &   0.0699 &          -0.0783 &          -0.0678 &  -0.0977 &  -0.0719 &  -0.0741 &  -0.0355 &          -0.0955 &  -0.0671 &  -0.1006 &          -0.0123 \\
6  &          -0.0958 &          -0.0843 &  -0.2166 &          -0.2132 &          -0.2548 &  -0.2739 &  -0.2598 &  -0.2628 &  -0.2685 &          -0.2194 &  -0.2587 &  -0.2809 &  \textbf{0.0104} \\
7  &           0.0746 &           0.0753 &   0.1127 &           0.1097 &           0.1141 &   0.1248 &   0.1181 &   0.1156 &   0.1514 &           0.1062 &   0.1029 &   0.1372 &  \textbf{0.0572} \\
8  &           0.1565 &           0.1466 &   0.1090 &           0.1011 &           0.1005 &   0.0972 &   0.0865 &   0.0850 &   0.0898 &           0.0946 &   0.0958 &   0.1152 &  \textbf{0.0729} \\
9  &          -0.5722 &          -0.5066 &  -0.8890 &          -0.3237 &          -0.4878 &  -0.5670 &  -0.0850 &  -0.0764 &  -0.1858 &          -0.1461 &  -0.1567 &  -0.2578 &          -0.5376 \\
10 &          -0.5134 &          -0.5120 &  -0.7648 &          -0.1604 &          -0.1790 &  -0.2061 &  -0.3251 &  -0.3273 &  -0.2509 &          -0.1731 &  -0.1712 &  -0.1875 &          -0.5848 \\
11 &          -0.1500 &          -0.0640 &  -0.3808 &           0.3297 &           0.3569 &   0.2032 &   3.9828 &   3.9992 &   4.7878 &           1.5458 &   1.5690 &   1.7544 &          -1.5191 \\
12 &           0.4234 &           0.5158 &   0.7573 &           0.5785 &           0.5916 &   0.7617 &   0.5461 &   0.5479 &   0.6864 &  \textbf{0.3640} &   0.4289 &   0.6427 &           0.4687 \\
13 &           0.0449 &           0.0428 &   0.0752 &  \textbf{0.0156} &           0.0195 &   0.0316 &   0.0316 &   0.0333 &   0.0461 &           0.0228 &   0.0254 &   0.0369 &           0.0228 \\
14 &          -7.3964 &          -7.0113 &  -4.6878 &          -4.0059 &          -6.1637 &  -7.5730 &  -3.7727 &  -3.7802 &  -4.3082 &          -3.0428 &  -2.9528 &  -3.8826 &          -8.5521 \\
15 &           0.0473 &           0.0503 &   0.0787 &           0.0034 &  \textbf{0.0022} &   0.0023 &   0.2065 &   0.2106 &   0.2158 &           0.0797 &   0.0797 &   0.0944 &           0.0226 \\
16 &          -0.1057 &          -0.1172 &  -0.1450 &          -0.0676 &          -0.0633 &  -0.0740 &   0.0744 &   0.0871 &   0.0936 &          -0.0177 &  -0.0207 &  -0.0245 &          -0.1177 \\
\bottomrule
\end{tabular}
  }
  \caption{Mean Bias of Estimates in PLR}
  \label{tab:biasplr}
\end{table}

\begin{table}[H]
  \centering
  \resizebox{\linewidth}{!}{
\begin{tabular}{lrrrrrrrrrrrrr}
\toprule
Learner & \multicolumn{3}{c}{FLAML} & \multicolumn{3}{c}{Lasso} & \multicolumn{3}{c}{Random Forest} & \multicolumn{3}{c}{XGBoost} & Linear \\
Scheme & Full Sample &    On Folds & Split Sample & Full Sample &    On Folds & Split Sample &    Full Sample &    On Folds & Split Sample & Full Sample &    On Folds & Split Sample & Full Sample \\
DGP \# &            &            &             &            &            &             &               &            &             &            &            &             \\
\midrule
1  &           0.1453 &  \textbf{0.1420} &  0.2050 &  0.1445 &  0.1430 &  0.1999 &           0.1463 &           0.1491 &  0.2041 &           0.1487 &           0.1456 &  0.1998 &           0.1495 \\
2  &           0.0847 &           0.0767 &  0.0936 &  0.0731 &  0.0725 &  0.1079 &  \textbf{0.0695} &           0.0724 &  0.0878 &           0.0713 &           0.0747 &  0.0887 &           0.0829 \\
3  &           0.0815 &  \textbf{0.0789} &  0.1192 &  0.0819 &  0.0807 &  0.1330 &           0.0823 &           0.0799 &  0.1226 &           0.0809 &           0.0847 &  0.1207 &           0.0848 \\
4  &           0.0848 &           0.0701 &  0.1086 &  0.0621 &  0.0703 &  0.0884 &           0.0605 &           0.0589 &  0.0826 &  \textbf{0.0583} &           0.0583 &  0.0833 &           0.0678 \\
5  &           0.5001 &           0.6484 &  0.9646 &  0.4509 &  0.4261 &  0.8772 &           0.4080 &           0.3399 &  0.6999 &  \textbf{0.3257} &           0.3900 &  0.4747 &           0.6717 \\
6  &           0.3675 &           0.3690 &  0.4730 &  0.4050 &  0.4002 &  0.5971 &           0.3951 &           0.3859 &  0.5058 &           0.4172 &  \textbf{0.3664} &  0.5202 &           0.4324 \\
7  &           0.0689 &           0.0681 &  0.1060 &  0.0866 &  0.0845 &  0.0970 &           0.0778 &  \textbf{0.0652} &  0.1077 &           0.1211 &           0.0859 &  0.1232 &           0.0685 \\
8  &           0.0496 &           0.0394 &  0.0364 &  0.0233 &  0.0260 &  0.0617 &           0.0223 &           0.0225 &  0.0332 &           0.0209 &  \textbf{0.0206} &  0.0311 &           0.0422 \\
9  &           0.3541 &           0.2784 &  0.5781 &  0.3181 &  0.2459 &  0.4320 &           0.2287 &           0.2188 &  0.2959 &           0.2121 &  \textbf{0.2102} &  0.3009 &           0.2809 \\
10 &           0.2472 &           0.1911 &  0.4545 &  0.1703 &  0.1474 &  0.2454 &           0.1534 &           0.1560 &  0.2314 &  \textbf{0.1402} &           0.1429 &  0.2297 &           0.2032 \\
11 &           0.4316 &           0.2969 &  0.7780 &  0.3704 &  0.3955 &  0.6770 &           0.4368 &           0.4274 &  0.6540 &           0.2552 &  \textbf{0.2312} &  0.3603 &           0.2387 \\
12 &  \textbf{0.5901} &           0.8615 &  2.1833 &  0.9240 &  0.8655 &  1.5371 &           0.8168 &           0.7960 &  1.6010 &           0.6127 &           0.8187 &  1.7961 &           0.9587 \\
13 &           0.0445 &           0.0442 &  0.0759 &  0.0444 &  0.0448 &  0.0963 &           0.0430 &           0.0419 &  0.0663 &  \textbf{0.0394} &           0.0406 &  0.0613 &           0.0450 \\
14 &           3.0356 &           2.7291 &  3.3949 &  3.3609 &  3.3169 &  4.9907 &           2.5251 &           2.5123 &  4.1216 &  \textbf{2.1265} &           2.1395 &  3.9963 &           3.4142 \\
15 &           0.0361 &           0.0357 &  0.0552 &  0.0397 &  0.0392 &  0.0683 &           0.0420 &           0.0438 &  0.0595 &  \textbf{0.0341} &           0.0356 &  0.0508 &           0.0359 \\
16 &           0.0972 &           0.1052 &  0.1567 &  0.0920 &  0.0914 &  0.1516 &           0.1519 &           0.1537 &  0.1952 &           0.1350 &           0.1311 &  0.1874 &  \textbf{0.0533} \\
\bottomrule
\end{tabular}
  }
  \caption{Standard Deviation of Estimates in PLR}
  \label{tab:stdplr}
\end{table}

\begin{table}[H]
  \centering
  \resizebox{\linewidth}{!}{
\begin{tabular}{lrrrrrrrrrrrrr}
\toprule
Learner & \multicolumn{3}{c}{FLAML} & \multicolumn{3}{c}{Lasso} & \multicolumn{3}{c}{Random Forest} & \multicolumn{3}{c}{XGBoost} & Linear \\
Scheme & Full Sample &    On Folds & Split Sample & Full Sample &    On Folds & Split Sample &    Full Sample &    On Folds & Split Sample & Full Sample &    On Folds & Split Sample & Full Sample \\
DGP \# &            &            &             &            &            &             &               &            &             &            &            &             \\
\midrule
1  &  0.9800 &           0.9700 &  0.9900 &  \textbf{0.9500} &           0.9600 &           0.9700 &  0.9700 &           0.9700 &           0.9900 &           0.9600 &           0.9600 &  0.9900 &           0.9600 \\
2  &  0.8900 &           0.9200 &  0.8800 &           0.8200 &           0.8400 &           0.8600 &  0.7700 &           0.7200 &           0.8300 &           0.8600 &           0.8200 &  0.8800 &  \textbf{0.9400} \\
3  &  0.4800 &           0.3600 &  0.6800 &           0.7500 &           0.6400 &           0.7400 &  0.3100 &           0.3800 &           0.7100 &           0.3900 &           0.3700 &  0.6400 &  \textbf{0.8500} \\
4  &  0.8500 &           0.9100 &  0.9000 &           0.9000 &           0.7700 &           0.7800 &  0.6900 &           0.7100 &           0.8000 &           0.8000 &           0.7800 &  0.8500 &  \textbf{0.9300} \\
5  &  0.9800 &  \textbf{0.9400} &  0.9800 &  \textbf{0.9400} &  \textbf{0.9600} &  \textbf{0.9600} &  0.9800 &           0.9700 &           0.9900 &           0.9700 &           0.9900 &  0.9900 &  \textbf{0.9600} \\
6  &  0.9300 &           0.9300 &  0.9800 &           0.8700 &           0.8700 &           0.8900 &  0.8700 &           0.8800 &  \textbf{0.9600} &           0.9300 &           0.8900 &  0.9100 &           0.9100 \\
7  &  0.8000 &           0.6900 &  0.7500 &           0.5600 &           0.5400 &           0.6900 &  0.5000 &           0.5100 &           0.6100 &           0.6300 &           0.6100 &  0.6800 &  \textbf{0.8600} \\
8  &  0.0000 &           0.0000 &  0.1900 &           0.0500 &           0.0500 &           0.1800 &  0.0700 &           0.0900 &           0.3500 &           0.0300 &           0.0300 &  0.1000 &  \textbf{0.4600} \\
9  &  0.3800 &           0.4100 &  0.3600 &           0.6800 &           0.4200 &           0.6500 &  0.8700 &  \textbf{0.9100} &  \textbf{0.9100} &           0.8400 &           0.8300 &  0.8600 &           0.3900 \\
10 &  0.1900 &           0.0800 &  0.2300 &           0.7100 &           0.7400 &  \textbf{0.7900} &  0.4000 &           0.3500 &           0.6800 &           0.7500 &           0.7400 &  0.7600 &           0.0600 \\
11 &  0.6600 &  \textbf{0.8500} &  0.5900 &           0.5700 &           0.5100 &           0.5800 &  0.0000 &           0.0000 &           0.0000 &           0.0000 &           0.0000 &  0.0000 &           0.0000 \\
12 &  0.5600 &           0.6600 &  0.6500 &           0.7100 &           0.6400 &           0.7200 &  0.6200 &           0.5900 &           0.8000 &           0.7700 &           0.7800 &  0.8200 &  \textbf{0.8600} \\
13 &  0.7700 &           0.8100 &  0.7600 &           0.9400 &           0.9300 &  \textbf{0.9500} &  0.9000 &           0.9100 &           0.8500 &           0.9300 &           0.9200 &  0.9300 &           0.9400 \\
14 &  0.1400 &           0.1800 &  0.6200 &           0.6300 &           0.4100 &           0.5300 &  0.5400 &           0.5500 &  \textbf{0.7200} &           0.6400 &           0.6900 &  0.6900 &           0.1800 \\
15 &  0.7500 &           0.7100 &  0.6900 &           0.9800 &           0.9800 &  \textbf{0.9600} &  0.0000 &           0.0000 &           0.0700 &           0.4000 &           0.4500 &  0.5900 &           0.9200 \\
16 &  0.7700 &           0.7300 &  0.7900 &           0.9100 &           0.9200 &           0.9200 &  0.9300 &           0.9200 &           0.9300 &  \textbf{0.9500} &  \textbf{0.9500} &  0.9300 &           0.5100 \\
\bottomrule
\end{tabular}
  }
  \caption{Coverage of a 95\% confidence interval of Estimates in PLR}
  \label{tab:covplr}
\end{table}

\subsection{Results for ACIC in IRM}
\begin{table}[H]
  \centering
  \resizebox{\linewidth}{!}{
\begin{tabular}{lrrrrrrrrrrrrr}
\toprule
Learner & \multicolumn{3}{c}{FLAML} & \multicolumn{3}{c}{Lasso} & \multicolumn{3}{c}{Random Forest} & \multicolumn{3}{c}{XGBoost} & Linear \\
Scheme & Full Sample &    On Folds & Split Sample & Full Sample &    On Folds & Split Sample &    Full Sample &    On Folds & Split Sample & Full Sample &    On Folds & Split Sample & Full Sample \\
DGP \# &            &            &             &            &            &             &               &            &             &            &            &             \\
\midrule
1  &           1.3588 &           1.3630 &   1.7276 &           1.1518 &   1.1371 &   2.0895 &    1.1122 &    1.1045 &    1.4990 &           1.1264 &  \textbf{1.0896} &   1.6069 &           2.5831 \\
2  &  \textbf{1.8481} &           1.9990 &   2.8468 &           2.2938 &   2.7010 &   4.9673 &    2.4966 &    2.5182 &    3.0464 &           1.9754 &           1.9987 &   2.6066 &           2.6247 \\
3  &           2.1507 &           2.3604 &   2.7521 &  \textbf{1.9395} &   2.3650 &   4.0566 &    2.6021 &    2.6391 &    2.9372 &           2.4988 &           2.4924 &   2.9337 &           2.3944 \\
4  &           1.8834 &           1.9894 &   2.6970 &  \textbf{1.8346} &   2.9436 &   3.7637 &    2.3158 &    2.3131 &    2.4912 &           2.1787 &           2.1952 &   2.4840 &           4.5590 \\
5  &           5.2058 &  \textbf{2.9685} &   4.7837 &           5.9087 &   5.2530 &   8.9381 &    3.5861 &    4.3097 &    7.6908 &           4.8240 &           4.1155 &   6.5163 &          14.1868 \\
6  &  \textbf{6.0683} &           9.4405 &   7.7748 &           7.2419 &   6.9156 &  42.8648 &    6.5744 &    6.3504 &    7.9838 &           6.1250 &           6.2455 &   7.8693 &          11.6220 \\
7  &           2.1790 &           2.4775 &   3.7704 &           3.1105 &   3.4190 &   4.0870 &    3.3558 &    3.3924 &    4.8019 &           3.2628 &           2.8861 &   5.1730 &  \textbf{1.9411} \\
8  &           2.1816 &           2.1540 &   2.9889 &           2.3516 &   2.6188 &   4.9129 &    2.5738 &    2.5683 &    3.6972 &  \textbf{1.9341} &           1.9384 &   2.8684 &           3.3512 \\
9  &           2.1780 &           2.1648 &   3.1616 &           3.3264 &   3.5310 &   6.5262 &    2.2096 &    2.2212 &    3.3503 &           2.1566 &  \textbf{2.1208} &   2.9709 &           6.0698 \\
10 &  \textbf{1.0583} &           1.1584 &   1.6751 &           1.2953 &   1.2360 &   1.6880 &    1.1315 &    1.1312 &    1.7191 &           1.0623 &           1.0628 &   1.7064 &           1.2540 \\
11 &          16.1035 &          17.1417 &  30.9374 &          14.8590 &  14.1398 &  26.1562 &  102.4055 &  102.6740 &  117.7050 &          44.9960 &          45.1375 &  61.1337 &  \textbf{6.8687} \\
12 &  \textbf{5.0122} &          17.5622 &  10.8482 &          13.3427 &  10.9391 &  14.3851 &    7.9385 &    8.8918 &   18.0516 &           6.0753 &           8.4058 &  19.1437 &          13.2814 \\
13 &           1.2434 &           1.2850 &   2.4153 &           1.4954 &   1.5418 &   2.1832 &    1.5515 &    1.5889 &    2.2084 &  \textbf{1.2405} &           1.2465 &   1.8756 &           1.9354 \\
14 &           2.4035 &           2.4894 &   3.0314 &           5.8649 &   5.1529 &  11.8739 &    2.7362 &    2.8075 &    3.8439 &  \textbf{1.9560} &           1.9653 &   3.0277 &           7.7059 \\
15 &           1.8830 &           1.9021 &   3.0548 &           1.8549 &   1.8952 &   4.2829 &    7.1471 &    7.3370 &    7.6195 &           3.3580 &           3.3432 &   4.0278 &  \textbf{1.5976} \\
16 &           2.2907 &           2.8196 &   4.7231 &           3.4038 &   3.0508 &   6.9476 &    3.6034 &    3.7914 &    4.5298 &           2.4305 &           2.4160 &   3.2699 &  \textbf{1.5786} \\
\bottomrule
\end{tabular}
  }
  \caption{Relative Root Mean Squared Error in IRM}
  \label{tab:rrmseirm}
\end{table}

\begin{table}[H]
  \centering
  \resizebox{\linewidth}{!}{
\begin{tabular}{lrrrrrrrrrrrrr}
\toprule
Learner & \multicolumn{3}{c}{FLAML} & \multicolumn{3}{c}{Lasso} & \multicolumn{3}{c}{Random Forest} & \multicolumn{3}{c}{XGBoost} & Linear \\
Scheme & Full Sample &    On Folds & Split Sample & Full Sample &    On Folds & Split Sample &    Full Sample &    On Folds & Split Sample & Full Sample &    On Folds & Split Sample & Full Sample \\
DGP \# &            &            &             &            &            &             &               &            &             &            &            &             \\
\midrule
1  &  \textbf{0.0017} &   0.0132 &   0.0121 &  -0.0129 &  -0.0158 &          -0.0174 &   0.0038 &   0.0070 &   0.0184 &   0.0041 &   0.0156 &   0.0245 &           0.0061 \\
2  &           0.0838 &   0.1014 &   0.1675 &   0.0932 &   0.1513 &           0.1754 &   0.1592 &   0.1607 &   0.1927 &   0.1183 &   0.1175 &   0.1545 &  \textbf{0.0294} \\
3  &           0.1160 &   0.1252 &   0.1573 &   0.0730 &   0.1269 &           0.1055 &   0.1734 &   0.1751 &   0.1788 &   0.1655 &   0.1640 &   0.1761 &  \textbf{0.0036} \\
4  &           0.0803 &   0.0908 &   0.1166 &   0.0693 &   0.1331 &           0.1139 &   0.1122 &   0.1128 &   0.1098 &   0.1043 &   0.1049 &   0.1096 &  \textbf{0.0091} \\
5  &          -0.0661 &  -0.0845 &  -0.1171 &  -0.1125 &  -0.0852 &          -0.0962 &  -0.0879 &  -0.0671 &  -0.0479 &  -0.0761 &  -0.0728 &  -0.0773 &          -0.0209 \\
6  &          -0.1540 &  -0.0734 &  -0.2870 &  -0.1898 &  -0.3038 &  \textbf{0.0318} &  -0.2756 &  -0.2828 &  -0.2810 &  -0.2699 &  -0.2814 &  -0.2962 &          -0.0363 \\
7  &           0.0852 &   0.0931 &   0.1401 &   0.1210 &   0.1440 &           0.1477 &   0.1567 &   0.1578 &   0.2161 &   0.1274 &   0.1241 &   0.1864 &  \textbf{0.0533} \\
8  &           0.0156 &   0.0121 &   0.0305 &   0.0221 &   0.0223 &           0.0370 &   0.0305 &   0.0304 &   0.0438 &   0.0193 &   0.0201 &   0.0281 &  \textbf{0.0060} \\
9  &          -0.0592 &  -0.0859 &  -0.0950 &  -0.1448 &  -0.1583 &          -0.1980 &  -0.0969 &  -0.1024 &  -0.1718 &  -0.0811 &  -0.0843 &  -0.1317 &          -0.1616 \\
10 &          -0.0384 &  -0.0357 &  -0.0057 &  -0.0143 &  -0.0219 &           0.0089 &  -0.0559 &  -0.0555 &  -0.0692 &  -0.0345 &  -0.0326 &  -0.0487 &           0.0134 \\
11 &           0.7748 &   0.9012 &   1.4762 &   0.3106 &   0.3859 &           0.4988 &   5.5669 &   5.5815 &   6.3789 &   2.4396 &   2.4493 &   3.3038 &          -0.0109 \\
12 &  \textbf{0.2658} &   0.4442 &   0.3857 &   0.6121 &   0.6047 &           0.5671 &   0.3786 &   0.4018 &   0.5647 &   0.2879 &   0.3216 &   0.5837 &           0.4215 \\
13 &           0.0239 &   0.0253 &   0.0273 &   0.0143 &   0.0210 &  \textbf{0.0075} &   0.0470 &   0.0484 &   0.0595 &   0.0296 &   0.0296 &   0.0389 &           0.0183 \\
14 &          -0.9836 &  -1.0578 &  -2.0111 &  -0.8780 &  -2.3449 &          -4.6327 &  -2.3683 &  -2.4044 &  -2.8725 &  -1.2906 &  -1.2137 &  -1.6891 &           2.3656 \\
15 &           0.0285 &   0.0318 &   0.0430 &  -0.0035 &  -0.0075 &          -0.0080 &   0.1764 &   0.1809 &   0.1830 &   0.0782 &   0.0775 &   0.0894 &           0.0075 \\
16 &          -0.0403 &  -0.0292 &  -0.0409 &  -0.0846 &  -0.0789 &          -0.0807 &   0.1266 &   0.1378 &   0.1451 &   0.0183 &   0.0198 &   0.0304 &  \textbf{0.0069} \\
\bottomrule
\end{tabular}
  }
  \caption{Mean Bias of Estimates in IRM}
  \label{tab:biasirm}
\end{table}

\begin{table}[H]
  \centering
  \resizebox{\linewidth}{!}{
\begin{tabular}{lrrrrrrrrrrrrr}
\toprule
Learner & \multicolumn{3}{c}{FLAML} & \multicolumn{3}{c}{Lasso} & \multicolumn{3}{c}{Random Forest} & \multicolumn{3}{c}{XGBoost} & Linear \\
Scheme & Full Sample &    On Folds & Split Sample & Full Sample &    On Folds & Split Sample &    Full Sample &    On Folds & Split Sample & Full Sample &    On Folds & Split Sample & Full Sample \\
DGP \# &            &            &             &            &            &             &               &            &             &            &            &             \\
\midrule
1  &           0.1802 &           0.1803 &  0.2288 &  0.1522 &  0.1500 &   0.2766 &           0.1475 &           0.1463 &  0.1979 &           0.1493 &  \textbf{0.1437} &  0.2117 &           0.3425 \\
2  &           0.0989 &           0.0968 &  0.1083 &  0.1312 &  0.1137 &   0.3014 &           0.0720 &           0.0724 &  0.0913 &  \textbf{0.0717} &           0.0761 &  0.0972 &           0.1821 \\
3  &           0.1077 &           0.1205 &  0.1276 &  0.1229 &  0.1191 &   0.2800 &           0.0807 &           0.0836 &  0.1212 &  \textbf{0.0797} &           0.0816 &  0.1247 &           0.1767 \\
4  &           0.0651 &           0.0606 &  0.0911 &  0.0732 &  0.0915 &   0.1727 &           0.0595 &  \textbf{0.0580} &  0.0815 &           0.0584 &           0.0591 &  0.0810 &           0.2508 \\
5  &           0.4451 &  \textbf{0.2422} &  0.3965 &  0.4982 &  0.4460 &   0.7667 &           0.2972 &           0.3664 &  0.6631 &           0.4100 &           0.3482 &  0.5580 &           1.2263 \\
6  &           0.4076 &           0.6742 &  0.4784 &  0.4841 &  0.3920 &   3.0795 &           0.3826 &           0.3569 &  0.4993 &  \textbf{0.3465} &           0.3484 &  0.4806 &           0.8342 \\
7  &           0.0713 &           0.0854 &  0.1317 &  0.1025 &  0.0981 &   0.1471 &  \textbf{0.0682} &           0.0702 &  0.1144 &           0.1070 &           0.0788 &  0.1867 &           0.0835 \\
8  &           0.0270 &           0.0283 &  0.0297 &  0.0252 &  0.0299 &   0.0595 &           0.0202 &           0.0203 &  0.0291 &           0.0196 &  \textbf{0.0190} &  0.0296 &           0.0475 \\
9  &           0.1986 &           0.1871 &  0.2854 &  0.2812 &  0.2961 &   0.5885 &           0.1864 &           0.1847 &  0.2681 &           0.1884 &  \textbf{0.1832} &  0.2499 &           0.5545 \\
10 &  \textbf{0.1217} &           0.1350 &  0.2020 &  0.1556 &  0.1474 &   0.2034 &           0.1244 &           0.1245 &  0.1953 &           0.1233 &           0.1239 &  0.1999 &           0.1507 \\
11 &           0.4149 &           0.2484 &  0.8199 &  0.7519 &  0.6707 &   1.3426 &           0.4231 &           0.4249 &  0.7007 &           0.2573 &  \textbf{0.2383} &  0.4411 &           0.3762 \\
12 &  \textbf{0.4294} &           1.7148 &  1.0236 &  1.1975 &  0.9214 &   1.3348 &           0.7048 &           0.8011 &  1.7307 &           0.5404 &           0.7841 &  1.8403 &           1.2713 \\
13 &           0.0438 &           0.0449 &  0.0930 &  0.0583 &  0.0582 &   0.0874 &           0.0406 &           0.0413 &  0.0655 &  \textbf{0.0400} &           0.0402 &  0.0644 &           0.0755 \\
14 &           2.8469 &           2.9348 &  3.2191 &  7.3008 &  6.0159 &  14.1417 &           2.4709 &           2.5599 &  3.8595 &  \textbf{2.0815} &           2.1411 &  3.3956 &           9.3652 \\
15 &           0.0385 &           0.0364 &  0.0647 &  0.0471 &  0.0477 &   0.1087 &           0.0409 &           0.0426 &  0.0617 &  \textbf{0.0337} &           0.0342 &  0.0494 &           0.0400 \\
16 &           0.1201 &           0.1532 &  0.2581 &  0.1681 &  0.1491 &   0.3758 &           0.1535 &           0.1576 &  0.2039 &           0.1332 &           0.1322 &  0.1784 &  \textbf{0.0871} \\
\bottomrule
\end{tabular}
  }
  \caption{Standard Deviation of Estimates in IRM}
  \label{tab:stdirm}
\end{table}

\begin{table}[H]
  \centering
  \resizebox{\linewidth}{!}{
\begin{tabular}{lrrrrrrrrrrrrr}
\toprule
Learner & \multicolumn{3}{c}{FLAML} & \multicolumn{3}{c}{Lasso} & \multicolumn{3}{c}{Random Forest} & \multicolumn{3}{c}{XGBoost} & Linear \\
Scheme & Full Sample &    On Folds & Split Sample & Full Sample &    On Folds & Split Sample &    Full Sample &    On Folds & Split Sample & Full Sample &    On Folds & Split Sample & Full Sample \\
DGP \# &            &            &             &            &            &             &               &            &             &            &            &             \\
\midrule
1  &           0.9800 &           0.9800 &  \textbf{0.9500} &  \textbf{0.9500} &           0.9800 &           0.9400 &  0.9800 &  0.9800 &  0.9900 &           0.9700 &  0.9800 &  0.9900 &           0.9800 \\
2  &           0.8300 &           0.8300 &           0.6600 &           0.8400 &           0.6300 &           0.7800 &  0.4200 &  0.3900 &  0.5500 &           0.6100 &  0.5800 &  0.7000 &  \textbf{0.9700} \\
3  &           0.7600 &           0.7800 &           0.7600 &           0.8200 &           0.5900 &           0.8800 &  0.3900 &  0.3900 &  0.6900 &           0.4300 &  0.4500 &  0.6800 &  \textbf{0.9400} \\
4  &           0.7300 &           0.7500 &           0.8400 &           0.8300 &           0.4600 &           0.8200 &  0.5100 &  0.5400 &  0.7500 &           0.5700 &  0.5800 &  0.7500 &  \textbf{0.9500} \\
5  &           1.0000 &           0.9700 &           1.0000 &           0.9700 &           0.9800 &           0.9900 &  0.9700 &  0.9700 &  1.0000 &           0.9800 &  0.9700 &  0.9900 &  \textbf{0.9500} \\
6  &           0.9300 &  \textbf{0.9400} &           0.9200 &           0.8900 &           0.8300 &           0.8900 &  0.8600 &  0.8500 &  0.9300 &           0.9100 &  0.9000 &  0.9100 &  \textbf{0.9600} \\
7  &           0.7700 &           0.7100 &           0.6800 &           0.6600 &           0.4700 &           0.6900 &  0.2500 &  0.2500 &  0.2800 &           0.4900 &  0.5100 &  0.5200 &  \textbf{0.9000} \\
8  &           0.9100 &           0.9100 &           0.8700 &           0.8300 &           0.8400 &           0.8000 &  0.7300 &  0.7300 &  0.7400 &           0.8400 &  0.8600 &  0.8400 &  \textbf{0.9600} \\
9  &  \textbf{0.9300} &           0.9200 &           0.9700 &           0.9200 &           0.8800 &  \textbf{0.9300} &  0.9000 &  0.9100 &  0.9200 &           0.8900 &  0.8900 &  0.9000 &           0.9700 \\
10 &           0.9400 &           0.9200 &  \textbf{0.9500} &           0.9400 &  \textbf{0.9500} &           0.9100 &  0.8800 &  0.8700 &  0.8700 &           0.9200 &  0.9000 &  0.8700 &  \textbf{0.9500} \\
11 &           0.2200 &           0.0800 &           0.2900 &           0.8400 &           0.7200 &           0.7400 &  0.0000 &  0.0000 &  0.0000 &           0.0000 &  0.0000 &  0.0000 &  \textbf{0.9400} \\
12 &           0.9300 &           0.9600 &           0.9800 &           0.8600 &           0.7500 &           0.9000 &  0.8600 &  0.8300 &  0.9100 &           0.8800 &  0.8700 &  0.9600 &  \textbf{0.9500} \\
13 &           0.9200 &           0.9600 &           0.9400 &           0.9200 &  \textbf{0.9500} &           0.9400 &  0.8500 &  0.8000 &  0.8300 &           0.8900 &  0.9000 &  0.9000 &           0.9200 \\
14 &           0.9000 &           0.9300 &           0.8200 &           0.8800 &           0.7400 &           0.7900 &  0.7100 &  0.6700 &  0.8200 &           0.8300 &  0.8300 &  0.8600 &  \textbf{0.9600} \\
15 &           0.8600 &           0.8600 &           0.8300 &           0.9600 &  \textbf{0.9500} &           0.9600 &  0.0100 &  0.0100 &  0.1300 &           0.4400 &  0.4300 &  0.6700 &           0.9600 \\
16 &           0.9600 &           0.9200 &           0.9400 &           0.9600 &           0.9400 &           0.9400 &  0.8500 &  0.8100 &  0.8900 &  \textbf{0.9500} &  0.9600 &  0.9600 &  \textbf{0.9500} \\
\bottomrule
\end{tabular}
  }
  \caption{Coverage of a 95\% confidence interval of Estimates in IRM}
  \label{tab:confintirm}
\end{table}

\begin{table}[ht]
    \begin{minipage}{.45\linewidth}
      \centering
        \begin{tabular}{l|ll}
                & IRM $\theta$ & PLR $\theta$ \\
                \hline
        IRM $Y_{loss}$ & 11.91\%                    & 24.42\%                   \\
        PLR $Y_{loss}$ & 30.25 \%                    & \textbf{33.42}\%  \\    
\end{tabular}
      \caption{Share of iterations in which each model was better in estimating $\theta$ and the predictive loss on $Y$ for the group of linear, additive DGP.}\label{tab:linadd}
    \end{minipage}\hfill
    \begin{minipage}{.45\linewidth}
      \centering
        \begin{tabular}{l|ll}
                & IRM $\theta$ & PLR $\theta$ \\
                \hline
        IRM $Y_{loss}$ & \textbf{75.65}\%                    & 5.95\%                     \\
        PLR $Y_{loss}$ & 15.75\%                     & 2.65\%                    
\end{tabular}
        \caption{Share of iterations in which each model was better in estimating $\theta$ and the predictive loss on $Y$ for the group of heterogeneous DGP.}\label{tab:het}
    \end{minipage} 
\end{table}
\section{Full Results in Figures}\label{app:fullresfig}
\subsection{Relative Root Mean Squared Error}
\label{app:rrmsefig}
\begin{figure}[H]
    \centering
    \includegraphics[height=.9\textheight]{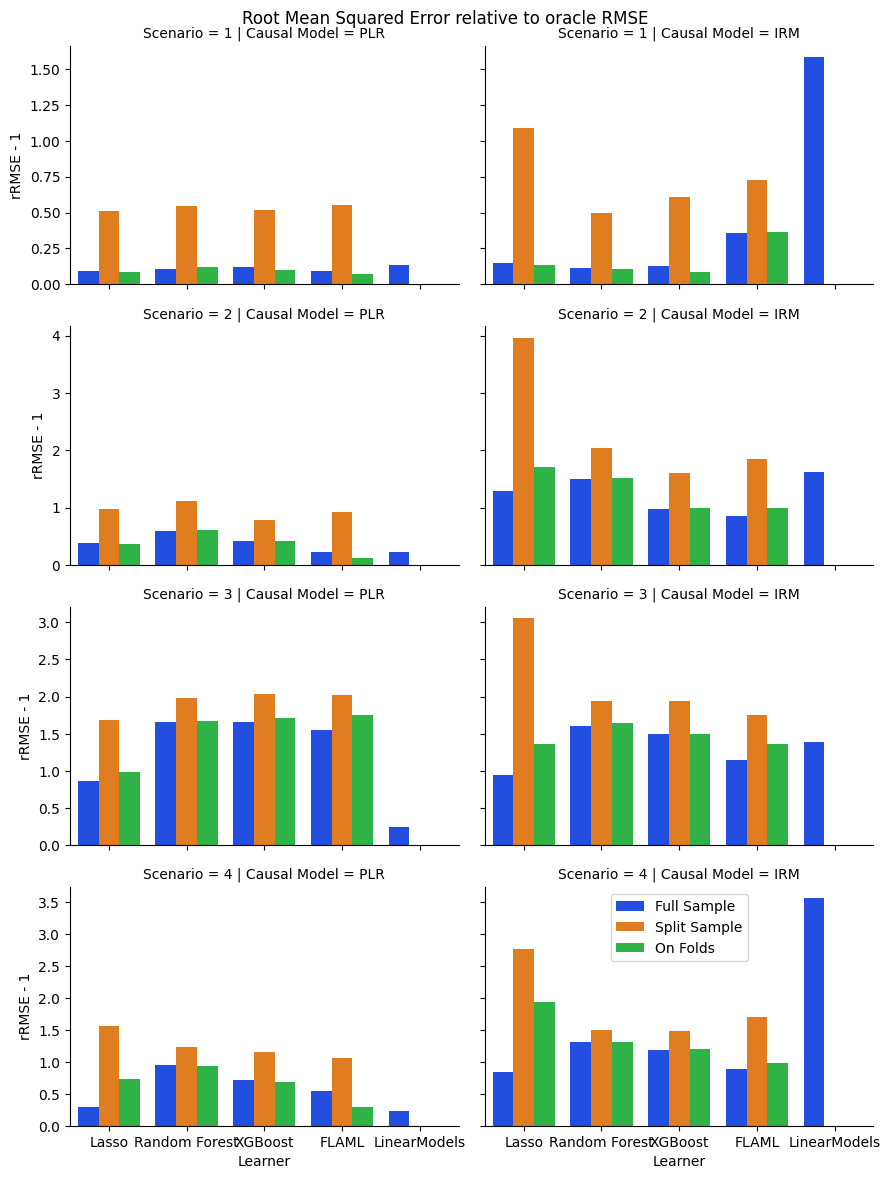}
    \caption{Relative Root Mean Squared Error of ACIC DGP 1-4 (adjusted by $-1$)}
    \label{fig:rrmseall1}
\end{figure}

\begin{figure}[H]
    \centering
    \includegraphics[height=.9\textheight]{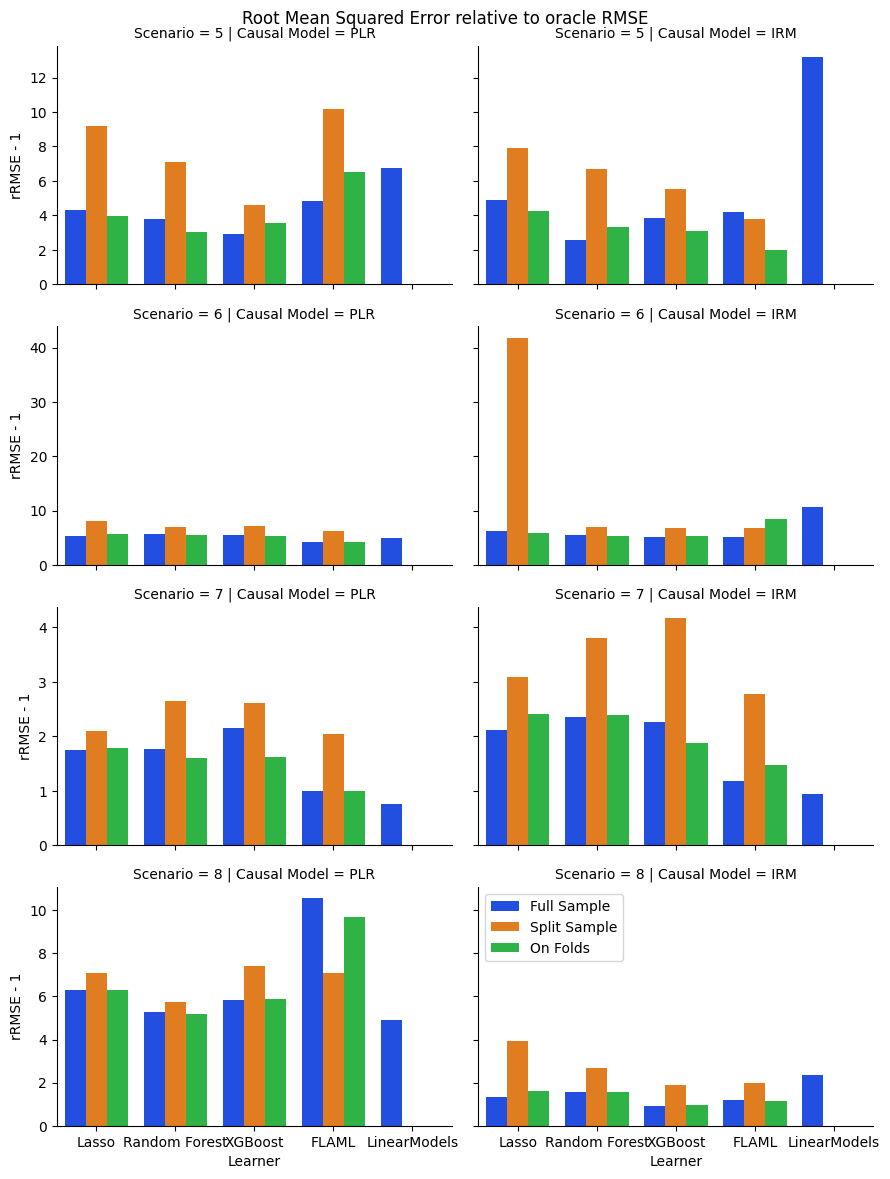}
    \caption{Relative Root Mean Squared Error of ACIC DGP 5-8 (adjusted by $-1$)}
    \label{fig:rrmseall2}
\end{figure}

\begin{figure}[H]
    \centering
    \includegraphics[height=.9\textheight]{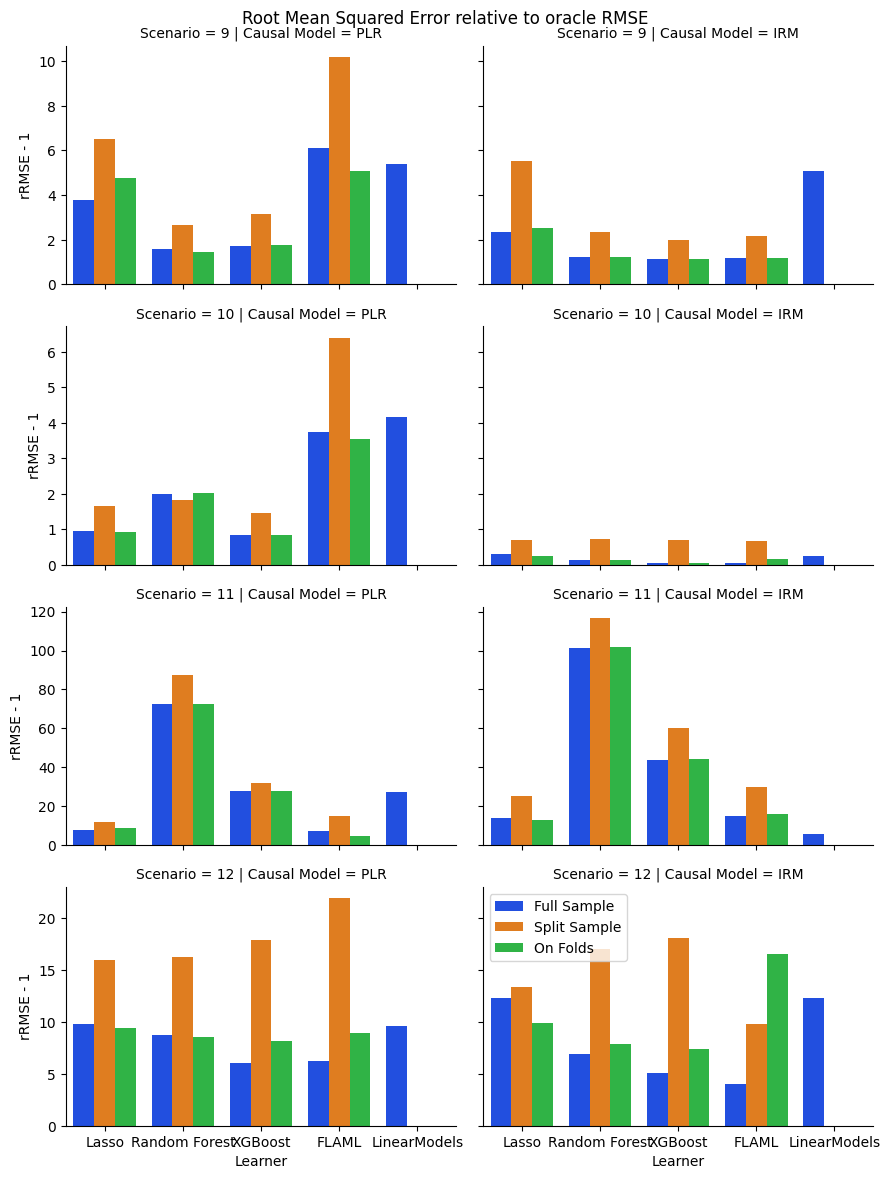}
    \caption{Relative Root Mean Squared Error of ACIC DGP 9-12 (adjusted by $-1$)}
    \label{fig:rrmseall3}
\end{figure}

\begin{figure}[H]
    \centering
    \includegraphics[height=.9\textheight]{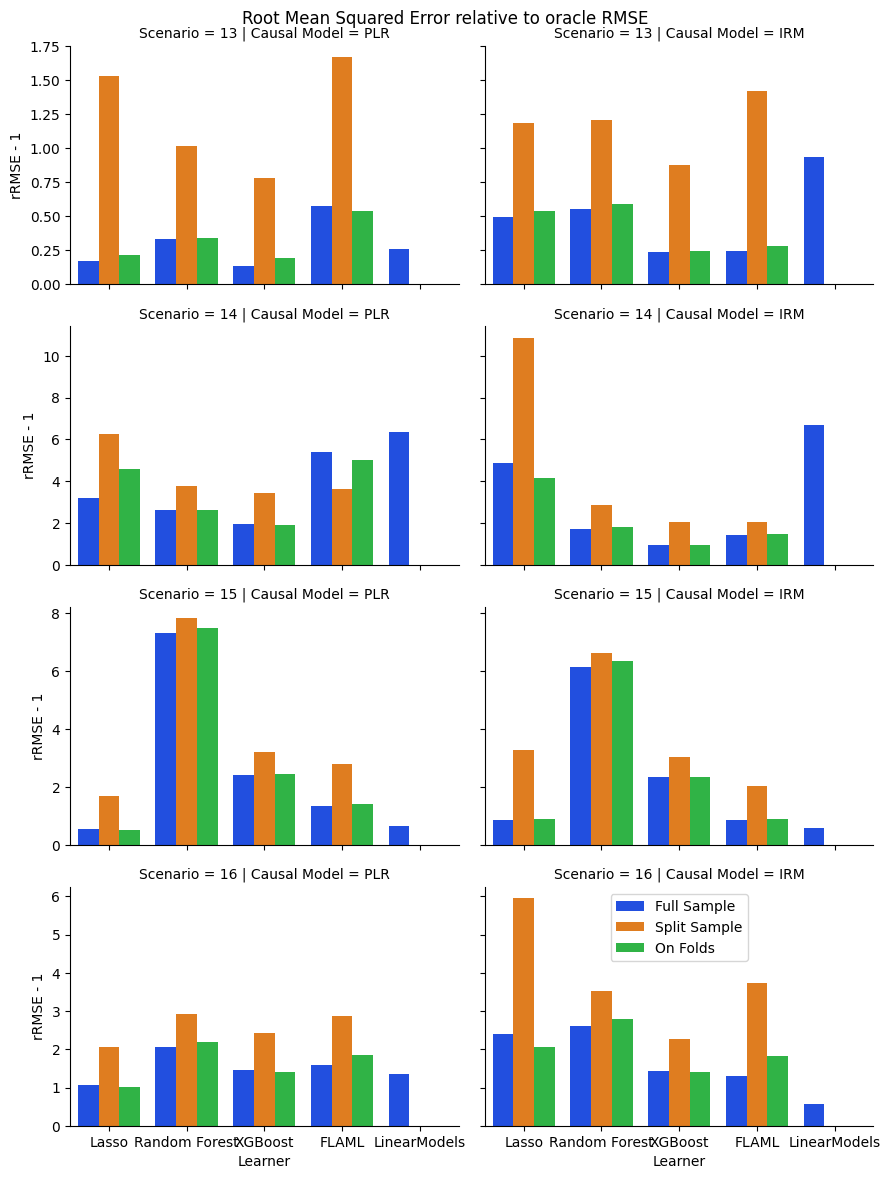}
    \caption{Relative Root Mean Squared Error of ACIC DGP 13-16 (adjusted by $-1$)}
    \label{fig:rrmseall4}
\end{figure}

\subsection{Bias of Estimates}
\label{app:biasfig}
\begin{figure}[H]
    \centering
    \includegraphics[height=.9\textheight]{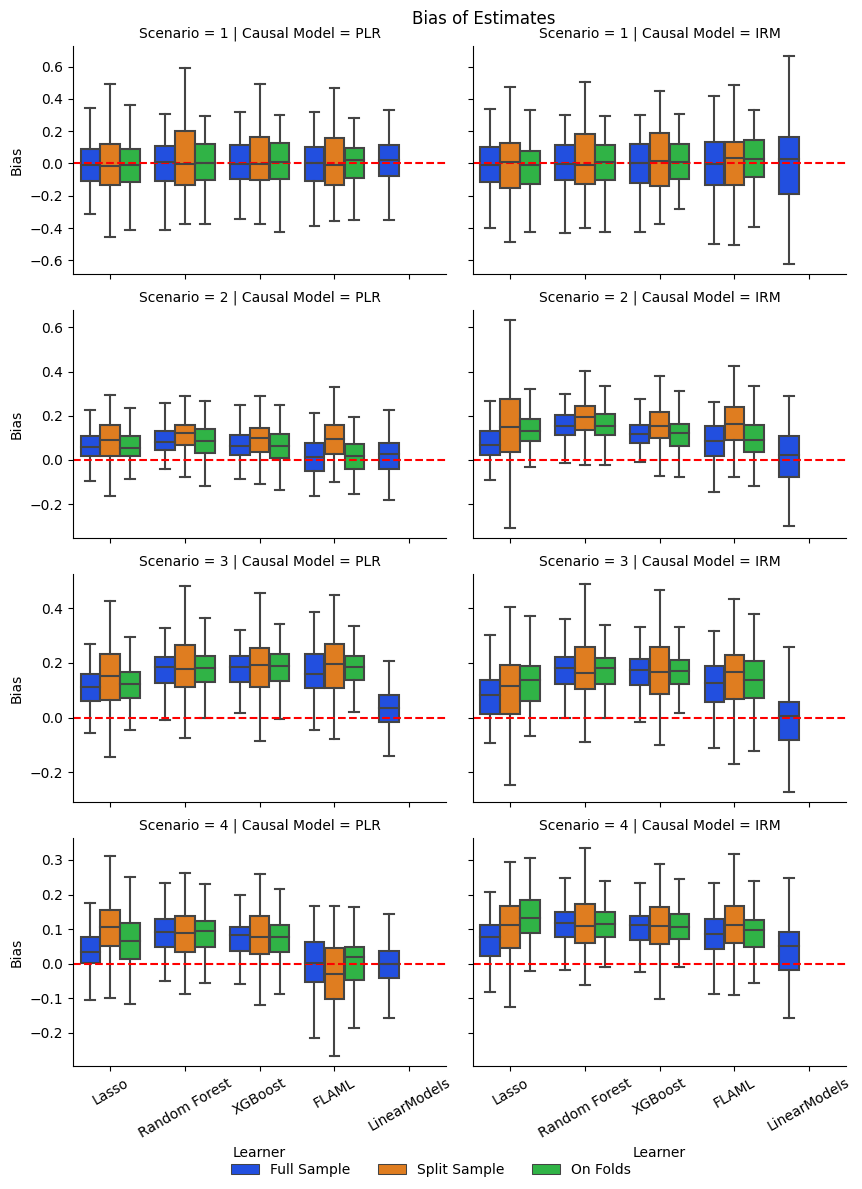}
    \caption{Boxplot of bias, ACIC DGP 1-4}
    \label{fig:biasall1}
\end{figure}

\begin{figure}[H]
    \centering
    \includegraphics[height=.9\textheight]{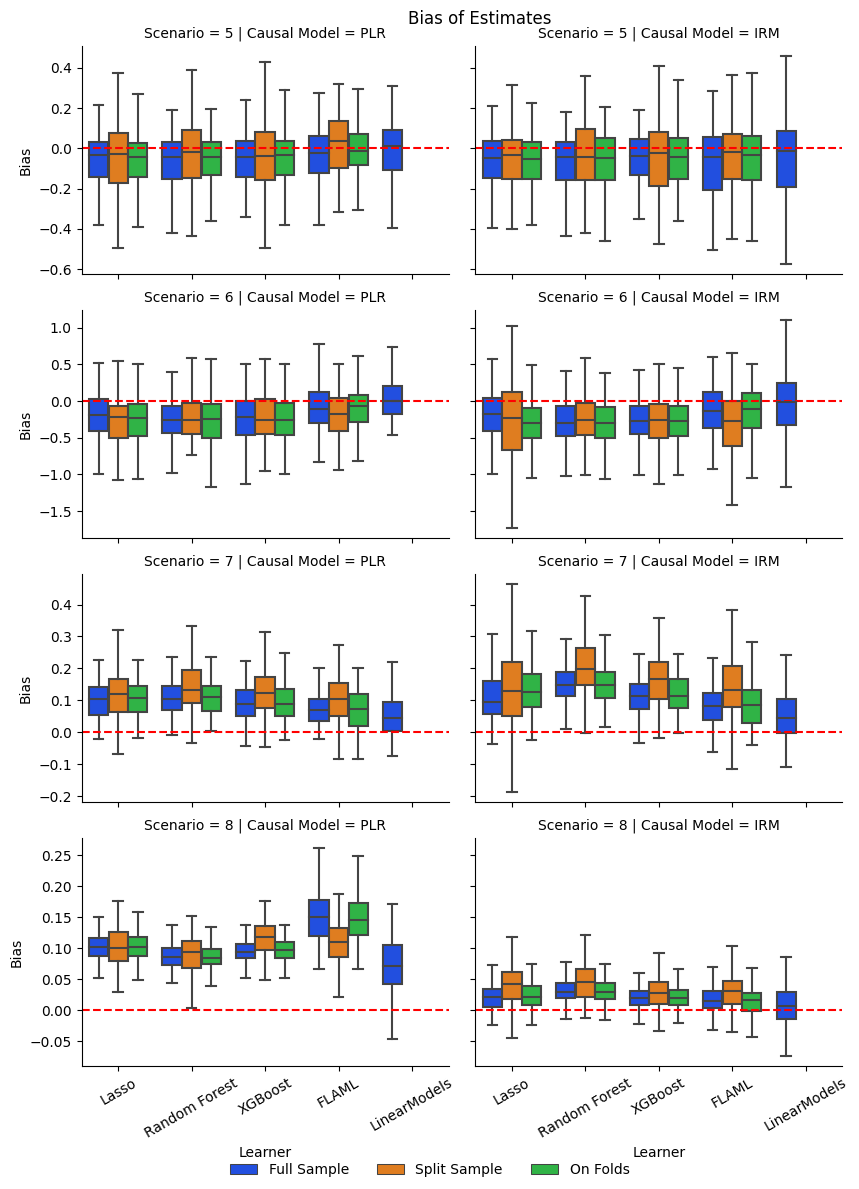}
    \caption{Boxplot of bias, ACIC DGP 5-8}
    \label{fig:biasall2}
\end{figure}

\begin{figure}[H]
    \centering
    \includegraphics[height=.9\textheight]{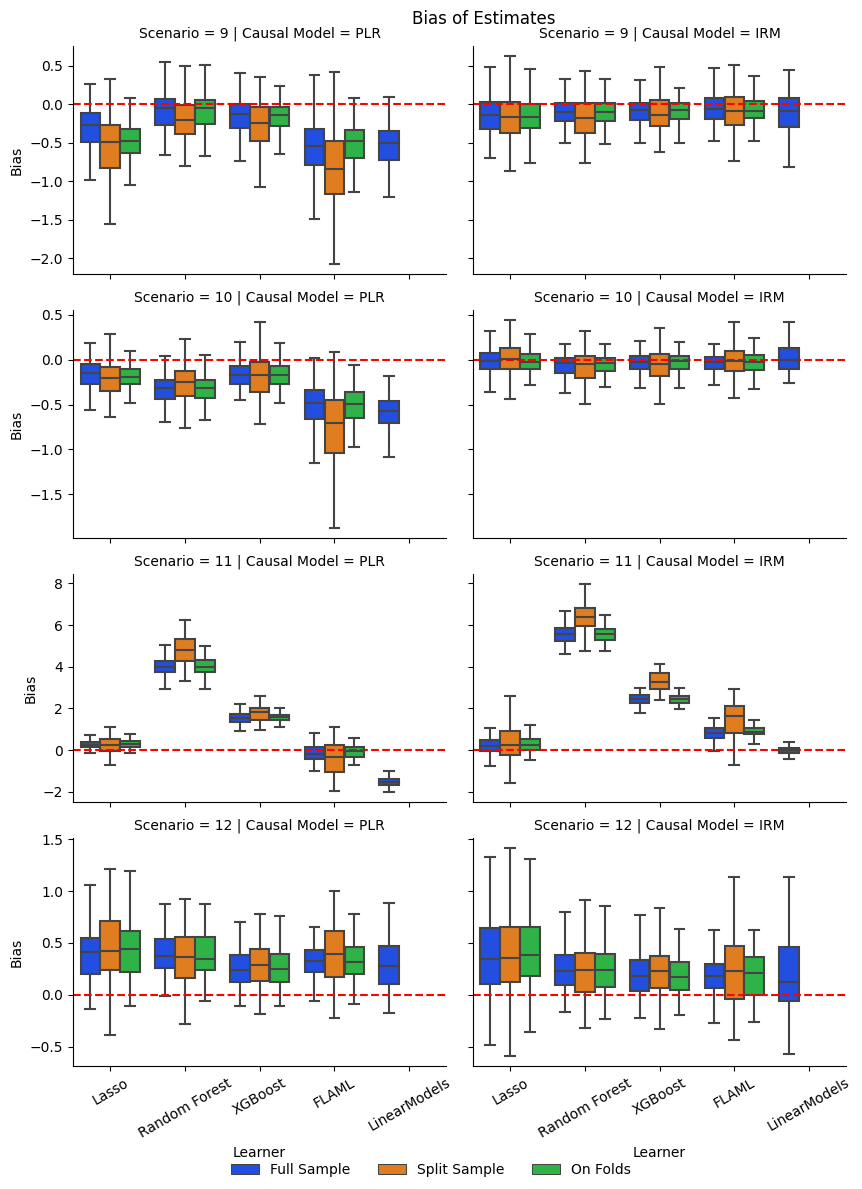}
    \caption{Boxplot of bias, ACIC DGP 9-12}
    \label{fig:biasall3}
\end{figure}

\begin{figure}[H]
    \centering
    \includegraphics[height=.9\textheight]{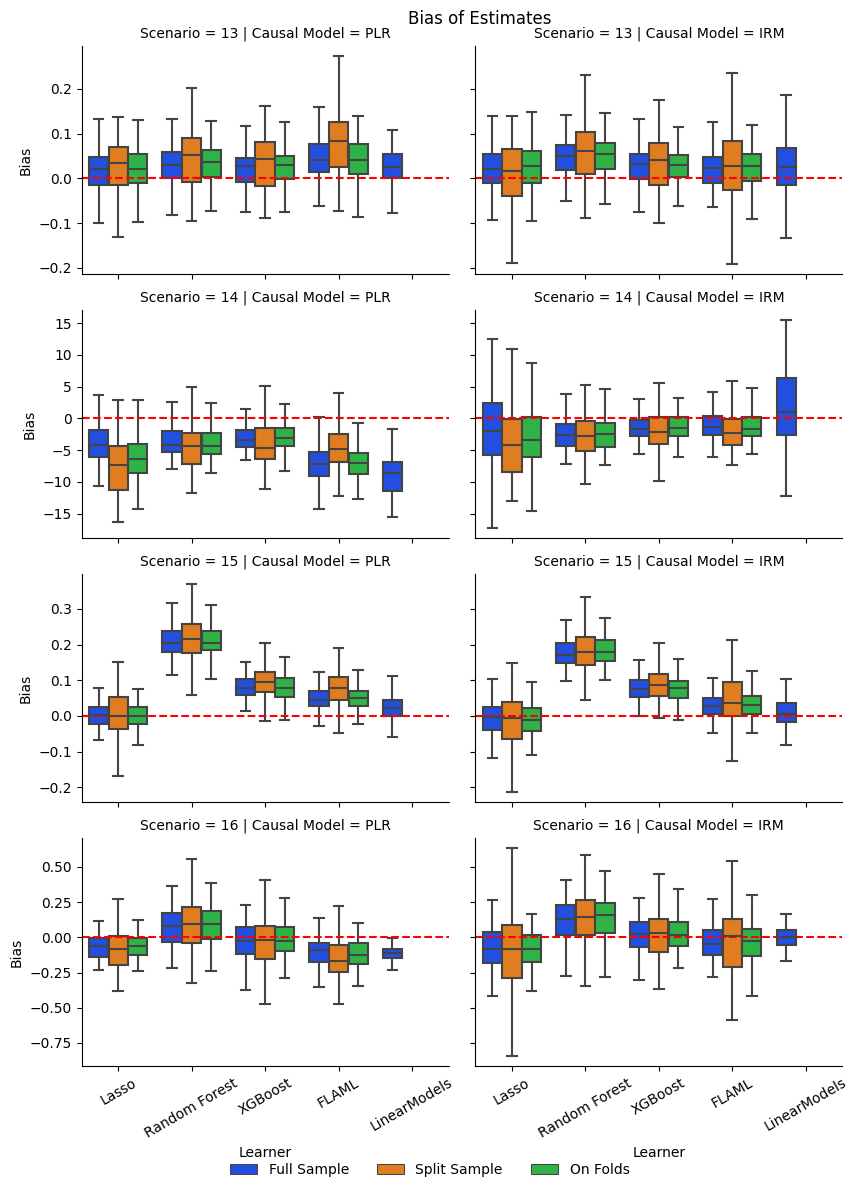}
    \caption{Boxplot of bias, ACIC DGP 13-15}
    \label{fig:biasall4}
\end{figure}

\subsection{Predictive Performance} \label{app:predperf}
\begin{figure}[H]
    \centering
    \includegraphics[height=.9\textheight]{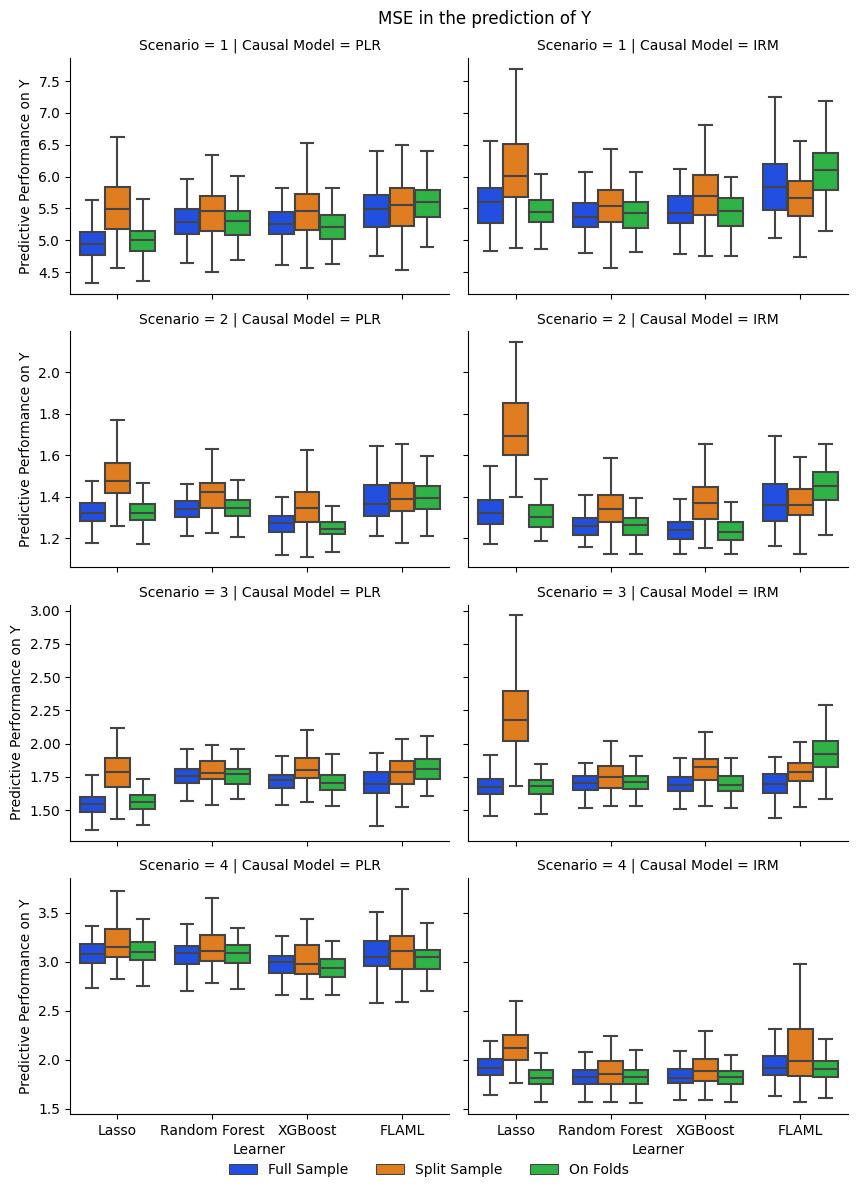}
    \caption{Predictive Performance on $Y$ of ACIC DGP 1-4}
    \label{fig:predyall1}
\end{figure}

\begin{figure}[H]
    \centering
    \includegraphics[height=.9\textheight]{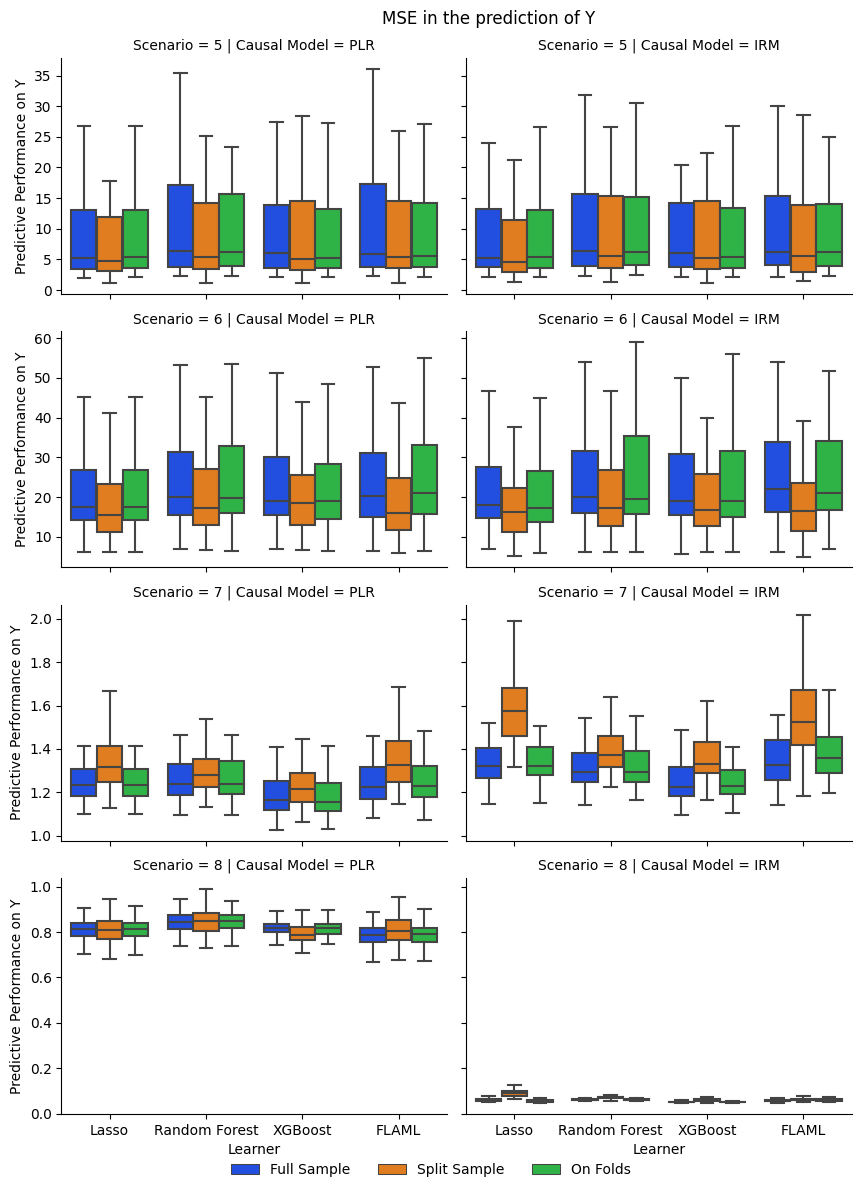}
    \caption{Predictive Performance on $Y$ of ACIC DGP 5-8}
    \label{fig:predyall2}
\end{figure}

\begin{figure}[H]
    \centering
    \includegraphics[height=.9\textheight]{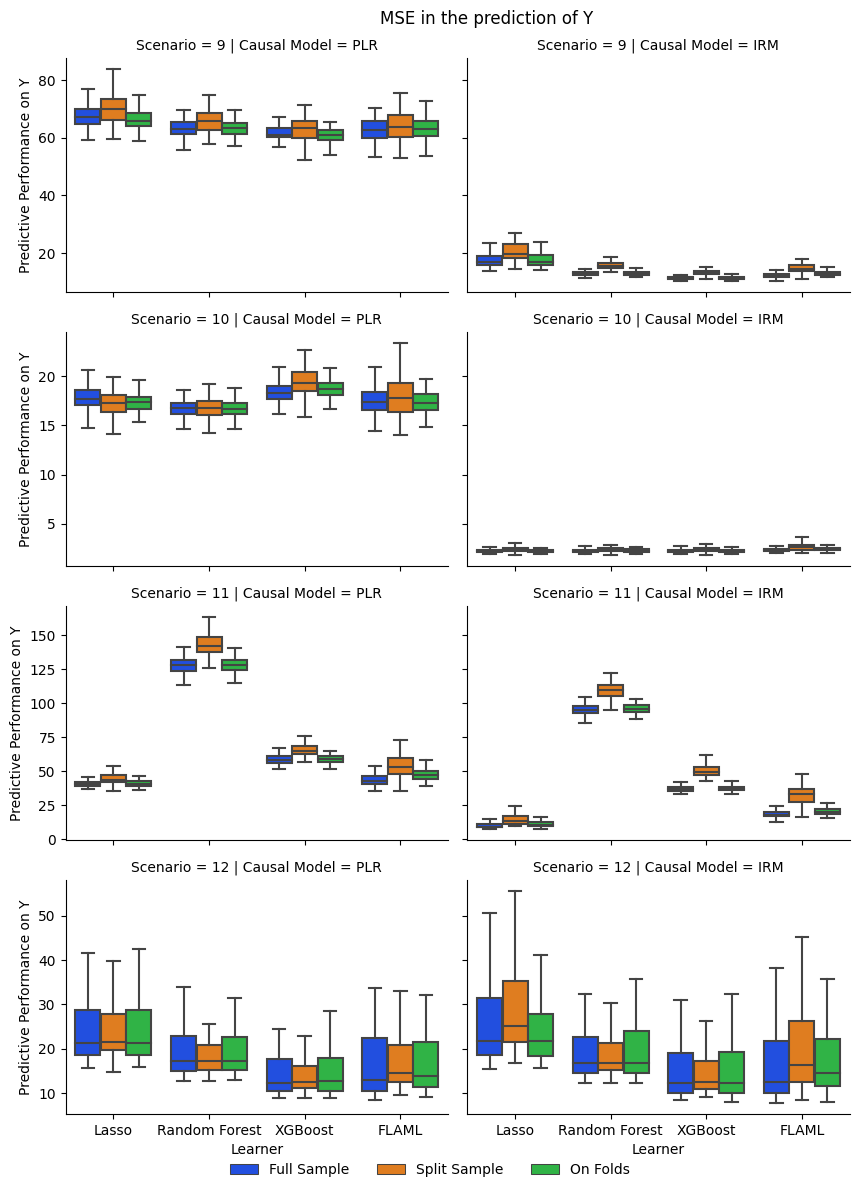}
    \caption{Predictive Performance on $Y$ of ACIC DGP 9-12}
    \label{fig:predyall3}
\end{figure}

\begin{figure}[H]
    \centering
    \includegraphics[height=.9\textheight]{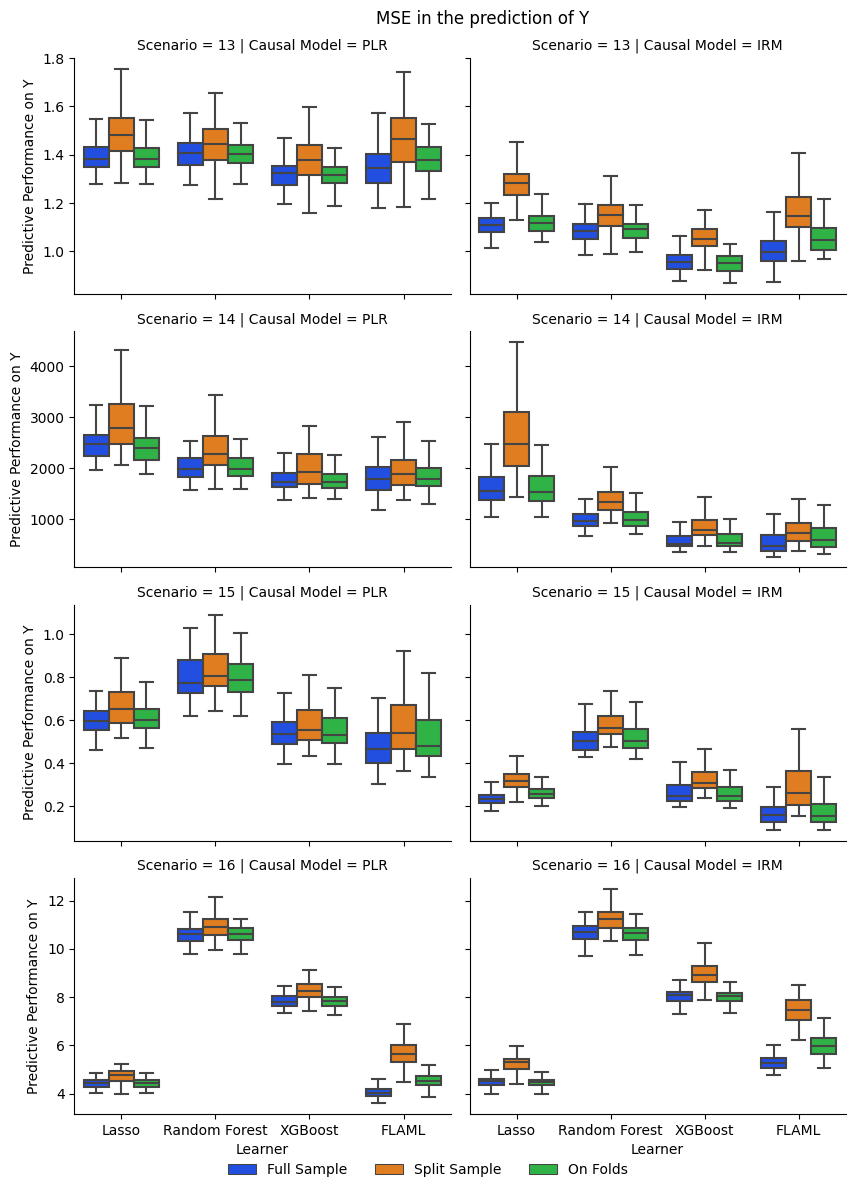}
    \caption{Predictive Performance on $Y$ of ACIC DGP 13-15}
    \label{fig:predyall4}
\end{figure}

\subsection{Combined Loss}\label{app:absbiasvloss}
\begin{figure}[H]
    \centering
    \includegraphics[height=.9\textheight]{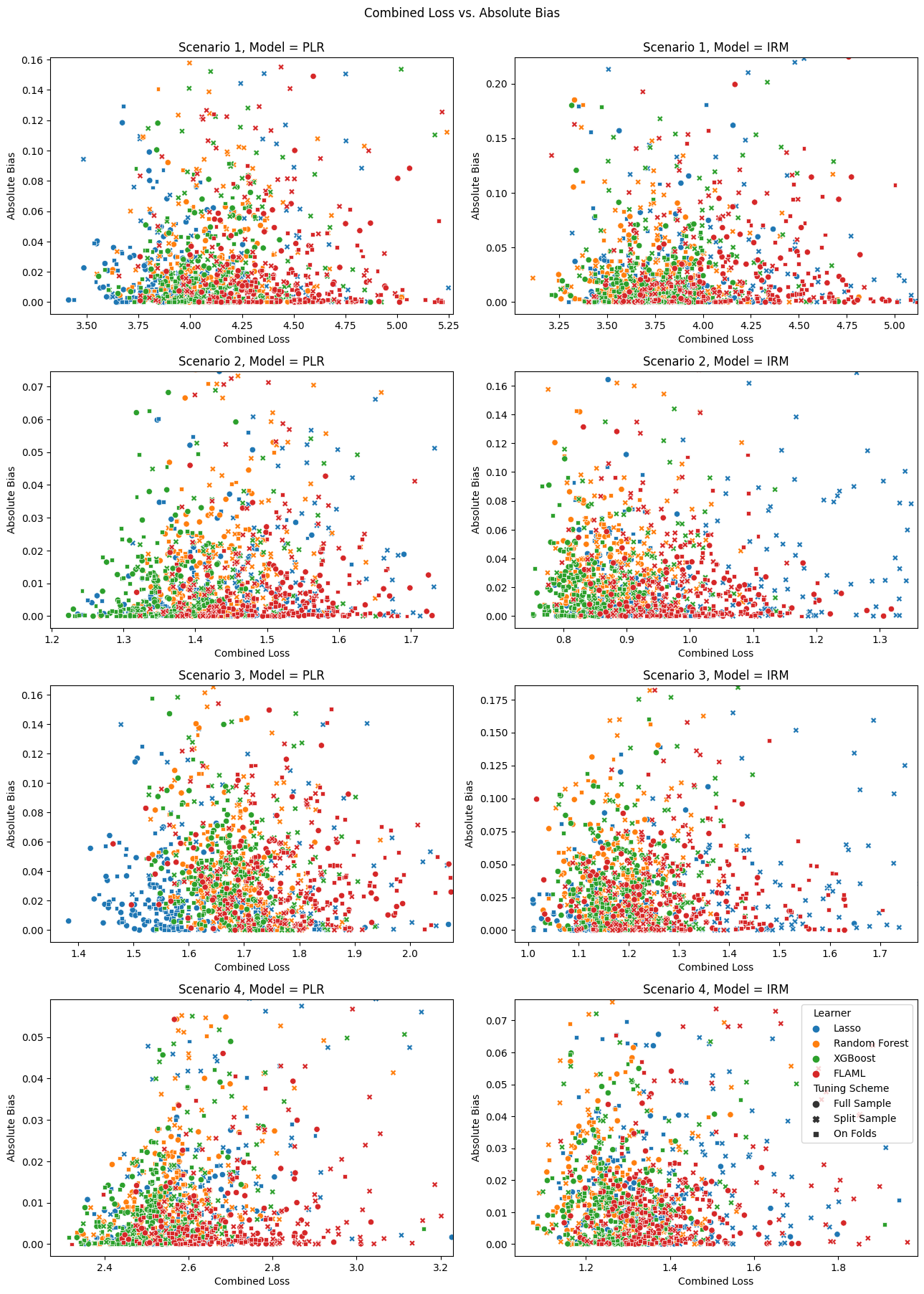}
    \caption{Combined Loss of ACIC DGP 1-4}
    \label{fig:comblossall1}
\end{figure}

\begin{figure}[H]
    \centering
    \includegraphics[height=.9\textheight]{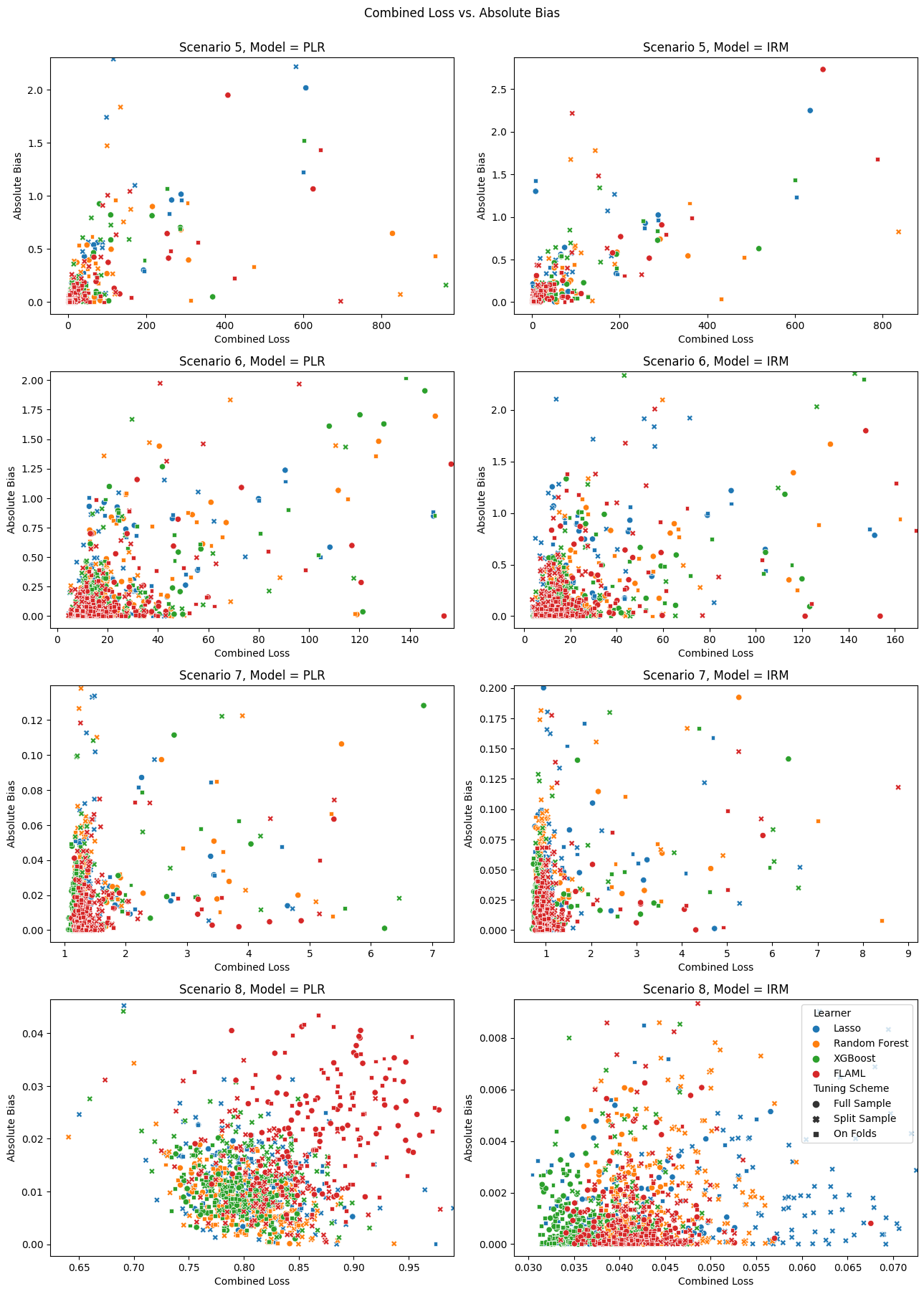}
    \caption{Combined Loss of ACIC DGP 5-8}
    \label{fig:comblossall2}
\end{figure}

\begin{figure}[H]
    \centering
    \includegraphics[height=.9\textheight]{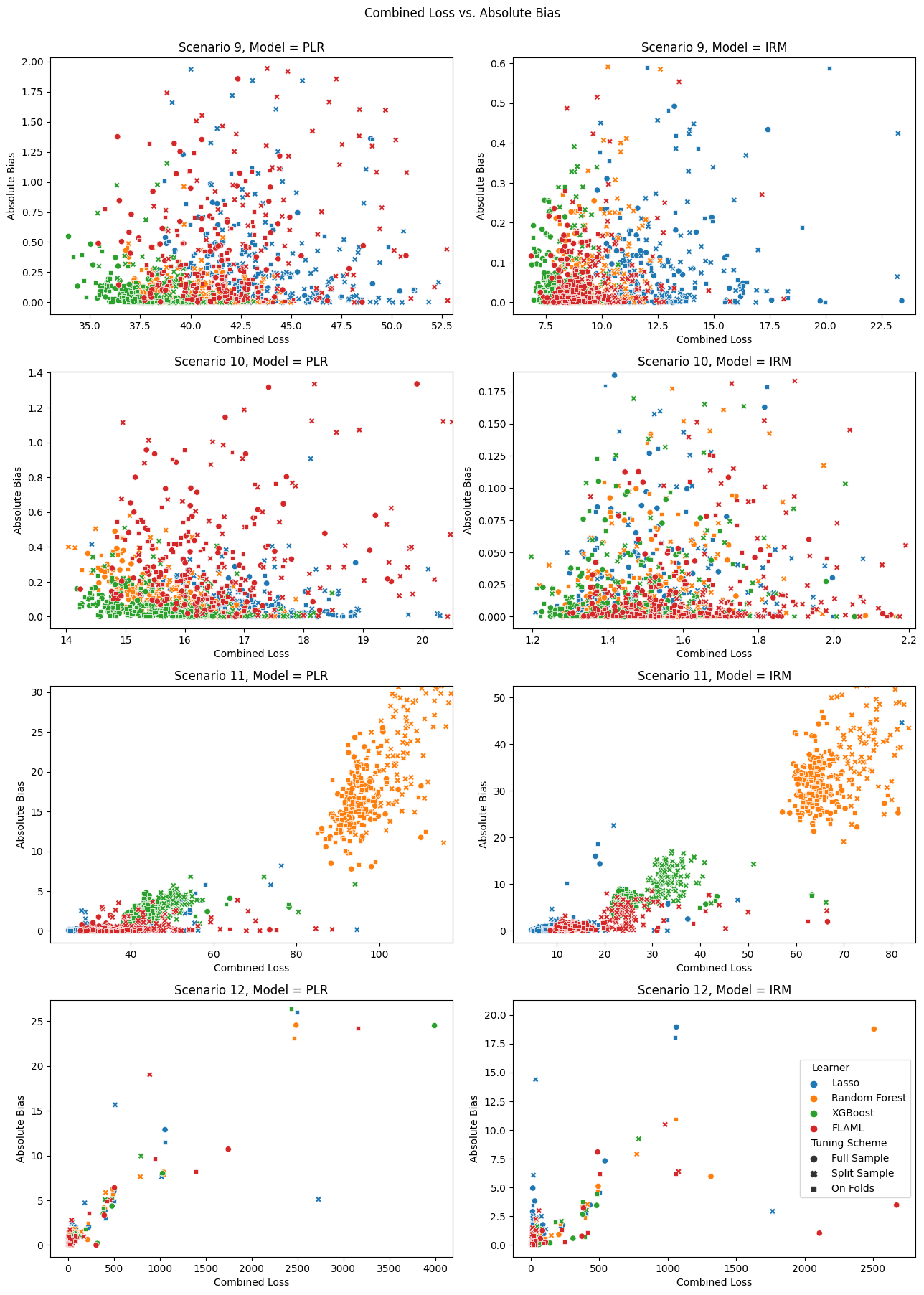}
    \caption{Combined Loss of ACIC DGP 9-12}
    \label{fig:comblossall3}
\end{figure}

\begin{figure}[H]
    \centering
    \includegraphics[height=.9\textheight]{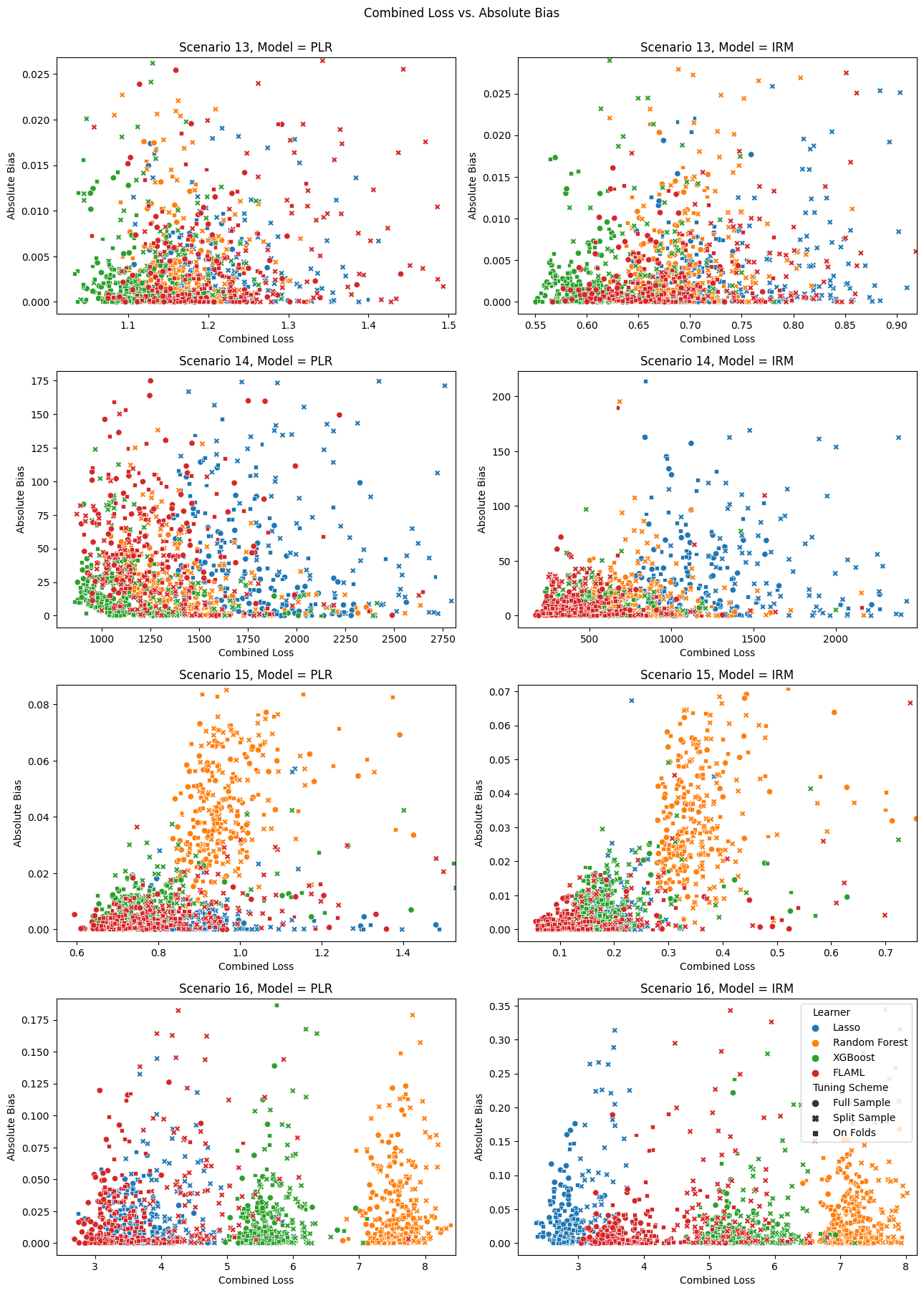}
    \caption{Combined Loss of ACIC DGP 13-15}
    \label{fig:comblossall4}
\end{figure}

\subsection{RRMSE-1 vs. Bias vs. Combined Loss PLR}\label{app:tripleplotplr}
\begin{figure}[H]
    \centering
    \includegraphics[height=.9\textheight]{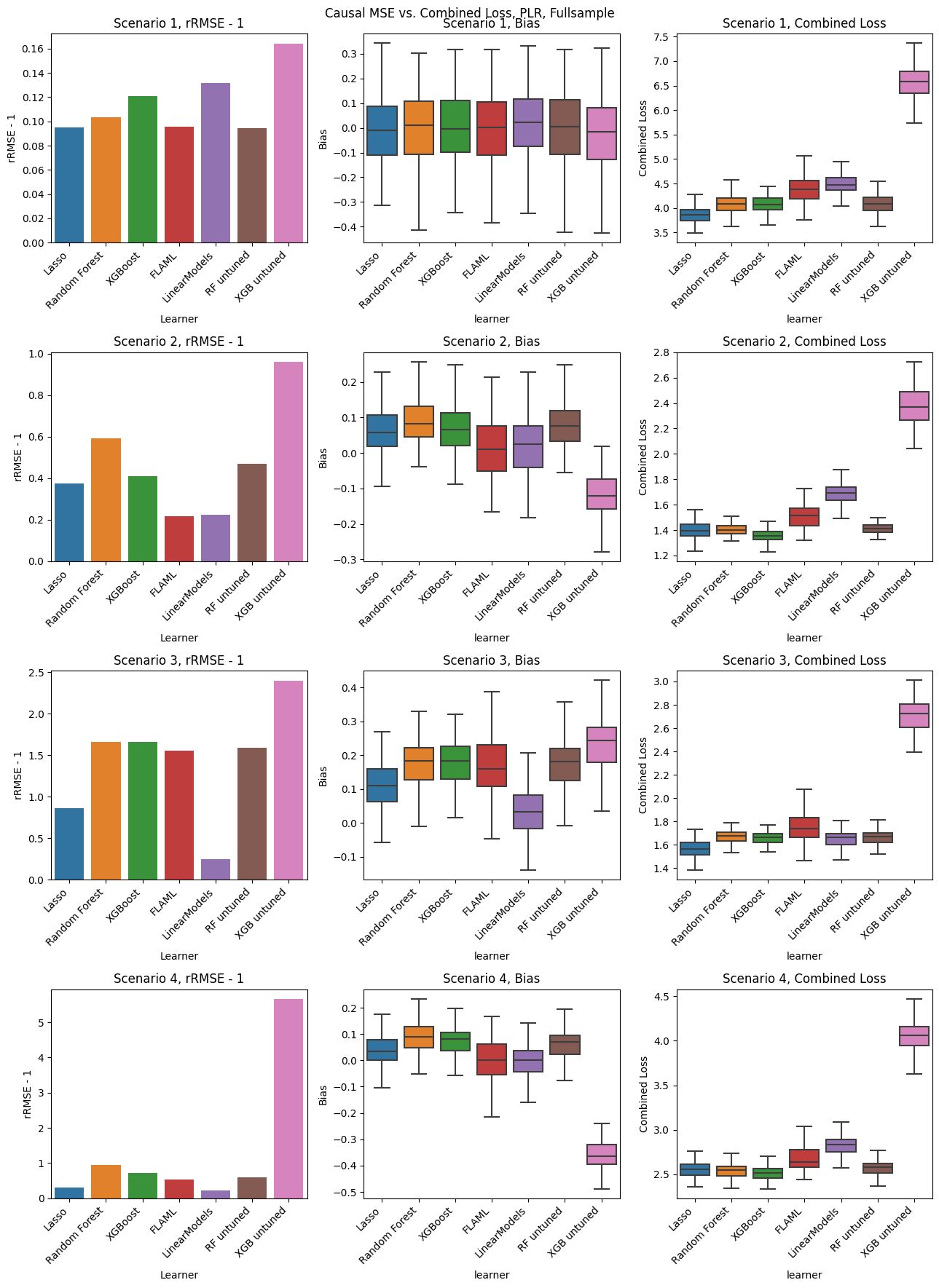}
    \caption{Combined Loss of ACIC DGP 1-4}
    \label{fig:tripleplotplr1}
\end{figure}

\begin{figure}[H]
    \centering
    \includegraphics[height=.9\textheight]{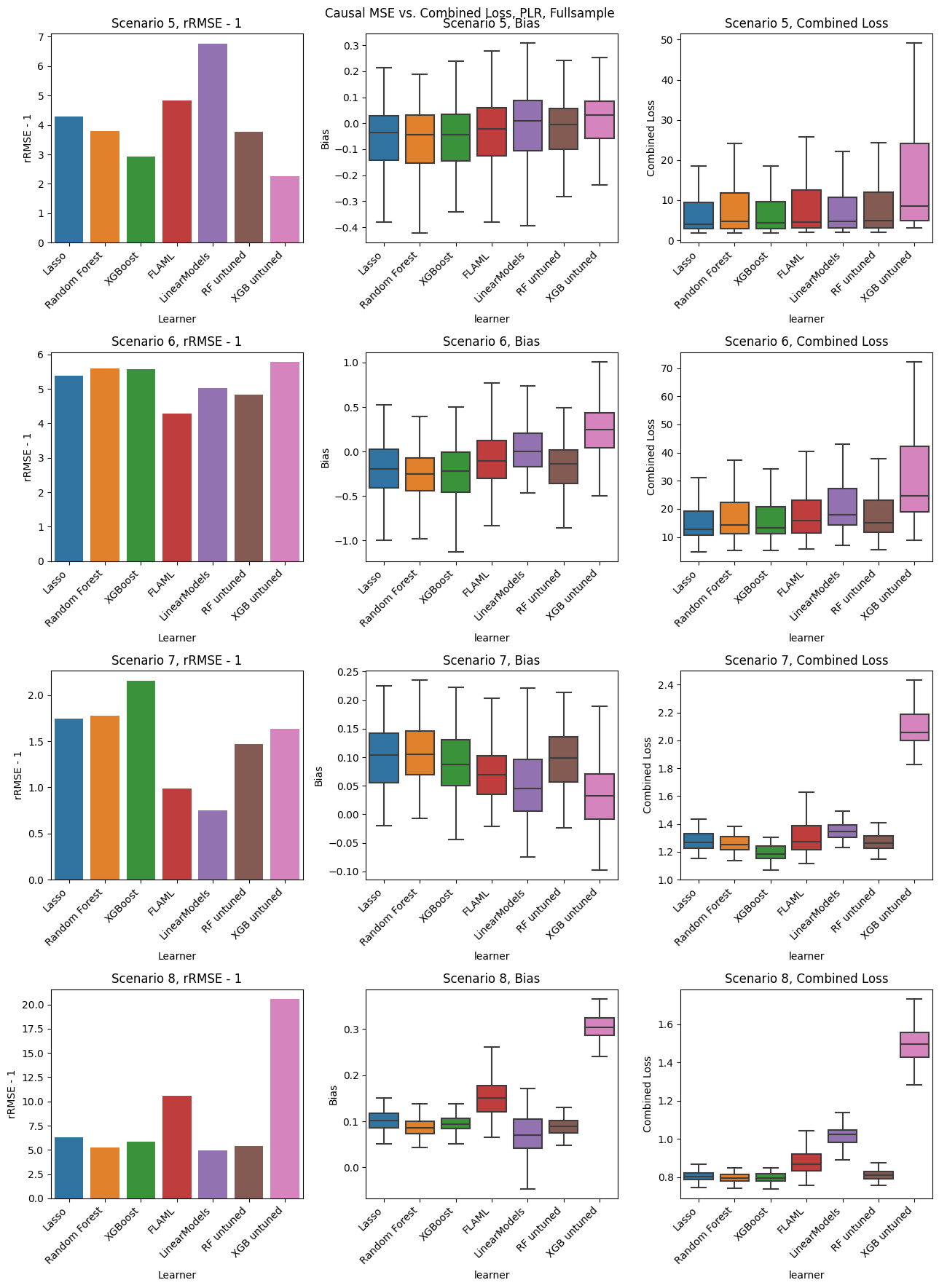}
    \caption{Combined Loss of ACIC DGP 5-8}
    \label{fig:tripleplotplr2}
\end{figure}

\begin{figure}[H]
    \centering
    \includegraphics[height=.9\textheight]{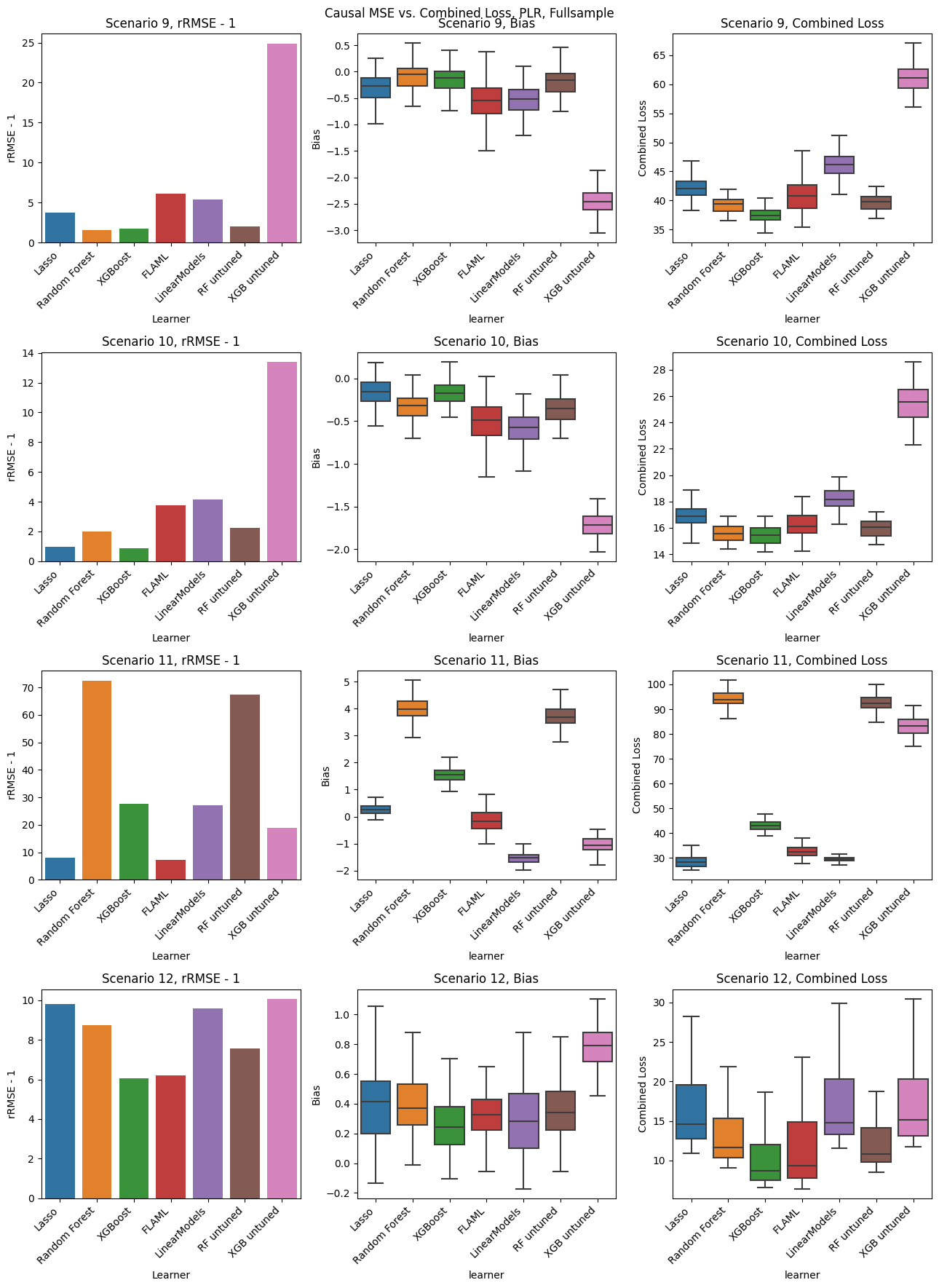}
    \caption{Combined Loss of ACIC DGP 9-12}
    \label{fig:tripleplotplr3}
\end{figure}

\begin{figure}[H]
    \centering
    \includegraphics[height=.9\textheight]{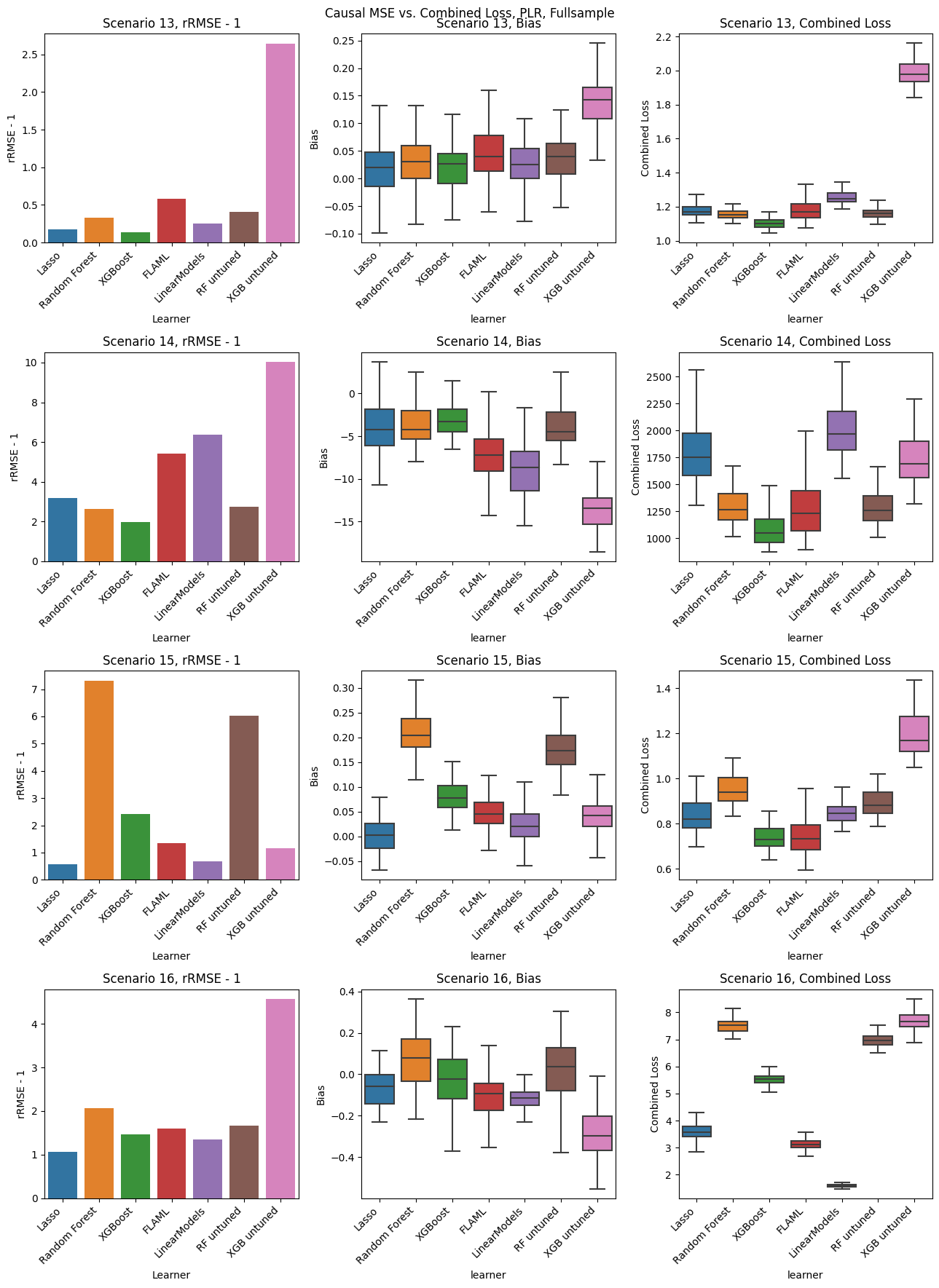}
    \caption{Combined Loss of ACIC DGP 13-15}
    \label{fig:tripleplotplr4}
\end{figure}

\subsection{RRMSE-1 vs. Bias vs. Combined Loss IRM}\label{app:tripleplotirm}
\begin{figure}[H]
    \centering
    \includegraphics[height=.9\textheight]{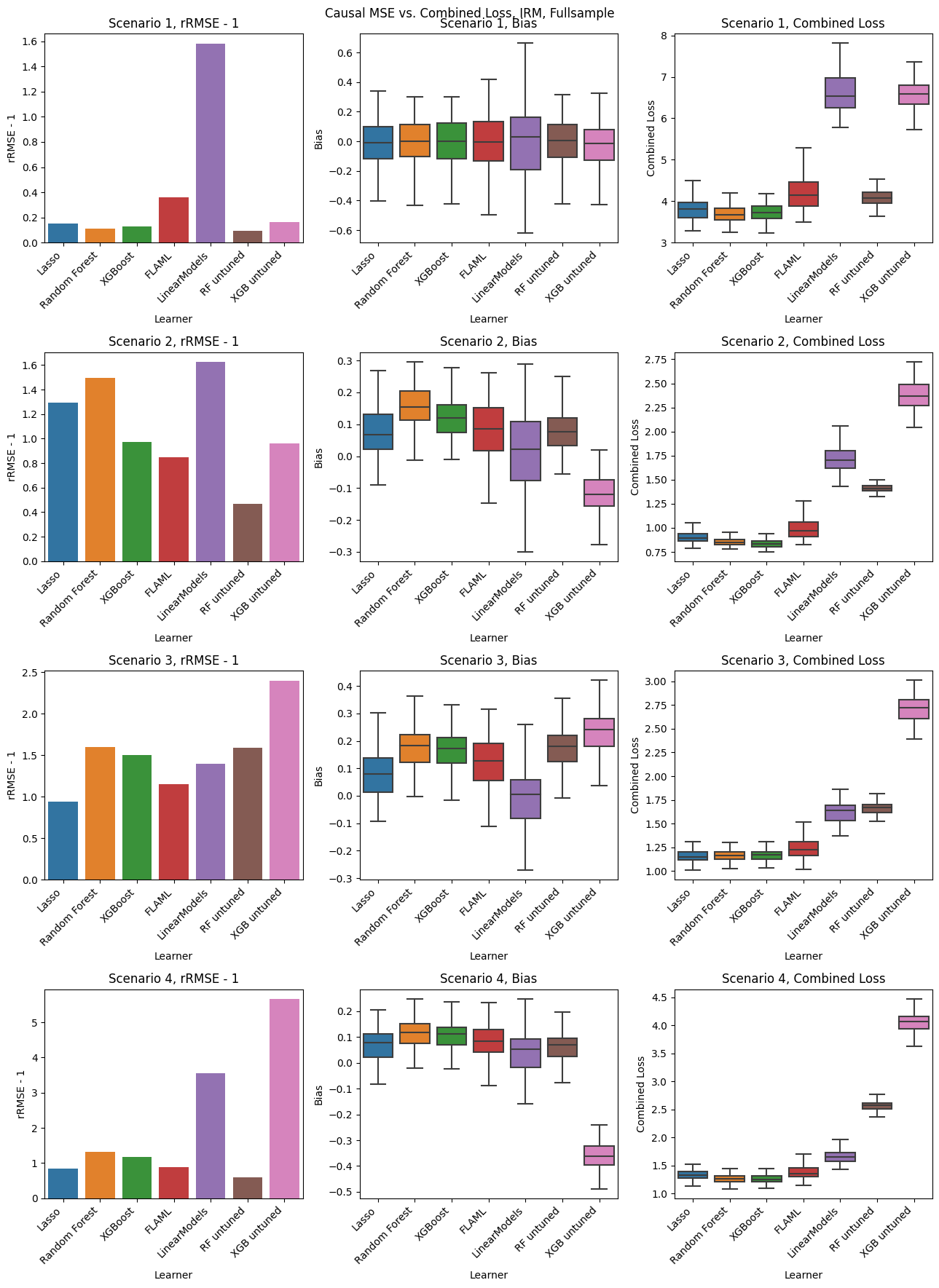}
    \caption{Comparison of RRMSE-1, Bias and Combined loss of ACIC DGP 1-4}
    \label{fig:tripleplotirm1}
\end{figure}

\begin{figure}[H]
    \centering
    \includegraphics[height=.9\textheight]{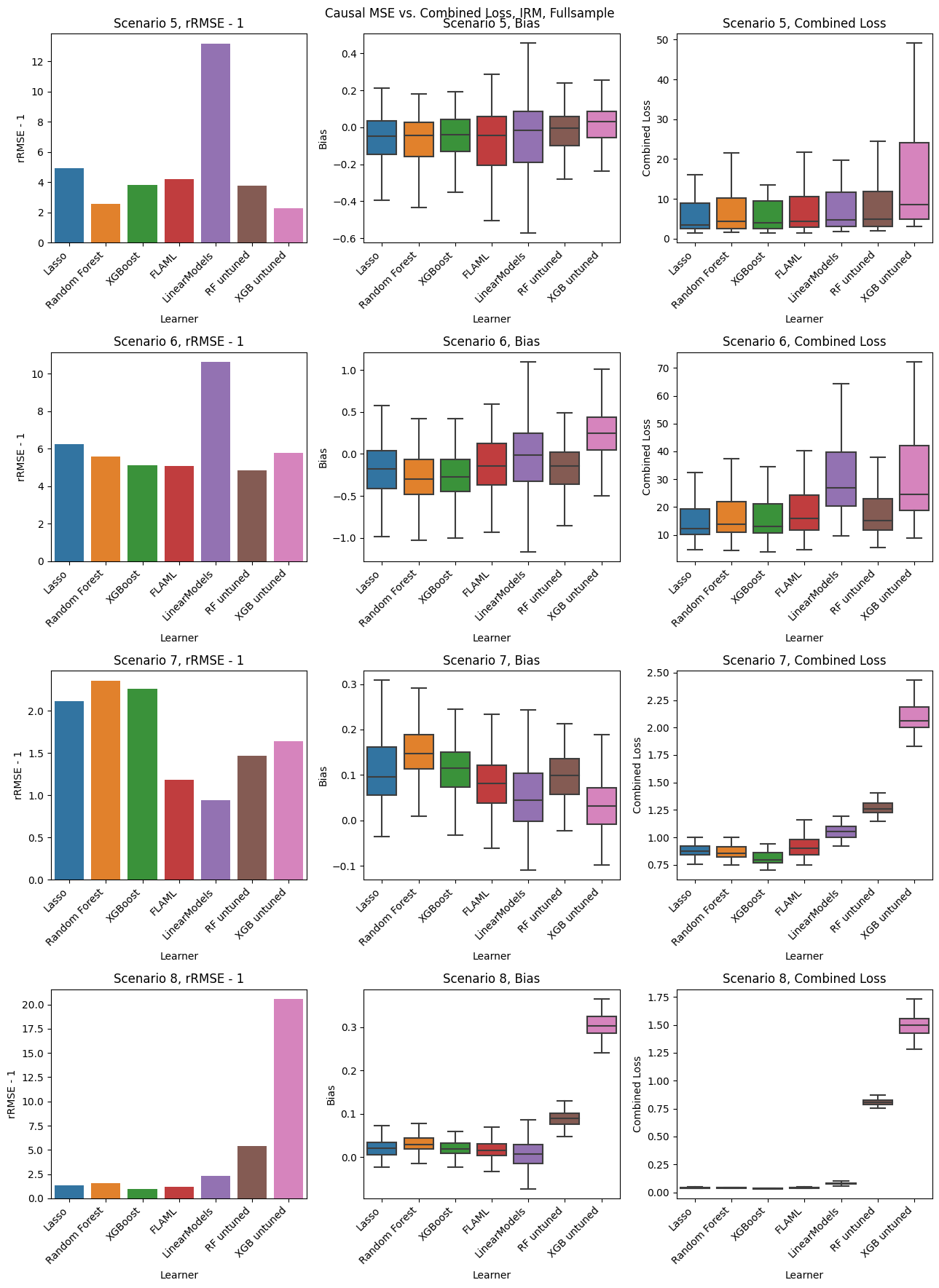}
    \caption{Comparison of RRMSE-1, Bias and Combined loss of ACIC DGP 5-8}
    \label{fig:tripleplotirm2}
\end{figure}

\begin{figure}[H]
    \centering
    \includegraphics[height=.9\textheight]{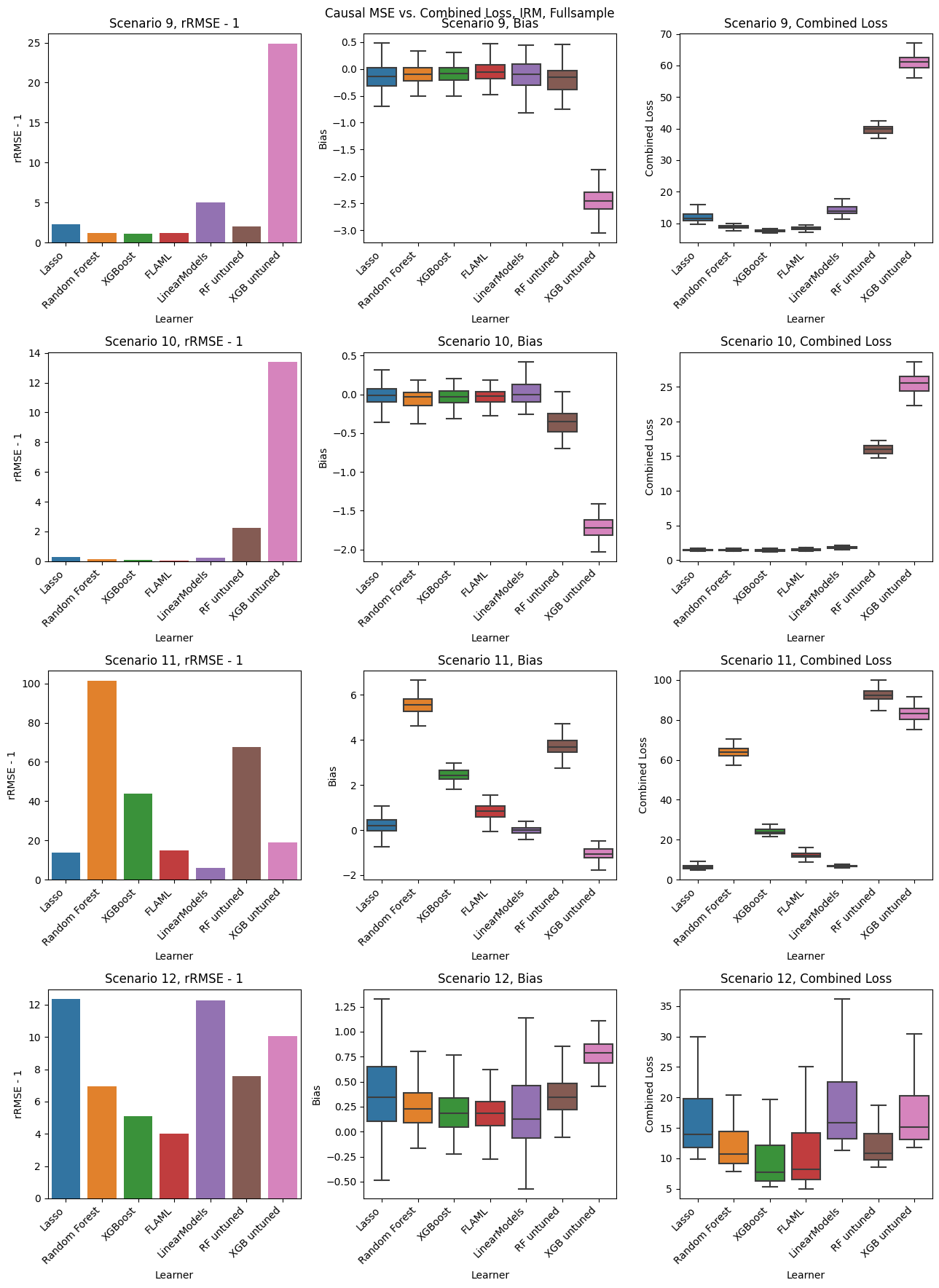}
    \caption{Comparison of RRMSE-1, Bias and Combined loss of ACIC DGP 9-12}
    \label{fig:tripleplotirm3}
\end{figure}

\begin{figure}[H]
    \centering
    \includegraphics[height=.9\textheight]{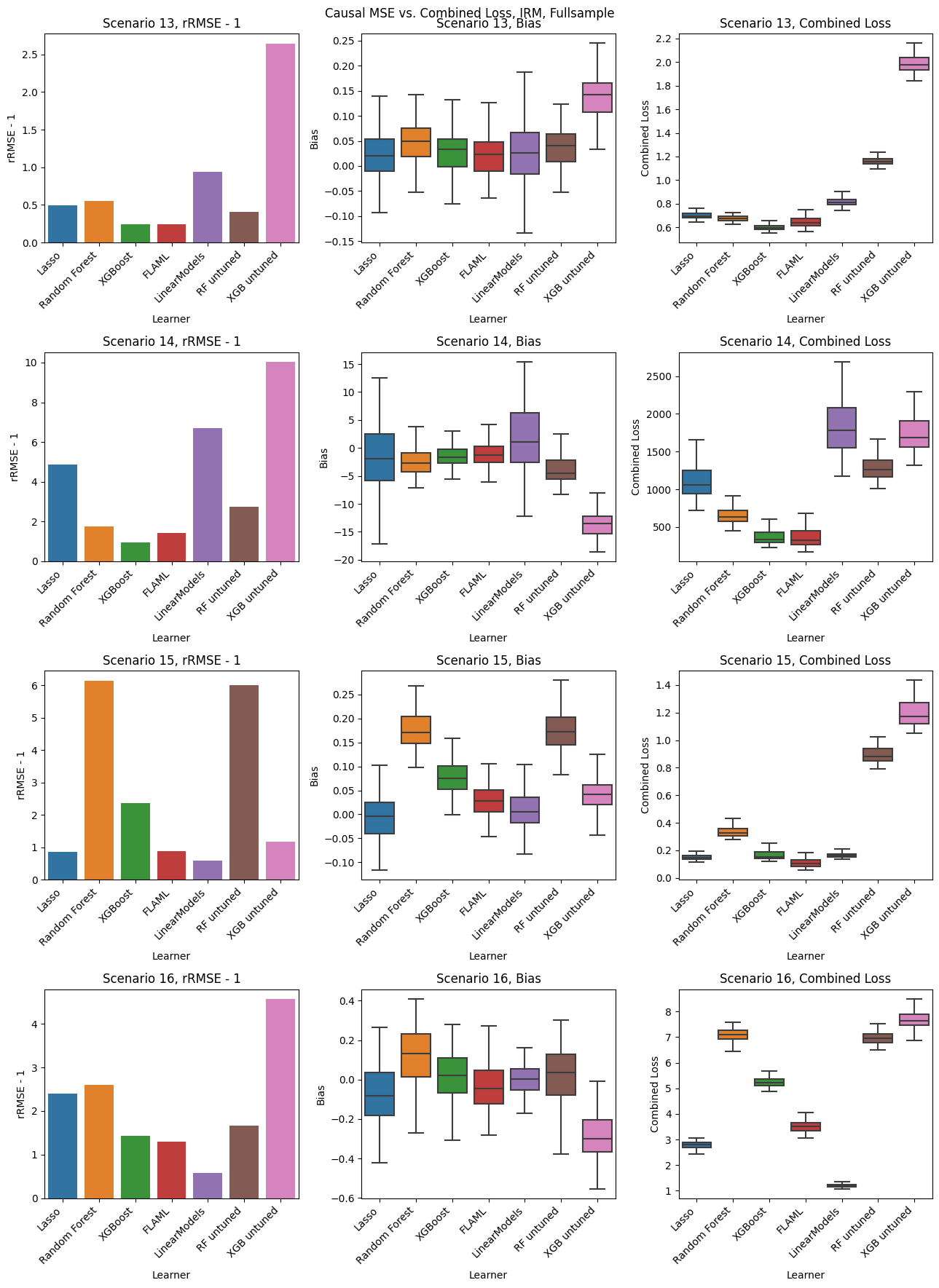}
    \caption{Comparison of RRMSE-1, Bias and Combined loss of ACIC DGP 13-15}
    \label{fig:tripleplotirm4}
\end{figure}
\section{ACIC 2019 DGP overview}\label{app:dgp}
In the following, we present a rough structure of the DGPs used. This is meant for an understanding of how the DGP look, presenting the exact data generating process with all coefficients would be infeasible. The $X_i$ are drawn from multiple distributions in multiple copula.
\subsection{DGP 1}
\begin{align*}
p &= \alpha_{0} + \sum_{j = 1}^{200} \alpha_{j} * X_{j} \\
A &\thicksim Bin(1, \textup{sig}(p)) \\
Y &= \theta*A +\sum_{i = 1}^{200} \beta_{i} * X_{i}+ \epsilon \\
\epsilon &\thicksim \mathcal{N}(0, 2)) 
\end{align*}
\subsection{DGP 2}
\begin{align*}
p &= \alpha_{0} + \alpha_1*\sqrt{X_1} + \alpha_2* X_5 + \alpha_3 * X_{32} + \alpha_4 X_5*X_{32} + \alpha_5* X_{70} \\
&+ \alpha_6 * X_{70}^2 + \alpha_7* [ \mathbbm{1} [X_{101} > 2.5]] + \alpha_8 * X_{150} + \alpha_9*[ \mathbbm{1} [X_{179} < -0.5]] * (X_{179}+ 1.5) \\
A &\thicksim Bin(1, \textup{sig}(p)) \\
Y &= \theta * A + \beta_{1} * \sqrt{X_{1}} + \beta_{2}*X_{5} + \beta_{3}*X_{23}+\beta_{4}* X_{32} + \beta_{5}*X_{5}*X_{32}+\beta_{6}*X_{70}+\beta_{7}*X_{70}^2\\
&+\beta_{8}* [ \mathbbm{1} [X_{101} > 2.5]] + \beta_{9}* X_{15}*X_{10}* [ \mathbbm{1} [X_{179} < -0.5]]*(X_{179}+1.5)+\beta_{11}*X_{179} + \epsilon\\
\epsilon &\thicksim \mathcal{N}(0, 1) \\  
\end{align*}
\subsection{DGP 3}
\begin{align*}
p &= \sum_{i=1}^{90} \alpha_{i} * X_{i} \quad \textup{i are 90  random  indices}\\
A &\thicksim Bin(1, \textup{sig}(p)) \\
Y &= \sum_{i=1}^{90} \alpha_{i} * X_{i}\\
\epsilon &\thicksim \mathcal{N}(0, 1) \\
\end{align*}
\subsection{DGP 4}
\begin{align*}
p &= \alpha_{0} + \alpha_{1}* X_{2} + \alpha_{2}* X_{5} + \alpha_{3}* X_{2}*X_{5}+ \alpha_{4}*X_{12}+ \alpha_{5}*X_{23}\\
&+ \alpha_{6} * X_{23}^{2} + \alpha_{7}* X_{12} *X_{23}^{2}+ \alpha_{8}* \sqrt{X_{67}}+\alpha_{9}*X_{77}+ \alpha_{10}* [ \mathbbm{1} [X_{89} > 19]]\\
&+ \alpha_{11} * [ \mathbbm{1} [X_{95} > 5]]*(X_{95}-3)+ \alpha_{12}* \exp(X_{106}) + \alpha_{13}* X_{122} + \alpha_{14} * X_{146}+\alpha_{15}* X_{122} * X_{146}\\ &+\alpha_{16}* X_{150} + \alpha_{17}* X_{168} +\alpha_{18}* [ \mathbbm{1} [X_{199} > 1]]  \\
A &\thicksim Bin(1, \textup{sig}(p))\\
Y &= (\theta * A) + \beta_{1}+ \beta_{2}* X_{2}+ \beta_{3}*X_{5}+ \beta_{4}* X_{2}*X_{5}+ \beta_{5}*X_{12}\\
&+\beta_{6}*X_{23}+\beta_{7}*X_{23}^{2}+\beta_{8}*X_{12}*X_{23}^{2}+\beta_{9}*X_{40}+ \beta_{10}* \sqrt{X_{67}}+ \beta_{11}*X_{77}\\
&+ \beta_{12}* [ \mathbbm{1} [X_{89} > 19]] + \beta_{13}*[ \mathbbm{1} [X_{95} > 5]]*(X_{95}-3)+ \beta_{14}* \exp(X_{106}) + \beta_{15}* X_{122}+ \beta_{16}* X_{133} \\
&+ \beta_{17}* X_{146} +  \beta_{18}* X_{122} * X_{146}+\beta_{19}* X_{150}* \beta_{20}* X_{168} + \beta_{21}* X_{198}\\
&+ \beta_{22}* [ \mathbbm{1} [X_{199} > 1]] + \epsilon\\
\epsilon &\thicksim t(5) \\
\end{align*}

\subsection{DGP 5}
\begin{align*}
P&=\alpha_{0}+\alpha_{1}*X_{23}+ \alpha_{2}*X_{23}^2+\alpha_{3}*\sqrt{X_{67}}+\alpha_{4}*X_{77}\\
&+\alpha_{5}* [ \mathbbm{1} [X_{89} > 19]]+\alpha_{6}* [ \mathbbm{1} [X_{95} > 5]]*(X_{95}-3)+ \alpha_{7}*\exp(X_{106})+\alpha_{8}*X_{122}\\
&+\alpha_{9}*X_{146}+\alpha_{10}*X_{122}*X_{146}+ \alpha_{11}*X_{150}+\alpha_{12}*X_{168}+\alpha_{13}* [ \mathbbm{1} [X_{199} > 1]]  \\
A &\thicksim Bin(1, \textup{sig}(p)) \\
Y &= \textup{exp}(\theta*A + \beta_{1} + \beta_{2}*\sqrt{X_{67}}+ \beta_{3}*X_{77}+\beta_{4}*[ \mathbbm{1} [X_{89} > 19]]\\
&+ \beta_{5}*\ \mathbbm{1} [X_{95} > 5]]*(X_{95}-3)+ \beta_{6}* \exp(X_{106}) + \beta_{7}*X_{122}+ \beta_{8}*X_{146}+ \beta_{9}*X_{146}*X_{122}\\
&+ \beta_{10}*X_{150}+ \beta_{11}*X_{168}+\beta_{12}* [ \mathbbm{1} [X_{199} > 1]] + \epsilon)\\
\epsilon &\thicksim t(12) \\
\end{align*}

\subsection{DGP 6}
\begin{align*}
p &= \alpha_{0}+\alpha_{1}*X_{4}+ \alpha_{2}*X_{19}+ \alpha_{3}*X_{44}\\
A &\thicksim Bin(1, \textup{sig}(p)) \\
Y &= \textup{exp}(A*(\theta+\tau*X_{55}) + \beta_{1} + \beta_{2}*X_{4} + \beta_{3}* X_{19}+ \beta_{4}* X_{44}+ \epsilon) \\
\epsilon &\thicksim t(19), \theta = 0.4, \tau = 0.2\\
\end{align*}

\subsection{DGP 7}
\begin{align*}
p&= \alpha_{0} + \alpha_{1}*X_{3} + \alpha_{2}*X_{6} + \alpha_{3}*X_{3}*X_{6} + \alpha_{4}*X_{24} + \alpha_{5}*X_{24}^2 + \alpha_{6}*X_{35} + \alpha_{7}*X_{68} \\
&+ \alpha_{8}* \sqrt{X_{68}} + \alpha_{9}*X_{35}* \sqrt{X_{68}} + \alpha_{10}*[ \mathbbm{1} [X_{96} < 2]]*(X_{96}-1) + \alpha_{11}*\exp{(X_{107})} + \alpha_{12}*X_{123} \\
&+ \alpha_{13}*X_{149} + \alpha_{14}*X_{123}*X_{149} + \alpha_{15}*X_{151} + \alpha_{16}*1/X_{169} + \alpha_{17}*X_{200}  \\
A &\thicksim Bin(1, \textup{sig}(p)) \\
Y &= \theta * A + \beta_{1} + \beta_{2}*X_{3} + \beta_{3}*X_{6} + \beta_{4}*X_{3}*X_{6} + \beta_{5}*X_{24} +\beta_{6}*X_{24}^2 \\
&+ \beta_{7}*X_{35} + \beta_{8}*X_{40} + \beta_{9}*X_{68} + \beta_{10}* \sqrt{X_{68}} + \beta_{11}*X_{35}* \sqrt{X_{68}} \\
&+ \beta_{12}*[\mathbbm{1} [X_{96} < 2]]*(X_{96}-1) + \beta_{13}*\exp{X_{107}} + \beta_{14}*X_{123} + \beta_{15}*X_{133} + \beta_{16}*X_{149} \\
&+ \beta_{17}*X_{123}*X_{149} + \beta_{18}*X_{151} + \beta_{19}*1/X_{169} + \beta_{20}*X_{198} +\beta_{21}*X_{200} + \epsilon \\
\epsilon &\thicksim \mathcal{N}(0,1) \\
\end{align*}

\subsection{DGP 8}
\begin{align*}
p &= \alpha_{0} + \alpha_{1}* \log(X_{7}+1) + \alpha_{2}*1/X_{16}+ \alpha_{3}* |X_{51}| +\alpha_{4}*X_{156} \\
&+ \alpha_{5}* X_{156}/\log(X_{7}+0.5) + \alpha_{6}*X_{163}^2 \\
A &\thicksim Bin(1, \textup{sig}(p)) \\
 \overline{y} &= \theta*A + \beta_{1} + \beta_{2}*X_{7} + \beta_{3}*X_{16} + \beta_{4}*X_{51} + \beta_{5}*X_{156}\\
 Y &= 10/(2 + \exp(-1*\overline{y})) + \epsilon\\
\epsilon &\thicksim \mathcal{N}(0,0.2)\\
\end{align*}

\subsection{DGP 9}
\begin{align*}
p &= \alpha_{0} + \alpha_{1}* \log(X_{8}) + \alpha_{2}*1/\sqrt{X_{17}} + \alpha_{3}* \exp(-X_{52}) +\alpha_{4}*|X_{157}| \\
&+ \alpha_{5}* X_{157}*\log(X_{8}) + \alpha_{6}*X_{164}+ \alpha_{7}*X_{165}+ \alpha_{8}*X_{164}*X_{165}\\
A &\thicksim Bin(1, \textup{sig}(p)) \\
 \overline{y_{1}} &= \theta*A + \beta_{1} + \beta_{2}*X_{8} + \beta_{3}*X_{17}\\
 \overline{y_{2}} &= \beta_{4}*X_{52} + \beta_{5}*X_{157}+ \beta_{6}*X_{166}\\
 \tilde{y} &= \exp(\overline{y_{1}})+(\overline{y_{2}} + 59)*(\overline{y_{2}}>-59.5)\\
Y &= \tilde{y} + \epsilon \\
\epsilon &\thicksim \mathcal{N}(0,3)\\
\end{align*}

\subsection{DGP 10}
\begin{align*} 
p &= \alpha_{0} + \alpha_{1}*X_{1} + \alpha_{2}* \log(X_{8}) + \alpha_{3}*1/\sqrt{X_{17}} + \alpha_{4}*\exp(-X_{52}) + \alpha_{5}* |X_{157}|\\
&+ \alpha_{6}*X_{157}*\log(X_{8})+ \alpha_{7}*X_{164}+ \alpha_{8}*X_{165}+ \alpha_{9}*|X_{157}|*X_{165} \\
A &\thicksim Bin(1, \textup{sig}(p)) \\
Y &= A*(\theta+X_{8}*\lambda+X_{157}*X_{166}*\delta) + \beta_{1} + \beta_{2}*X_{8} + \beta_{3}*X_{17}+ \beta_{4}*X_{52}\\
&+ \beta_{5}*X_{157}+ \beta_{6}*X_{166}+\beta_{7}*X_{157}*X_{166} + \epsilon\\
\epsilon &\thicksim t(4),
\theta = 5, \lambda = 5, \delta = 0.5\\
\end{align*}

\subsection{DGP 11}
\begin{align*} 
p &= \sum_{i=1}^{18} \alpha_{i} * X_{i} \quad \textup{(18  random drawn indices)} \\
A &\thicksim Bin(1, \textup{sig}(p))\\
\Tilde{Y} &= \sum_{i=1}^{23} \alpha_{i} * X_{i}+\theta*A\\
Y &= \Tilde{Y}^2 +\epsilon\\
\epsilon &\thicksim N(0,2) \\
\end{align*}

\subsection{DGP 12}
\begin{align*}
p &= \alpha_{0} + \alpha_{1}* (X_{5}-200)*(X_{5}>204)+ \alpha_{2}* \log(X_{7}+1) + \alpha_{3}*1/(X_{16})+ \alpha_{4}* \sqrt{X_{25}} \\
&+ \alpha_{5}*|X_{51}|+ \alpha_{6}*X_{96} + \alpha_{7}*X_{96}/(\log(X_{7}+1)+0.5) + \alpha_{8}*X_{156} + \alpha_{9}*X_{163}^2 \\
A &\thicksim Bin(1, \textup{sig}(p)) \\
 \overline{y} &= -0.5*A+ \beta_{1} + \beta_{2}* (X_{5}-200)*(X_{5}>204) + \beta_{3}*\sqrt{X_{25}}+ \beta_{4}*X_{96}\\
 &+ \beta_{5}*X_{163}^2+ \beta_{6}*|X_{169}|+\beta_{7}*X_{188}^3 \\
 Y &= 3/(\exp(-0.6*\overline{y})) + \epsilon\\
\epsilon &\thicksim \mathcal{N}(0,2)\\
\end{align*}

\subsection{DGP 13}
\begin{align*}
p &= \alpha_{0} + \alpha_{1}* (X_{5}-200)*(X_{5}>204)+ \alpha_{2}* \sqrt{X_{25}} + \alpha_{3}* (X_{5}-200)*(X_{5}>204)*\sqrt{X_{25}} \\
&+ \alpha_{4}*X_{96}+ \alpha_{5}*X_{163}+ \alpha_{6}*X_{96}*X_{163} + \alpha_{7}*X_{163}^2 + \alpha_{8}*|X_{169}| + \alpha_{9}*X_{188}^3 \\
&+ \alpha_{10}*|X_{169}|*X_{188}^3\\
A &\thicksim Bin(1, \textup{sig}(p)) \\
\tilde{y} &= A*(-0.5 -0.25*X_{163}^2) + \beta_{1} + \beta_{2}* (X_{5}-200)*(X_{5}>204) + \beta_{3}*\log(X_{7}+1) \\
&+ \beta_{4}*1/X_{16}+ \beta_{5}*|X_{51}|+ \beta_{6}*X_{96}+\beta_{7}*X_{96}/(\log(X_{7}+1)+0.5) + \beta_{8}*X_{156} \\
&+ \beta_{9}*X_{163}^2 + \beta_{10}*X_{188}^2\\
Y &= \tilde{y} + \epsilon\\
\epsilon &=  (1-A)*\epsilon_{1} + A*\epsilon_{2}\\\
\epsilon_{1} &\thicksim \mathcal{N}(0,1) \\
\epsilon_{2} &\thicksim \mathcal{N}(0,0.5) \\
\end{align*}

\subsection{DGP 14}
\begin{align*}
p &= \alpha_{0} + \alpha_{1}* \log(X_{8})+ \alpha_{2}* 1/\sqrt{X_{17}} + \alpha_{3}* \exp(X_{52}) + \alpha_{4}*X_{100}+ \alpha_{5}*|X_{157}| \\
&+ \alpha_{6}*|X_{157}|*\log(X_{8}) + \alpha_{7}*X_{162}^2 + \alpha_{8}*X_{165} + \alpha_{9}*X_{100}*X_{165}\\
 \overline{y_{1}} &= 0.75*A + \beta_{1} + \beta_{2}*\log(X_{8}) + \beta_{3}*\sqrt{X_{17}} + \beta_{4}*X_{52}+ \beta_{5}*|X_{157}| \\
 \overline{y_{2}} &= 10*A + | \beta_{6}* X_{162} + \beta_{7}*X_{166} + \beta_{8}*(X_{188}^2)*(X_{188} > 0.4)|\\
 Y &= \exp(\overline{y_{1}}) + \overline{y_{2}} + \epsilon\\
\epsilon &\thicksim \mathcal{N}(0,4) \\
\end{align*}

\subsection{DGP 15}
\begin{align*}
p &= \alpha_{0} + \alpha_{1}*(X_{10}-5)*(X_{10}>5)+ \alpha_{2}*\sqrt{X_{25}} + \alpha_{3}*(X_{10}-5)*(X_{10}>5)*\sqrt{X_{25}} \\
&+ \alpha_{4}*X_{42}+ \alpha_{5}*\log(X_{52}) + \alpha_{6}*|X_{70}|^{1/3} + \alpha_{7}*X_{96} \\
&+ \alpha_{8}*\log(X_{52})*X_{96} + \alpha_{9}*X_{158} +\alpha_{10}*X_{96}*X_{158} \\
&+ \alpha_{11}*X_{163}^2 + \alpha_{12}* |X_{169}| + \alpha_{13}*X_{188}^3 + \alpha_{14}*X_{192} + \alpha_{15}*X_{188}^3*X_{192}\\
A &\thicksim Bin(1, \textup{sig}(p)) \\
Y &= A*(-0.75 -0.25*X_{192} + 0.6*X_{51}) + \beta_{1} + \beta_{2}*\log(X_{7}+1) + \beta_{3}*(1/{X_{16}}) + \beta_{4}*\sqrt{X_{25}}\\
&+ \beta_{5}*\log(X_{52}) + \beta_{6}*X_{96}+ \beta_{7}*X_{96}/((\log(X_{7}+1)+0.5) \\
&+\beta_{8}*X_{158} + \beta_{9}*X_{163}^2 + \beta_{10}*X_{188}^2 + \beta_{11}*X_{192} + \beta_{12}*X_{96} \\
&+ \beta_{13}*X_{188}^3*X_{192} + \epsilon \\
\epsilon &=  (1-A)*\epsilon_{1} + A*\epsilon_{2}\\
\epsilon_{1} &\thicksim \mathcal{N}(0,0.05) \\
\epsilon_{2} &\thicksim \mathcal{N}(0,0.1) \\
\end{align*}

\subsection{DGP 16}
\begin{align*} 
p &= \alpha_{0} + \sum_{j = 1}^{200} \alpha_{j} * X_{j} ] \\
A &\thicksim Bin(1, \textup{sig}(p)) \\
Y &= (\theta*A) +\sum_{i = 1}^{200} \beta_{i} * X_{i} + \epsilon\\
\epsilon &\thicksim t(8) \\
\end{align*}
\end{document}